%% file: paper.tex
\documentclass{IEEEtran}

\usepackage{cite}
\usepackage{amsmath,amssymb,amsfonts}
\usepackage{algorithmic}
\usepackage{graphicx}
\usepackage{siunitx}
\usepackage{tikz}
\usepackage{cleveref}
\usepackage{pgf,tikz}
\usepackage{mathrsfs}
\usepackage[disable]{todonotes}
\usetikzlibrary{arrows}
\usepackage{pgfplotstable}
\usepackage{pgfplots}
\usetikzlibrary{pgfplots.groupplots}
\usepackage{booktabs}
\usepackage{rotating} 
\pgfplotsset{compat=1.16}

\usepackage{textcomp}
\usepackage{comment}
\usepackage{ulem}
\def\BibTeX{{\rm B\kern-.05em{\sc i\kern-.025em b}\kern-.08em
 T\kern-.1667em\lower.7ex\hbox{E}\kern-.125emX}}
\usepackage{subfig} 
\usepackage{multirow}
 
\newcommand{\td}[1]{\todo[inline,size=\small]{#1}} 
\newcommand{\tdb}[1]{\todo[inline,size=\small,backgroundcolor=blue!25]{#1}} 
\newcommand{\tdg}[1]{\todo[inline,size=\small,backgroundcolor=green!25]{#1}} 
\newcommand{\vt}[1]{\boldsymbol{#1}} 
\newcommand{\uv}[1]{\hat{\boldsymbol{#1}}} 
\newcommand{\mat}[1]{\boldsymbol{\mathsf{#1}}} 
\newcommand{\identity}[0]{\bf{I}} 
\newcommand{\operator}[1]{\mathcal{#1}} 
\newcommand{\transpose}[1]{{#1}^\mathrm{T}} 
\newcommand{\diff}{\,\mathrm{d}} 
\newcommand\underrel[2]{\mathrel{\mathop{#2}\limits_{#1}}}
\newcommand{\cond}[1]{\>\text{cond}\left(#1\right)}
\newcommand{\vr}{\vt{r}} 
\newcommand{\norm}[1]{\left\lVert#1\right\rVert}
\newcommand{\mi}{\mathrm{j}} 
\newcommand{\deltaKappa}{\mathcal{\delta \kappa}} 
\newcommand{\rvline}{\hspace*{-\arraycolsep}\vline\hspace*{-\arraycolsep}} 
\newcommand{\matN}[1]{\boldsymbol{\mathsf{\overline{\overline{#1}}}}} 
\newcommand{\matH}[1]{\hat{\mat{{#1}}}} 
\newcommand{\matO}{\mat{0}} 
\newcommand{\Ci}{\mathbb{C}} 
\newcommand{\R}{\mathbb{R}} 

\definecolor{ao}{rgb}{0.0, 0.0, 0.0} 
\definecolor{amethyst}{rgb}{0.0, 0.0, 0.0} 
\xdefinecolor{melon}{rgb}{0.0,0.0,0.0} 
\definecolor{alizarin}{rgb}{0.82,0.1,0.26} 
\definecolor{asparagus}{rgb}{0.53, 0.66, 0.42}

\begin{document}
\title{On a Low-Frequency and Contrast Stabilized Full-Wave Volume Integral Equation Solver for Lossy Media}
\author{Clément Henry,~\IEEEmembership{Student Member}, IEEE, Adrien Merlini,~\IEEEmembership{Member, IEEE}, Lyes Rahmouni,~\IEEEmembership{Member, IEEE}, \\ and Francesco P. Andriulli, ~\IEEEmembership{Senior Member, IEEE}}

\maketitle

\setlength{\abovedisplayskip}{4pt}
\setlength{\belowdisplayskip}{4pt}


\begin{abstract}
In this paper we present a new regularized electric flux volume integral equation (D-VIE) for modeling high-contrast conductive dielectric objects in a broad frequency range. This new formulation is particularly suitable for modeling biological tissues at low frequencies, as it is required by brain epileptogenic area imaging, but also at higher ones, as it is required by several applications including, but not limited to, transcranial magnetic and deep brain stimulation (TMS and DBS, respectively). When modeling inhomogeneous objects with high complex permittivities at low frequencies, the traditional D-VIE is ill-conditioned and suffers from numerical instabilities that result in slower convergence and in less accurate solutions. In this work we address these shortcomings by leveraging a new set of volume quasi-Helmholtz projectors. Their scaling by the material permittivity matrix allows for the re-balancing of the equation when applied to inhomogeneous scatterers and thereby makes the proposed method accurate and stable even for high complex permittivity objects until arbitrarily low frequencies. Numerical results, canonical and realistic, corroborate the theory and confirm the stability and the accuracy of this new method both in the quasi-static regime and at higher frequencies.
\end{abstract}

\begin{IEEEkeywords}
Volume integral equations, preconditioning, high-contrast (HC) conductive media, bio-electromagnetism.
\end{IEEEkeywords}


\section{Introduction}

The electromagnetic modeling of human tissues has numerous applications that include brain source localization \cite{source_loc_baillet}, dosimetry \cite{aguirre2012evaluation}, deep brain stimulation (DBS) \cite{KUNCEL20042431}, transcranial magnetic stimulation (TMS) \cite{barker1985non}, electric impedance tomography \cite{EIT}, and hyperthermic cancer therapy \cite{4121557}. All these procedures require an accurate modeling of the interactions between electromagnetic fields and the human body. Depending on the application, numerical solvers can be either full-wave solvers directly derived from Maxwell's equations or static solvers based on Poisson's equation, which are only valid in the quasi-static regime \cite{Bossetti_2007}. While static solvers are sufficient to model resistive effects occurring in biological tissues, full-wave solvers are required when capacitive, inductive, and propagation effects should also be taken into consideration \cite{Bossetti_2007, plonsey1967considerations}. For instance, full-wave models are needed when the source has a larger spectral content (e.g. the magnetic pulse emitted by TMS coils \cite{TMS_integral_equation} or the electrical current injected by an electrode in neurostimulation \cite{howell2015effects}). For these applications, the modeling should be done at various frequencies and, hence, the solver used should be able to perform accurately at arbitrary frequency.

Two main families of numerical solvers are widely employed for frequency domain bio-electromagnetic modeling: integral equation (IE) and differential equation solvers. In particular, integral equation solvers can be either surfacic (SIE) or volumic (VIE) depending on the nature of the tissue to model \cite{yla2014surface}. Although these solvers give rise to dense matrices, the higher computational cost incurred can be significantly reduced using acceleration techniques such as the fast multipole method (FMM) \cite{GREENGARD1997280} or the adaptive cross approximation (ACA) \cite{bebendorf2000approximation}. Moreover, since IE formulations automatically enforce radiation conditions, no discretization is required outside the object. Unfortunately, however, bio-electromagnetic modeling is a challenging task for solvers based on integral equations because of the high complex permittivity contrast between the different tissues and their background \cite{Gabriel_1996}. Indeed, both a high-contrast object and/or a low operating frequency introduce a severe ill-conditioning in the discretized IE operator, which yields a slower convergence and a loss of accuracy in the solution \cite{qH_Andriulli}. These two phenomena are often referred to as the low-frequency (LF) and the high-contrast (HC) breakdowns respectively \cite{lfBreakdown, vecchi, dezutter, budko}. 
In addition to the conditioning issue, the ill-scaling between the different components of the solution of the discretized system causes the latter to have fewer digits of accuracy due to finite machine precision \cite{chew_scalings}. 

The high-contrast breakdown in piecewise homogeneous scatterers has been cured in the case of the Poggio-Miller-Chang-Harrington-Wu-Tsai (PMCHWT) formulation \cite{NishimuraPMCHWT} and in a novel single-source integral equation \cite{dezutter} by leveraging the Calder\'on identities while the low-frequency breakdown of the PMCHWT was tackled by its preconditioning with quasi-Helmholtz projectors \cite{beghein2017low}. While these stabilized surface formulations have numerous advantages, they are limited to piecewise homogeneous models of biological tissues. Volume integral equations, instead, can model objects with a high degree of inhomogeneity. Unfortunately, as their surface counterparts, they suffer from the HC breakdown \cite{markkanen_contrast,MARKKANEN2016269, zouros,budko,COSTABEL2012193, van2008well, DOBBELAERE2015355} and fail to converge rapidly in applications with high-permittivity contrast scatterers. Another limitation of traditional VIE is that, even though they are immune from the low-frequency breakdown in purely dielectric objects \cite{chewVIE}, a frequency ill-scaling between the different parts of the VIE can occur when the object under study has a complex permittivity which depends on the frequency \cite{rubinacci}. Therefore, in these cases, the LF breakdown can be considered an intrinsic part of the HC breakdown in the VIE. These limitations prevent the standard volume formulations to perform well in realistic biomedical applications where the modeling of high-contrast conductive tissues from the quasi-static regime to the microwave regime is required. 

Regularization techniques have been introduced for solving the LF and the HC problems in the electric current VIE (J-VIE), the electric field VIE (E-VIE), and the electric flux VIE (D-VIE). The LF breakdown has been cured in the J-VIE for anisotropic and inhomogeneous scatterers using a Loop-Star-Facet decomposition for re-scaling properly the unknown \cite{forestiere}. However, solving this problem using a Helmholtz discretization deteriorates the dense discretization behavior of the VIE, i.e., it causes the conditioning of the system matrix to deteriorate when the average edge length $h$ of the discretized geometry decreases \cite{andriulli_multigrid}. Instead, the quasi-Helmholtz projectors \cite{qH_Andriulli} allow for the removal of the ill-scaling in the formulation while keeping its dense mesh behavior unchanged. These projectors have been adapted to the J-VIE and used for curing the HC limitations of this equation for isotropic inhomogeneous scatterers \cite{markkanen}. Another approach to solve the HC problem is presented in \cite{zouros}, where the E-VIE is regularized using symbol calculus and the Calder\'on identities. Its application to the J-VIE is discussed in \cite{POLIMERIDIS2014280}. While they are free from the HC breakdown, the two above-mentioned methods do not consider the numerical stability of the J-VIE or the E-VIE at low frequencies when modeling lossy dielectric objects, which is an important feature for a solver operating in low-frequency biomedical applications.
Regarding the D-VIE, an effective solution to both the LF and HC breakdowns has been proposed in \cite{michielssen, gomez_letter} where an additional surface integral equation is used to adjust the background permittivity and lower the dielectric contrast. This technique is quite effective when the main problem is the contrast between the background and the object but it does not remove directly the internal contrasts between the different media within the simulated object, and an ill-conditioning thus remains for inhomogeneous objects. 

The contribution of this paper is twofold, on the one hand we propose a new set of quasi-Helmholtz projectors that, differently from those proposed in the past, are the first of their kind to be oblique and to be particularly suited to manipulate the solenoidal and non-solenoidal parts the D-VIE. On the other hand, we leverage these new projectors to obtain a new regularized D-VIE which is immune from both ill-conditioning and the loss of accuracy occurring at low frequencies and in high-contrast objects. 
More specifically, the new projectors, built from a weighted graph Laplacian matrix, when combined with the appropriate re-scaling, allow for the re-balancing the D-VIE in both the LF and HC regimes. The regularized D-VIE obtained is free from both the HC and the LF breakdowns and exhibits a solution that is valid until arbitrarily low frequencies, unlike standard full-wave solvers. This versatility makes it an appropriate formulation for biomedical applications where solvers that can operate in a broad frequency range and in high-contrast objects are required. The reader should however note that, although in our numerical validations we focus on biomedical applications and brain modeling in particular, the solver we propose is a completely general purpose one and several other application scenarios such as the modeling of lossy interconnects in printed circuit boards \cite{rubinacci} could benefit from its use. Very preliminary results of this work have been presented in the conference contribution \cite{aps2020}.

This paper is organized as follows: in \Cref{sec:bg} we set the background and notation, including the definition of the D-VIE along with its discrete Helmholtz decomposition. The low-frequency and high-contrast behaviors of the D-VIE are analyzed in \Cref{sec:breakdown-analysis}. These analyses are followed by the presentation of a new set of quasi-Helmholtz projectors in \Cref{sec:oblique-projectors} which is then employed in \Cref{sec:regularized_eq} to regularize the D-VIE at low frequencies and for high-contrast lossy scatterers. Bounds for the condition number of this new regularized formulation are derived in \Cref{sec:bounds}. \Cref{sec:implementation} is dedicated to the computational considerations and the introduction of another effective scheme for regularizing the D-VIE and \Cref{sec:results} presents numerical examples demonstrating the stability and accuracy of these new formulations in a broad frequency range and for high-contrast media. Finally, for the sake of readability, we have omitted some of the mathematical technicalities in the main text. However, the interested reader will find them in the Appendices.


\section{Notation and Background} \label{sec:bg}

Let $\Omega \subset \mathbb{R}^3$ be a simply connected object composed of a lossy dielectric and illuminated by a time-harmonic incident electric field $\vt{E}_i$ in a background of permittivity $\epsilon_0$ and permeability $\mu_0$. The scatterer is characterized by its complex isotropic relative permittivity $\epsilon_r(\vt{r}) = \epsilon_r'(\vt{r}) - \mi {\sigma(\vt{r})}/{\omega \epsilon_0}$ where $\vr \in \Omega$, $\epsilon_r'(\vr)$ is the relative permittivity, $\sigma(\vr)$ is the conductivity, and $\omega$ is the angular frequency of $\vt{E}_i$. The permeability of the scatterer is further assumed to be the permeability of vacuum $\mu_0$. Leveraging the volume equivalence principle, the scatterer can be substituted by a volume current density distribution
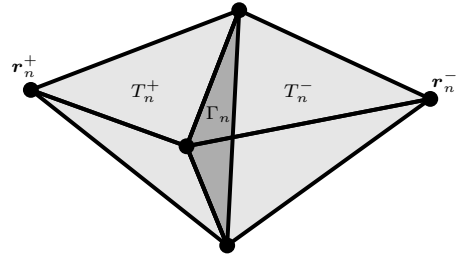
\begin{figure}
 \centering
	 \input{figures/geogebra-swg2.tex}
	 \caption{\label{fig:swg} Convention used to define the SWG and the star functions: the SWG basis function is defined on the two tetrahedra $T_n^+$ and $T_n^-$ that are formed with their common face $\Gamma_n$ and the vertices $\vr_n^+$ and $\vr_n^-$, respectively. The half SWG basis functions are defined on $T_n^+$ only.}
\end{figure}	
$ \vt{J}(\vt{r}) = \mi \omega \kappa(\vt{r}) \vt{D}(\vt{r})$
where $\kappa(\vt{r}) = ({\epsilon}(\vt{r}) - \epsilon_0)/{{\epsilon}(\vt{r})}$ is the dielectric contrast and $\vt{D}$ is the electric flux density. The D-VIE which relates $\vt{E}_i$ and $\vt{D}$ is expressed as \cite{VIE_cohen}
\begin{equation}\label{eq:VIE}
  \frac{\vt{D}(\vr)}{\epsilon(\vr)} - \frac{k_0^2}{\epsilon_0} \left(\operator{T}_A \vt{D}\right)(\vr) - \frac{1}{\epsilon_0} \left(\operator{T}_\Phi \vt{D}\right)(\vr) = \vt{E}_i(\vr)\,,\quad \vr \in \Omega,
\end{equation}
where the vector potential $\operator{T}_A$ and the scalar potential $\operator{T}_\Phi$ are defined as
\begin{gather} 
  \left(\operator{T}_A \vt{D}\right)(\vr) = \int_\Omega G_0(\vr,\vr') \kappa(\vr') \vt{D}(\vr') \diff v' \label{eq:operators_DVIE_Ta}\,,\\
  \left(\operator{T}_\Phi \vt{D}\right)(\vr) = \nabla \int_\Omega G_0(\vr,\vr') \nabla' \cdot \left( \kappa(\vr') \vt{D}(\vr') \right) \diff v'\,, \label{eq:operators_DVIE_Tphi}
\end{gather}
in which $G_0(\vr,\vr') = \exp{(-\mi k_0 |\vr - \vr'|)}/|\vr - \vr'|$ is the free space 3D Green's function and $k_0 = \omega / c_0$ is the wavenumber of $\vt{E}_i$ in free space. 

In general, the electric field radiated by an electric current distribution $\vt{J}_i$, which is commonly used to model the brain's electric activity, is
\begin{equation} \label{eq:dipole_field}
  \vt{E}_i(\vr) = \left(\operator{T}_\Phi' \vt{J}_i\right)(\vr) + \left(\operator{T}_A' \vt{J}_i\right)(\vr)\,,
\end{equation}
where $\operator{T}_A' \vt{J}_i = -\mi k_0 \eta_0 \int_\Omega G_0(\vr,\vr') \vt{J}_i(\vr') \diff v'$ and $\operator{T}_\Phi' \vt{J}_i = {\eta_0}/{\mi k_0} \nabla \int_\Omega G_0(\vr,\vr') \nabla' \cdot \vt{J}_i(\vr') \diff v'$.

Equation \eqref{eq:VIE} is numerically solved by applying a Galerkin approach on a tetrahedral discretization of the geometry. Because the unknown $\vt{D}$ must exhibit a continuous normal component through material discontinuities, it is discretized with SWG basis functions $\{\vt{f_n}\}$ \cite{swg}. The SWG function associated to the face $\Gamma_n$ is defined as
\begin{equation} \label{eq:refSWGbg}
	\vt{f_n}(\vr) = 
	\begin{cases}
		\phantom{-}\frac{1}{3 V_n^+} (\vr - \vt{r_{n}^+}), \quad \vr \in T_n^+ \\
		-\frac{1}{3 V_n^-} (\vr -\vt{r_{n}^-}), \quad \vr \in T_n^-
	\end{cases}\,,
\end{equation}
where we have used the notations of Fig.~\ref{fig:swg} and where $V_n^\pm$ is the volume of the tetrahedron $T_n^\pm$.
\tdg{AM: add the definition of the SWGs?}
\tdg{AM: @ clement please check the modifs above}
Using these functions, the unknown electric flux can be expanded as
$
  \vt{D}(\vt{r}) \approx \sum_{n=1}^{N_F} [\mat{\alpha}]_n \vt{f_n}(\vt{r})
$
in which $N_F$ is the number of faces in the discretized geometry. The set of source functions $\{\vt{f_n}\}$ also contains the half SWG basis functions (supported by a single tetrahedron) that are defined at the boundary of the object to properly model the surface charges. The resulting equation, tested with the functions $\{ \vt{f_m}\}$ yields the linear system
\begin{gather} 
  \mat{Z} \mat{\alpha} = \left( \mat{G_{\epsilon}} + \mat{Z_A} + \mat{Z_\Phi} \right) \mat{\alpha} = \mat{v}\,, \text{ where}
  \label{eq:discretized_eq} \\
   \left[\mat{G_{\epsilon}}\right]_{mn} = \langle \vt{f_m},\vt{f_n} / \epsilon \rangle_\Omega \label{eq:def_gram}\,,\\
  \left[ \mat{Z_A} \right]_{mn} = - {k_0^2}{\epsilon_0}^{-1} \langle \vt{f_m},\operator{T}_A \vt{f_n} \rangle_\Omega \label{def_vec_pot}\,,\\
  \left[ \mat{Z_\Phi} \right]_{mn} = - {\epsilon_0}^{-1} \langle \vt{f_m},\operator{T}_\Phi \vt{f_n} \rangle_\Omega\,, 
\end{gather}
and $\left[ \mat{v} \right]_{m} = \langle \vt{f_m},\vt{E}_i \rangle_\Omega$ with $\langle \vt{a} , \vt{b} \rangle_\Omega = \int_\Omega \vt{a} \cdot \vt{b} \diff v$. 

In the following, for the sake of completeness, we briefly review some results on the discrete Helmholtz decomposition of the electric flux in the D-VIE. After being expanded with divergence conforming functions, the electric flux can be further decomposed into a linear combination of solenoidal (i.e. divergence free) and non-solenoidal (i.e. non-divergence free) components \cite{MARKKANEN2016269}. This decomposition is crucial because the solenoidal and non-solenoidal components of the D-VIE behave differently with respect to the frequency or the permittivity, hence they will have to be separated in the analysis and the regularization of the D-VIE. Such a separation can be performed using a Loop-Star decomposition on the discretized D-VIE, similarly to what is done in surface formulations \cite{vecchi}.
\tdg{AM : explain what loops and stars are. Also, the notion of scaling of loop star is not clear and not defined.}
\tdg{AM: @clement check modifs + what do you mean decomposing the solution? You need to decompose the whole system}
The unknown $\mat{\alpha}$ in \eqref{eq:discretized_eq} can thus be decomposed into a sum of its Loop (solenoidal) and Star (non-solenoidal) components such that
$
  \mat{\alpha} = \mat{\Lambda} \mat{l} + \mat{\Sigma} \mat{s}
$ \cite{MARKKANEN2016269}
where $\mat{\Lambda} \in \R^{N_F \times N_L}$ is the loop-to-SWG transformation matrix, $\mat{\Sigma} \in \R^{N_F \times N_S}$ is the star-to-SWG transformation matrix, $\mat{l}$ are the expansion coefficients of the unknown in the solenoidal basis, and $\mat{s}$ are the expansion coefficients in the non-solenoidal basis. The dimensions $N_L$ and $N_S$ are the numbers of independent loops and stars in the discretized geometry, respectively. In a tetrahedral discretization of a simply connected object, the number of stars and loops are \cite{li2006applying,euler_volume,MARKKANEN2016269}
\begin{equation} 
   N_S = N_T + N_{eF} ;~
   N_L = N_{iE}- N_{iV}\,,
\end{equation}
where $N_T$, $N_{eF}$, $N_{iE}$, and $N_{iV}$ are the number of tetrahedra, external faces, internal edges, and internal vertices, respectively.
The mapping $\mat{\Sigma_v}\in \R^{N_F \times N_T}$ relates the stars and the SWG functions as
\begin{equation} 
  \left[\mat{\Sigma_v}\right]_{mn} =
  \begin{cases}
   \phantom{-}1 & \text{ if tetrahedron } n \text{ is tetrahedron } T_m^+ \\
   -1 & \text{ if tetrahedron } n \text{ is tetrahedron } T_m^-\\
   \phantom{-}0 & \text{ otherwise,}
  \end{cases}
  \label{eq:def_stars}
\end{equation}
where $T_m^+$ and $T_m^-$ represent the two tetrahedra on which the basis function $\vt{f_m}$ is defined (see Fig.~\ref{fig:swg}). For half basis functions, only the entries corresponding to tetrahedron $T_m^+$ are filled in $\mat{\Sigma_v}$. Since these specific basis functions also model surface charges at the boundary of the object, another transformation matrix $\mat{\Sigma_s}\in \R^{N_F \times N_{eF}}$ needs to be filled for these basis functions
\begin{equation} 
  \left[\mat{\Sigma_s}\right]_{mn} = 
  \begin{cases} -1 & \text{ if face } n \text{ is face } \Gamma_m \\ 
  \phantom{-}0 & \text{ otherwise} 
  \end{cases}\,.
  \label{eq:def_stars_hbf}
\end{equation}
Note here that, by convention, the half basis function $\vt{f_m}$ is always supported by tetrahedron $T_m^+$. 
The transformation matrix from star-to-SWG is then defined as $\mat{\Sigma} = \left[ \mat{\Sigma_v}~\mat{\Sigma_s} \right]$. After eliminating one column from $\mat{\Sigma}$ (zero total charge in $\Omega$) we obtain the full-column-rank matrix $\mat{\tilde{\Sigma}}$. The Loop functions are defined on the edges of the mesh as linear combinations of SWG basis functions \cite{li2006applying}. Although schemes for the creation of an independent set of loops in a tetrahedral mesh exist \cite{li2006applying, mendes_loops}, we do not build them explicitly here, they are only introduced for supporting the discussion. For the purpose of this paper, we will employ the properties $\mat{\Lambda}^\mathrm{T} \mat{\tilde{\Sigma}} = \mat{0}$ and $\mat{\tilde{\Sigma}}^\mathrm{T} \mat{\Lambda} = \mat{0}$, that $\mat{\Lambda}$ and $\mat{\tilde{\Sigma}}$ satisfy by construction (refer to \cite{bossavit1999computational} and references therein).

\section{Low-frequency and High-contrast Analyses of the D-VIE} \label{sec:breakdown-analysis}
In this section we will first present an analysis of the low-frequency problems of the D-VIE. This will then be followed by an analysis of the high-contrast problems of the D-VIE. Leveraging these analyses and the nature of the criticalities observed, a cure for both problems will then be presented in \Cref{sec:regularized_eq}.

\subsection{Low-frequency Analysis of the D-VIE} \label{sec:lf_part}

We analyze here the low-frequency behavior of the D-VIE when modeling lossy dielectric objects surrounded by free space. At low frequencies VIEs are subject to the low-frequency breakdown due to the frequency dependence of the complex permittivity of the object \cite{rubinacci}. In this subsection we will confirm this with an analysis coherent with our framework that will then be used in the next section to construct the new formulation we propose in this work.
Here we will also show that for the discretized D-VIE, the equation ill-scaling not only causes the ill-conditioning of the associated linear system matrix, but also the loss of significant digits in the solution coefficients. 
Our strategy will be to expose the low-frequency behavior of the D-VIE through a suitably normalized Loop-Star decomposition. 

We propose here a new generalization of the Loop-Star decomposition, that differently from the standard decomposition, is not coefficient orthogonal, but rather an oblique decomposition. In its normalized form, it is defined as follows
\begin{equation} \label{eq:normDecomposition}
	\mat{B^{A^{-1}}_{\Lambda \Sigma}} = \left[ \mat{\Lambda} (\transpose{\mat{\Lambda}} \mat{\Lambda})^{-\frac{1}{2}} ,	\mat{A} \mat{\tilde{\Sigma}} (\transpose{\mat{\tilde{\Sigma}}} \mat{A}^{2}\mat{\tilde{\Sigma}})^{-\frac{1}{2}}\right]\,,
\end{equation}
\tdg{You do not explain why this is a decomposition: the proof for invertibility of this matrix works for real matrices only}
\tdg{AM: the usage of U like this is a bit sloppy, and you talk about 1/2 and you have only -1/2}
 where $\mat{A}$ is an invertible real symmetric matrix and for any invertible matrix $\mat{U}$, $\mat{U}^{-\frac{1}{2}}$ is the inverse of one of the (potentially many) square root matrices of $\mat{U}$. The fact that $\mat{B^{A^{-1}}_{\Lambda \Sigma}}$ is invertible is proven in Appendix~\ref{app:appB}. The reader should note that, due to the presence of the matrix square roots, the above decomposition is quite inefficient to implement. In this work, however, the decomposition will only be used as a theoretical tool in the conditioning analysis and, subsequently, as the starting point for defining our new oblique projectors. In other words, the computation of \eqref{eq:normDecomposition} will never be required for the methods proposed in this paper and thus the often inefficient-to-compute matrix square roots will have no computational impact.
Since the matrix \eqref{eq:normDecomposition} will be used to study the low-frequency limit, it must have a well-defined static limit. This is obtained by selecting $\mat{A}^{-1} = \mat{\tilde{G}_{\epsilon}}$ with $ \mat{\tilde{G}_\epsilon}= \mat{G_\epsilon}/\left(\mi \omega \right)$. It should also be noted that $\mat{B^{\tilde{G}_\epsilon}_{\Lambda \Sigma}}$ block-diagonalizes the D-VIE Gram matrix $\mat{G_\epsilon}$. In fact, since $(\transpose{\mat{\Lambda}} \mat{\Lambda})^{-\frac{1}{2}} \transpose{\mat{\Lambda}}
j\omega\mat{\tilde{G}_{\epsilon}}\mat{\tilde{G}_\epsilon}^{-1} \mat{\tilde{\Sigma}} (\transpose{\mat{\tilde{\Sigma}}} \mat{\tilde{G}_\epsilon}^{-2}\mat{\tilde{\Sigma}})^{-\frac{1}{2}}=\matO$ (because $\transpose{\mat{\Lambda}}\mat{\tilde{\Sigma}}=\matO$), we have
\begin{equation} \label{eq:identity_scaled_ls}
		\transpose{\mat{B^{\tilde{G}_\epsilon}_{\Lambda \Sigma}}} \mat{{G}_{\epsilon}} \mat{B^{\tilde{G}_\epsilon}_{\Lambda \Sigma}} 
			=\begin{bmatrix}
		\transpose{\matN{\Lambda}}\mat{{G}_{\epsilon}}\matN{\Lambda}&\matO \\
			\matO &\transpose{\matN{\Sigma}} \mat{{G}_{\epsilon}}\matN{\Sigma}
		\end{bmatrix}\,,
	\end{equation}
where $\matN{\Lambda} = \mat{\Lambda} (\transpose{\mat{\Lambda}} \mat{\Lambda})^{-\frac{1}{2}}$ and 
$\matN{\Sigma} = \mat{\tilde{G}_\epsilon}^{-1} \mat{\tilde{\Sigma}} (\transpose{\mat{\tilde{\Sigma}}} \mat{\tilde{G}_\epsilon}^{-2}\mat{\tilde{\Sigma}})^{-\frac{1}{2}}$. Using $\mat{B^{\tilde{G}_{{\epsilon}}}_{\Lambda \Sigma}}$, the Loop-Star decomposition of the D-VIE can be expressed as
\begin{equation} \label{eq:decompositio}
\mat{Z_{\Lambda \Sigma}} \mat{\alpha_{\Lambda \Sigma}} =\transpose{\mat{B^{\tilde{G}_{{\epsilon}}}_{\Lambda \Sigma}}}(\mat{G_{\epsilon}} + \mat{Z_A} + \mat{Z_{\Phi}})\mat{B^{\tilde{G}_{{\epsilon}}}_{\Lambda \Sigma}} \mat{\alpha_{\Lambda \Sigma}}\,,
\end{equation}
where $\mat{B^{\tilde{G}_{{\epsilon}}}_{\Lambda \Sigma}} \mat{\alpha_{\Lambda \Sigma}} = \mat{\alpha}$. We further decompose $\mat{Z_{\Phi}}$ as
$
\mat{Z_{\Phi}} = \mat{Z_{\Phi, 11}} + \mat{Z_{\Phi,1 \epsilon}}
$
in which $\mat{Z_{\Phi, 11}}$ only accounts for the volumic contributions in the object and the surfacic contributions on its boundary ($\partial \Omega$)
\begin{equation} \label{eq:Z11-ns}
\begin{split}
	&[\mat{Z_{\Phi,11}}]_{mn} =  \epsilon_0^{-1} . \\
	&\left[ \int_\Omega \nabla \cdot \vt{f_m}(\vt{r}) \int_\Omega G_0(\vr,\vr') \kappa(\vr') \nabla \cdot \vt{f_n}(\vt{r'}) \diff v' \diff v \right. \\
	&\left. - \int_\Omega \nabla \cdot \vt{f_m}(\vt{r}) \int_{\partial \Omega} G_0(\vr,\vr') \kappa^+_n \uv{n}_n \cdot \vt{f_n}(\vt{r'}) \diff s' \diff v \right.\\ 
	&\left. - \int_{\partial \Omega} \uv{n}_m \cdot \vt{f_m}(\vt{r}) \int_{\Omega} G_0(\vr,\vr') \kappa(\vr')\nabla \cdot \vt{f_n}(\vt{r'}) \diff v' \diff s \right.\\
	&\left.+ \int_{\partial \Omega} \uv{n}_m \cdot \vt{f_m}(\vt{r}) \int_{\partial \Omega} G_0(\vr,\vr') \kappa^+_n \uv{n}_n \cdot \vt{f_n}(\vt{r'}) \diff s' \diff s \right]\,,
\end{split}
\end{equation}
with $\kappa^+_n$ being the dielectric contrast in tetrahedron $T^+_n$ and $\uv{n}_n$ the outward unit normal of the triangle of $T^+_n$ which pertains to $\partial \Omega$; instead $\mat{Z_{\Phi, 1 \epsilon}}$ includes the purely surfacic contributions internal to the object (at the interface $\Gamma_n$ between two tetrahedra $T^-_n$ and $T^+_n$ of different material contrasts $\kappa^-_n$ and $\kappa^+_n$)
\begin{equation} \label{eq:Z1eps-ns}
\begin{split}
	&[\mat{Z_{\Phi,1\epsilon}}]_{mn} = \epsilon_0^{-1} . \\
	&\left[\int_\Omega \nabla \cdot \vt{f_m}(\vt{r}) \int_{\Gamma_n} G_0(\vr,\vr') \deltaKappa_n \uv{n}_n \cdot \vt{f_n}(\vt{r'}) \diff s' \diff v \right.\\
	&\left.- \int_{\partial \Omega} \uv{n}_m \cdot \vt{f_m}(\vt{r}) \int_{\Gamma_n} G_0(\vr,\vr') \deltaKappa_n \uv{n}_n \cdot \vt{f_n}(\vt{r'}) \diff s' \diff s \right]\,,
\end{split}
\end{equation}
in which $\deltaKappa_n = \kappa^-_n - \kappa^+_n$ and $\uv{n}_n$ is a unit vector normal to $\Gamma_n$ oriented from $T^+_n$ to $T^-_n$.
When applying $\matN{\Lambda}^\mathrm{T}$ and $\matN{\Lambda}$ to the scalar potential matrices, we obtain the relations
$ \matN{\Lambda}^\mathrm{T} \mat{Z_{\Phi, 11}}= \matO$, $\matN{\Lambda}^\mathrm{T} \mat{Z_{\Phi, 1\epsilon}}= \matO$, and $\mat{Z_{\Phi, 11}} \matN{\Lambda} = \matO$.
The fact that $\mat{\Lambda}$ (and thus $\matN{\Lambda}$) cancels the surface terms defined on $\partial \Omega$ in $\mat{Z_{\Phi, 11}}$ and $\mat{Z_{\Phi, 1\epsilon}}$ results from the absence of Loop functions on $\partial \Omega$. This property is proven in Appendix \ref{app:appC}. 

We then represent $\mat{Z_{\Lambda \Sigma}}$ in \eqref{eq:decompositio} as a 2-by-2 block matrix
\begin{gather} 
	\mat{Z_{\Lambda \Sigma}}= \transpose{\mat{B^{\tilde{G}_\epsilon}_{\Lambda \Sigma}}} \mat{Z} \mat{B^{\tilde{G}_\epsilon}_{\Lambda \Sigma}}
	= \begin{bmatrix}
		\matN{\Lambda}^\mathrm{T} \mat{Z} \matN{\Lambda} & \matN{\Lambda}^\mathrm{T} \mat{Z} \matN{\Sigma}\\
		 \matN{\Sigma}^\mathrm{T}\mat{Z} \matN{\Lambda} & \matN{\Sigma}^\mathrm{T} \mat{Z} \matN{\Sigma}
	\end{bmatrix}\,, \text{ in which}\label{eq:block_mat_loop-star-ns} \\
\matN{\Lambda}^\mathrm{T} \mat{Z} \matN{\Lambda} = \matN{\Lambda}^\mathrm{T} (\mat{G_{\epsilon}} + \mat{Z_A})\matN{\Lambda} \label{eq:decomp_LL_ns}\,,\\
\matN{\Lambda}^\mathrm{T} \mat{Z}\matN{\Sigma} = \matN{\Lambda}^\mathrm{T} \mat{Z_A} \matN{\Sigma} \label{eq:decomp_LS_ns}\,,\\
 \matN{\Sigma}^\mathrm{T} \mat{Z} \matN{\Lambda} = \matN{\Sigma}^\mathrm{T} (\mat{Z_A} + \mat{Z_{\Phi,1 \epsilon}}) \matN{\Lambda} \label{eq:decomp_SL_ns}\,,\\
 \matN{\Sigma}^\mathrm{T} \mat{Z} \matN{\Sigma} = \matN{\Sigma}^\mathrm{T} (\mat{G_{\epsilon}} + \mat{Z_A} + \mat{Z_{\Phi}}) \matN{\Sigma}\,.
\label{eq:decomp_SS_ns}
\end{gather}
To identity the frequency behavior of the terms in \eqref{eq:decomp_LL_ns}, \eqref{eq:decomp_LS_ns}, \eqref{eq:decomp_SL_ns}, and \eqref{eq:decomp_SS_ns}, we first need to introduce the low-frequency behavior of the real and imaginary parts of the material parameters $\epsilon$, $\kappa$, and $\deltaKappa$ 
\tdg{AM: introduce the notation $\Re$ and $\Im$?}
\begin{gather}
\Re(\epsilon) \underrel{\omega \to 0}{=} \mathcal{O}(1), ~ 
\Im(\epsilon) \underrel{\omega \to 0}{=} \mathcal{O}(1/\omega) \label{eq:eps_scaling}\,,\\
\Re(1/\epsilon) \underrel{\omega \to 0}{=} \mathcal{O}(\omega^2), ~ 
\Im(1/\epsilon) \underrel{\omega \to 0}{=} \mathcal{O}(\omega) \label{eq:eps_inv_scaling}\,,\\
\Re(\kappa) \underrel{\omega \to 0}{=} \mathcal{O}(1), ~ 
\Im(\kappa) \underrel{\omega \to 0}{=} \mathcal{O}(\omega) \label{eq:kappa_scaling}\,,\\
\Re(\deltaKappa) \underrel{\omega \to 0}{=} \mathcal{O}(\omega^2), ~ 
\Im(\deltaKappa)\underrel{\omega \to 0}{=} \mathcal{O}(\omega)\,, \label{eq:delta_kappa_scaling}
\end{gather}
which derives from the definitions of the complex permittivity and the dielectric contrast. Then, using the definitions of $\mat{G_\epsilon}$, $\mat{Z_A}$, $\mat{Z_{\Phi,11}}$, and $\mat{Z_{\Phi,1\epsilon}}$ in \eqref{eq:def_gram}, \eqref{def_vec_pot}, \eqref{eq:operators_DVIE_Ta}, \eqref{eq:Z11-ns}, and \eqref{eq:Z1eps-ns}, we can deduce the frequency scalings of these matrices from \eqref{eq:eps_scaling}, \eqref{eq:eps_inv_scaling}, \eqref{eq:kappa_scaling}, and \eqref{eq:delta_kappa_scaling}. We finally obtain the following low-frequency behavior for the real and imaginary parts of $\mat{G_\epsilon}$, $\mat{Z_A}$, $\mat{Z_{\Phi,11}}$, and $\mat{Z_{\Phi,1\epsilon}}$
\tdg{these are not the decomposition terms since you do not have $\Lambda$ and $\Sigma$}
\begin{gather}
\Re(\mat{G_{\epsilon}}) \underrel{\omega \to 0}{=} \mathcal{O}(\omega^2), ~ 
\Im(\mat{G_{\epsilon}}) \underrel{\omega \to 0}{=} \mathcal{O}(\omega) \label{eq:scaling-gram-ns}\,,\\
\Re(\mat{Z_A}) \underrel{\omega \to 0}{=} \mathcal{O}(\omega^2), ~ 
\Im(\mat{Z_A}) \underrel{\omega \to 0}{=} \mathcal{O}(\omega^3) \label{eq:scaling-A-ns}\,,\\\
\Re(\mat{Z_{\Phi,11}}) \underrel{\omega \to 0}{=} \mathcal{O}(1), ~ 
\Im(\mat{Z_{\Phi,11}}) \underrel{\omega \to 0}{=} \mathcal{O}(\omega) \label{eq:scaling-11-ns}\,,\\
\Re(\mat{Z_{\Phi,1 \epsilon}}) \underrel{\omega \to 0}{=} \mathcal{O}(\omega^2), ~ 
\Im(\mat{Z_{\Phi,1 \epsilon}}) \underrel{\omega \to 0}{=} \mathcal{O}(\omega)\,. \label{eq:scaling-1eps-ns}
\end{gather}
Using \eqref{eq:scaling-gram-ns} to \eqref{eq:scaling-1eps-ns}, we deduce the scalings for $\mat{Z_{\Lambda \Sigma}}$ at low frequencies
\begin{gather} 
\Re\left(\mat{Z_{\Lambda \Sigma}} \right) \underrel{\omega \to 0}{=} \begin{bmatrix}
	\mathcal{O}(\omega^2) &\mathcal{O}(\omega^2) \\
	\mathcal{O}(\omega^2) & \mathcal{O}(1)
\end{bmatrix} \label{eq:scaling_mat_real_ns}\,,\\
\Im\left( \mat{Z_{\Lambda \Sigma}} \right) \underrel{\omega \to 0}{=} \begin{bmatrix}
	\mathcal{O}(\omega) &\mathcal{O}(\omega^3) \\
	\mathcal{O}(\omega) & \mathcal{O}(\omega)
\end{bmatrix} \label{eq:scaling_mat_imag_ns}\,,
\end{gather}
which, according to the Gershgorin circle theorem, confirm that $\mat{Z_{\Lambda \Sigma}}$ is ill-conditioned for $\omega\to 0$. Since 
$\mat{B^{\tilde{G}_{{\epsilon}}}_{\Lambda \Sigma}}$ is a well-conditioned matrix (refer to Appendix~\ref{app:appB}), the ill-conditioning of $\mat{Z_{\Lambda \Sigma}}$ implies the ill-conditioning of $\mat{Z}$ for $\omega\to 0$. 
Besides the conditioning of the system matrix, it is also important to determine whether or not the solution coefficients and the right hand side vectors are preserved in the static limit. We first provide the scalings of the right hand side for plane wave and dipole excitations, which are frequently employed in bioelectromagnetic applications. For a plane wave excitation, the Loop and Star components of the right hand side $\mat{v_{\textnormal{{PW}}}}$ have the following scalings when $\omega\rightarrow 0$
\begin{equation} \label{eq:scaling_rhs_pw_ns}
\Re \left(\transpose{\mat{B^{\tilde{G}_\epsilon}_{\Lambda \Sigma}}} \mat{v_{\textnormal{{PW}}}}\right) \underrel{\omega \to 0}{=} \begin{bmatrix}
	\mathcal{O}(\omega^2) \\
	\mathcal{O}(1) 
\end{bmatrix};
\Im \left(\transpose{\mat{B^{\tilde{G}_\epsilon}_{\Lambda \Sigma}}} \mat{v_{\textnormal{{PW}}}}\right) \underrel{\omega \to 0}{=} \begin{bmatrix}
	\mathcal{O}(\omega) \\
	\mathcal{O}(\omega) 
\end{bmatrix},
\end{equation}
\tdg{AM: In 3d, the integration of a loop is still 0, correct?}
in which $\left[ \mat{v_{\textnormal{PW}}} \right]_{m} = \int_\Omega \vt{f_m}(\vr) \cdot \vt{E}_0 \exp{(-\mi \vt{k} \cdot \vr )} \diff v$ with $\vt{E}_0$ being the polarization of the plane wave and $\vt{k}$ its wave vector. Note that the right hand side in the D-VIE for a plane wave excitation scales similarly to the plane wave right hand side of the surface electric field integral equation \cite{chew_scalings}. To derive the frequency dependence of the Loop-Star decomposition of a dipole excitation, we start from the expression of the field it radiates \eqref{eq:dipole_field} in which $\vt{J}_i(\vr, \vr_0) = \mi \omega \delta(\vr - \vr_0)\vt{p}$ is the current dipole with $\delta$, $\vt{p}$, and $\vt{r_0}$ being the Dirac delta function, the dipole moment, and the dipole position, respectively. We then test \eqref{eq:dipole_field} with $\{\vt{f_m}\}$ to obtain the discretized right hand side
\begin{equation}
\begin{split}
	\left[ \mat{v_{\textnormal{dip}}} \right]_{m} &= \left[ \mat{v_{\mathrm{A}}} \right]_{m} + \left[ \mat{v_{\mathrm{\Phi}}} \right]_{m}\\
	&= \frac{k_0^2}{\epsilon_0} \vt{p} \cdot \int_\Omega \vt{f_m}(\vr) G_0(\vr, \vr_0) \diff v\\
	&-\frac{1}{ \epsilon_0}\int_{\Omega} \nabla \cdot \vt{f_m}(\vr) \vt{p} \cdot \nabla G_0(\vr, \vr_0) \diff v\,.
\end{split}
\end{equation}
From the Taylor expansion of $G_0$ and its gradient, we obtain that $\Re(G_0(\vr,\vr_0)) = \mathcal{O}(1)$, $\Im(G_0(\vr,\vr_0)) = \mathcal{O}(\omega)$, $\Re(\nabla G_0(\vr,\vr_0)) = \mathcal{O}(1)$, and $\Im(\nabla G_0(\vr,\vr_0)) = \mathcal{O}(\omega^3)$ at low frequencies. This yields the following frequency dependencies for the real and imaginary parts of the Loop-Star decomposition of $\mat{v_{\mathrm{A}}}$ and $\mat{v_{\mathrm{\Phi}}}$
\begin{gather}
\Re(\matN{\Lambda}^\mathrm{T} \mat{v_{\mathrm{A}}}) \underrel{\omega \to 0}{=} \mathcal{O}(\omega^2), ~ 
\Im(\matN{\Lambda}^\mathrm{T} \mat{v_{\mathrm{A}}}) \underrel{\omega \to 0}{=} \mathcal{O}(\omega^5)\,,\\
\Re(\matN{\Sigma}^\mathrm{T} \mat{v_{\mathrm{A}}}) \underrel{\omega \to 0}{=} \mathcal{O}(\omega^2), ~ 
\Im(\matN{\Sigma}^\mathrm{T} \mat{v_{\mathrm{A}}}) \underrel{\omega \to 0}{=} \mathcal{O}(\omega^3)\,,\\
\Re(\matN{\Lambda}^\mathrm{T} \mat{v_{\mathrm{\Phi}}}) = 0, ~
\Im(\matN{\Lambda}^\mathrm{T} \mat{v_{\mathrm{\Phi}}}) = 0\,,\\
\Re(\matN{\Sigma}^\mathrm{T} \mat{v_{\mathrm{\Phi}}} ) \underrel{\omega \to 0}{=} \mathcal{O}(1), ~ 
\Im(\matN{\Sigma}^\mathrm{T} \mat{v_{\mathrm{\Phi}}} ) \underrel{\omega \to 0}{=} \mathcal{O}(\omega^3)\,,
\end{gather}
\tdg{AM: $\matN{\Lambda}^\mathrm{T} \mat{v_{\mathrm{\Phi}}}$ is not an exact 0? Why the limit?}
from which we obtain the scalings of $\mat{v_{\mathrm{dip}}}$ at low frequencies
\begin{equation} \label{eq:scaling_rhs_dipole_ns}
\Re \left(\transpose{\mat{\mat{B^{\tilde{G}_\epsilon}_{\Lambda \Sigma}}}} \mat{v_{\textnormal{dip}}}\right) \underrel{\omega \to 0}{=} \begin{bmatrix}
	\mathcal{O}(\omega^2) \\
	\mathcal{O}(1) 
\end{bmatrix};
\Im \left(\transpose{\mat{\mat{B^{\tilde{G}_\epsilon}_{\Lambda \Sigma}}}} \mat{v_{\textnormal{dip}}}\right) \underrel{\omega \to 0}{=} \begin{bmatrix}
	\mathcal{O}(\omega^5) \\
	\mathcal{O}(\omega^3) 
\end{bmatrix}.
\end{equation}
This concludes the scaling analysis of the right hand side for plane wave and dipole excitations. The scalings obtained are summarized in \Cref{tab:table_v}, in which $\mat{v_\Lambda}$ denotes the Loop part of the right hand side vector and $\mat{v_\Sigma}$ its Star part. In the Loop-Star D-VIE, all the terms in \eqref{eq:scaling_rhs_dipole_ns} are preserved since they are stored separately. However, in the standard D-VIE, the terms recovered correspond only to the dominant terms (real and imaginary) in \eqref{eq:scaling_rhs_dipole_ns}, all the other terms are lost due to finite precision arithmetic.
\tdg{AM: this does not sound correct, if the scalings are not the same, the dominant one remains}
\tdg{AM: You should reming the reader that this will only happen if you do not use loop-star} 
To identify the impact of this loss on the solution coefficients, we subsequently retrieve the scalings of $\mat{\alpha_{\Lambda \Sigma}}$ for plane wave and dipole excitations. This requires the knowledge of the frequency scalings of the inverse of $\mat{Z_{\Lambda \Sigma}}$, which are derived by inverting \eqref{eq:block_mat_loop-star-ns} using Schur complement formulas \cite{henderson1981deriving}
\begin{gather} 
\Re\left(\mat{Z_{\Lambda \Sigma}}^{-1} \right) \underrel{\omega \to 0}{=} \begin{bmatrix}
	\mathcal{O}(1) &\mathcal{O}(\omega^2) \\
	\mathcal{O}(1) & \mathcal{O}(1)
\end{bmatrix} \label{eq:scaling_mat_inv_real_ns}\,,\\
\Im\left( \mat{Z_{\Lambda \Sigma}}^{-1} \right) \underrel{\omega \to 0}{=} \begin{bmatrix}
	\mathcal{O}(1/\omega) &\mathcal{O}(\omega) \\
	\mathcal{O}(\omega) & \mathcal{O}(\omega)
\end{bmatrix} \label{eq:scaling_mat_inv_imag_ns}.
\end{gather}
Finally by multiplying the scaling matrix of $\mat{Z_{\Lambda \Sigma}}^{-1}$ (\eqref{eq:scaling_mat_inv_real_ns} and \eqref{eq:scaling_mat_inv_imag_ns}) and the scaling vector of $\transpose{\mat{B^{\tilde{G}_\epsilon}_{\Lambda \Sigma}}} \mat{v}$ (\eqref{eq:scaling_rhs_pw_ns} for the plane wave or \eqref{eq:scaling_rhs_dipole_ns} for the dipole), we obtain the following scalings of $\mat{\alpha_{\Lambda \Sigma}}$
\begin{gather} 
\Re \left(\mat{\alpha_{\Lambda \Sigma}^\text{PW}} \right) \underrel{\omega \to 0}{=} \begin{bmatrix}
	\mathcal{O}(1) \\
	\mathcal{O}(1) 
\end{bmatrix};
\Im \left(\mat{\alpha_{\Lambda \Sigma}^\text{PW}} \right) \underrel{\omega \to 0}{=} \begin{bmatrix}
	\mathcal{O}(\omega) \\
	\mathcal{O}(\omega) 
\end{bmatrix}\,,  \label{eq:scaling_solution_pw}\\
	\Re \left(\mat{\alpha_{\Lambda \Sigma}^\text{dip}} \right) \underrel{\omega \to 0}{=} \begin{bmatrix}
		\mathcal{O}(\omega^2) \\
		\mathcal{O}(1) 
	\end{bmatrix};
	\Im \left(\mat{\alpha_{\Lambda \Sigma}^\text{dip}} \right) \underrel{\omega \to 0}{=} \begin{bmatrix}
		\mathcal{O}(\omega) \\
		\mathcal{O}(\omega) 
	\end{bmatrix}\label{eq:scaling_solution_dipole}\,, 
\end{gather}
for plane wave and dipole excitations, respectively.
\setlength{\tabcolsep}{1.2pt} 
\begin{table}
\centering
\caption{Frequency scalings of the real and imaginary parts of the right hand side vectors $\mat{v}$ and $\mat{\tilde{v}}$ for plane wave and dipole excitations.}
\subfloat[Right hand side $\mat{v}$\label{tab:table_v}]{
	\begin{tabular}{c}
	\end{tabular}
	\begin{tabular}{ccccc} 
		\toprule
		\multirow{2}{*}{Source} 
		& \multirow{2}{*}{($\Re$ , $\Im$) $(\mat{v_\Lambda} )$}
		&\multirow{2}{*}{$(\Re , \Im)( \mat{v_\Sigma})$} 
		& Terms & Terms required \\
		&&& recovered & for a correct solution\\
		\midrule
		\multirow{2}{*}{Plane Wave} 
		& \multirow{2}{*}{$(\omega^2 , \omega)$} 
		& \multirow{2}{*}{$(1 ,\omega)$} 
		& $\Re(\mat{v_\Sigma})$
		&$\Re(\mat{v_\Lambda}),\Re(\mat{v_\Sigma})$ \\ &&&$\Im(\mat{v_\Lambda}),\Im(\mat{v_\Sigma})$ &$\Im(\mat{v_\Lambda}),\Im(\mat{v_\Sigma})$ \\
		\midrule
		Dipole
		& $(\omega^2 , \omega^5)$ 
		& $(1 , \omega^3)$ 
		& $\Re(\mat{v_\Sigma}),\Im(\mat{v_\Sigma})$
		&$\Re(\mat{v_\Lambda}),\Re(\mat{v_\Sigma})$ \\ 
		\bottomrule
	\end{tabular}
}\\
\subfloat[Scaled right hand side $\mat{\tilde{v}}$ \label{tab:table_vtilde}]{
	\begin{tabular}{c}
	\end{tabular}
	\begin{tabular}{ccccc} 
		\toprule
		\multirow{2}{*}{Source} 
		& \multirow{2}{*}{($\Re$ , $\Im$) $(\mat{\tilde{v}_\Lambda} )$}
		&\multirow{2}{*}{$(\Re , \Im)( \mat{\tilde{v}_\Sigma})$}
		& Terms & Terms required \\
		&&& recovered & for a correct solution \\
		\midrule
		\multirow{2}{*}{Plane Wave} 
		& \multirow{2}{*}{$(1 ,\omega)$}
		& \multirow{2}{*}{$(1 ,\omega)$} 
		& $\Re(\mat{\tilde{v}_\Lambda}),\Re(\mat{\tilde{v}_\Sigma})$ & $\Re(\mat{\tilde{v}_\Lambda}),\Re(\mat{\tilde{v}_\Sigma})$ \\
		&&&$\Im(\mat{\tilde{v}_\Lambda}),\Im(\mat{\tilde{v}_\Sigma})$ & $\Im(\mat{\tilde{v}_\Lambda}),\Im(\mat{\tilde{v}_\Sigma})$\\
		\midrule
		Dipole
		& $(\omega^4 ,\omega)$
		& $(1 ,\omega^3)$ 
		& $\Re(\mat{\tilde{v}_\Sigma}),\Im(\mat{\tilde{v}_\Lambda})$ 
		& $\Re(\mat{\tilde{v}_\Sigma}),\Im(\mat{\tilde{v}_\Lambda})$ \\
		\bottomrule
	\end{tabular}
}
\end{table}
At low frequencies, the real (or imaginary) part of the Loop and Star components of the solution coefficients in \eqref{eq:scaling_solution_pw} have the same frequency scalings, hence all the components of the solution for a plane wave excitation are preserved in this regime. However, the real part of the Loop component of the solution for a dipole excitation in \eqref{eq:scaling_solution_dipole} becomes much smaller than the real part of its Star component at low frequencies and is lost due to finite precision arithmetic. In the scope of this paper, the D-VIE is applied in scenarios where the electric field inside the object is required. Since there is a simple scalar relation between the unknown of the D-VIE (i.e. the electric flux) and the electric field, the re-amplification of a lost (Loop or Star) part is not possible, and thus it is not required to preserve the components of the solution lost due to finite precision at low frequencies.

Nevertheless, a possible loss of accuracy in the solution occurs when a term of the right hand side, which is lost at low frequencies, contributes to one of the dominant terms of the solution.
\tdg{AM: we might want to make this discussion more vague, since it might not be the only effect into play: Re-ask} In \cref{tab:table_v}, the terms contributing to the solution for both types of excitation are provided. For the dipole and the plane wave, the real part of the Loop component of the right hand side is lost due to finite precision while it is supposed to contribute to the dominant terms of the solution. For this reason, there is a loss of accuracy in the solution at low frequencies with plane wave and dipole excitations.

\subsection{High-Contrast Analysis of the D-VIE} \label{sec:hc-part}
In the following we analyze the high-contrast behavior of the D-VIE for complex permittivity objects with dominant imaginary part ($\sigma / (\omega \epsilon_0 \epsilon_r') \gg 1$) at low frequencies. In this regime, the dielectric contrast $\kappa$ between the object and the background is approximately \num{1} and the Green's function $G_0$ can be bounded from above by a frequency-independent term.
%

In our theoretical treatment, the scatterer $\Omega$ is supposed to be a piecewise homogeneous object composed of a region $\Omega_M$ which has the maximum conductivity $\sigma_{\text{max}}$ of $\Omega$ (associated to the complex permittivity $\epsilon_{\text{max}}$) while the conductivity is bounded in its complementary domain $\Omega \backslash \Omega_M$. In the following, the boundary between $\Omega_M$ and the rest of the object is denoted by $\partial \Omega_M$ and the maximum conductivity ratio is defined as $r_\sigma = \sigma_\text{max}/\sigma_\text{min}$ with $\sigma_\text{min}$ being the minimum conductivity in $\Omega$. The following high-contrast analysis will be performed for $r_\sigma$ going toward infinity with $\sigma_\text{min}$ fixed. Note that this scenario represents the high internal contrast occurring in lossy dielectric objects in several application scenarios. 
\tdg{AM: it sounds a bit fishy to claim that the ratio going to infinity is relevant for biomedical application.\\ Also, in this section the frequency is fixed, correct? If so it mght need clarifying: SHould be fixed now (removed source loc application)}

Similarly to the low-frequency breakdown, the HC problem originates from the solenoidal part of $\mat{Z}$ that is ill-scaled due to the permittivity scaling of the Gram matrix $\mat{G_\epsilon}$. While the approach carried out in the low-frequency analysis only required the determination of the frequency scaling of $\|\mat{G_\epsilon}\|$, some additional considerations on the material dependence of the minimum singular value of the Gram matrix are needed in the HC analysis. To this aim, we leverage the block structure of $\mat{G_\epsilon}$ for the above-mentioned scatterer. The entries of $\mat{G_\epsilon}$ can be written as
$
 \left[ \mat{G_{\epsilon}} \right]_{mn} = \int_{T_m^+} \vt{f_m} \cdot \epsilon^{-1} \vt{f_n} \diff v + 
 \int_{T_m^-} \vt{f_m} \cdot \epsilon^{-1}  \vt{f_n} \diff v\,,
$
with $T^\pm_m$ being the tetrahedra on which $\vt{f_m}$ is defined. In the case of a half basis function, only $T^+_m$ is required. 
From this definition, we can build the following block matrix
\begin{gather} 
\mat{G_\epsilon}
 = \begin{bmatrix}
 \mat{G_M} & \transpose{\mat{B_{M}}} \\
 \mat{B_{M}} & \mat{G_R} 
 \end{bmatrix}\,, \text{ in which} \label{eq:block-structure} \\
  \mat{G_{M}}  = \mat{G_\epsilon}[a_1,\dots, a_{N_{FM}};a_1,\dots, a_{N_{FM}}]\,, \\
  \mat{B_{M}}  = \mat{G_\epsilon}[a_{N_{FM} + 1},\dots, a_{N_F};a_1,\dots, a_{N_{FM}}] \,,\\
  \mat{G_{R}}  = \mat{G_\epsilon}[a_{N_{FM} + 1},\dots, a_{N_F};a_{N_{FM} + 1},\dots, a_{N_F}]\,, 
\end{gather}
with $\{a_1,\dots, a_{N_{FM}}\}$ being the indices of the rows (and columns) of $ \mat{G_\epsilon}$ corresponding to the $N_{FM}$ SWG basis functions that have both of their supporting tetrahedra in $\Omega_M$ and $\{a_{N_{FM} + 1},\dots, a_{N_F}\}$ being the ${N_F-N_{FM}}$ indices of the remaining SWG basis functions.
\tdg{AM: the matrix indices are not consistent}
\td{AM: much better, ask francesco if the matlab notation should be introduced though (CH: source: https://proofwiki.org/wiki/Definition:Submatrix/Notation)}

Using the block structure given in \eqref{eq:block-structure} and the fact $\Re(1/\epsilon_{\text{max}} ) = \mathcal{O}(r_{\sigma}^{-2})$, $\Im(1/\epsilon_{\text{max}} ) = \mathcal{O}(r_{\sigma}^{-1})$, $\mat{G_R} =\mathcal{O}(1/\epsilon)$ (bounded complex permittivity), $\mat{B_M} =\mathcal{O}(1/\epsilon_{\text{max}} )$, and $\mat{G_M} =\mathcal{O}(1/\epsilon_{\text{max}} )$ when $r_\sigma \rightarrow \infty$, we obtain the following scalings for the blocks of $\mat{G_\epsilon}$ when the contrast goes to infinity
\begin{gather}
 \Re \left(\mat{G_\epsilon}\right) \underrel{r_\sigma \to \infty}{=} \begin{bmatrix}
  \mathcal{O}(1 / r_\sigma^2) & \mathcal{O}(1 / r_\sigma^2) \\
  \mathcal{O}(1 / r_\sigma^2) & \mathcal{O}(1)
 \end{bmatrix} \label{eq:scaling_Gram_Re}\,,\\
 \Im \left(\mat{G_\epsilon}\right) \underrel{r_\sigma \to \infty}{=} \begin{bmatrix}
  \mathcal{O}(1 / r_\sigma) & \mathcal{O}(1 / r_\sigma) \\
  \mathcal{O}(1 / r_\sigma) & \mathcal{O}(1) 
 \end{bmatrix} ,\label{eq:scaling_Gram_Im}
\end{gather}
which, according to the Gershgorin circle theorem, show that $\mat{G_\epsilon}$ suffers from ill-conditioning in objects with high internal contrast. This HC problem in the Gram matrix can be a source of ill-conditioning in the discretized D-VIE, and hence should be considered when regularizing the D-VIE for high-contrast.
\tdg{AM: this statement seems a bit abrupt.}

\section{Oblique Quasi-Helmholtz Projectors} \label{sec:oblique-projectors}
Just like the orthogonal quasi-Helmholtz projectors \cite{andriulli_multigrid,qH_Andriulli} arise from the inversion of the standard Loop-Star decomposition, the oblique Loop-Star decomposition introduced in equation \eqref{eq:normDecomposition} can give rise to an entire new family of projectors that, in general, will not be orthogonal, but oblique.
So starting from the decomposition described in \eqref{eq:normDecomposition} (omitting the normalization which is not required for the following and using $\mat{\Sigma}$ instead of $\mat{\tilde{\Sigma}}$), we write the solution coefficient vector as
\begin{equation}\label{eq:obliquedec}
	\mat{\alpha} = \mat{\Lambda} \mat{\tilde{l}} + \mat{A} \mat{\Sigma} \mat{\tilde{s}}\,,
\end{equation}
\tdg{pb coherence when building projectors: $\mat{\Sigma}$ has its columns linearly independent}
in which as before $\mat{A}$ is an invertible symmetric matrix.
Note that \eqref{eq:obliquedec} is a valid Helmholtz decomposition, as demonstrated in Appendix \ref{app:appA}. 
We can now obtain a new set of projectors by solving for $\mat{\Lambda} \mat{\tilde{l}}$ and for $\mat{A} \mat{\Sigma} \mat{\tilde{s}}$ separately. To this end, we first multiply \eqref{eq:obliquedec} by $\mat{\Lambda}^\mathrm{T} \mat{A}^{-1}$ and $\mat{\Sigma}^\mathrm{T} $, and we obtain the following equations
\begin{equation}
 \mat{\Lambda}^\mathrm{T} \mat{A}^{-1} \mat{\alpha} = \mat{\Lambda}^\mathrm{T} \mat{A}^{-1} \mat{\Lambda} \mat{\tilde{l}};~
 \mat{\Sigma}^\mathrm{T} \mat{\alpha} = \mat{\Sigma}^\mathrm{T} \mat{A} \mat{\Sigma} \mat{\tilde{s}}\label{eq:app1_decomp_lp}\,,
\end{equation}
in which we used the properties that $\mat{\Lambda}^\mathrm{T} \mat{A}^{-1} \mat{A} \mat{\Sigma} = \mat{0}$ and $\mat{\Sigma}^\mathrm{T} \mat{\Lambda} = \mat{0}$. From \eqref{eq:app1_decomp_lp}, we can now express the coefficients of the oblique Loop and Star basis functions from the coefficients of the original basis functions
\begin{equation} 
 \mat{\tilde{l}} = (\mat{\Lambda}^\mathrm{T} \mat{A}^{-1} \mat{\Lambda})^{-1} \mat{\Lambda}^\mathrm{T} \mat{A}^{-1} \mat{\alpha};~
 \mat{\tilde{s}} = (\mat{\Sigma}^\mathrm{T} \mat{A} \mat{\Sigma})^+ \mat{\Sigma}^\mathrm{T} \mat{\alpha} \label{eq:coeff-loopstar}\,,
\end{equation}
where $^+$ is the Moore-Penrose pseudo inverse, which is required in \eqref{eq:coeff-loopstar} due to the one-dimensional null space of $\mat{\Sigma}^\mathrm{T} \mat{A} \mat{\Sigma}$.
\tdg{AM: pinv notation not introduced}
The next step is to obtain the solenoidal and non-solenoidal parts of $\mat{\alpha}$ in terms of $\mat{\alpha}$ itself by applying $\mat{\Lambda} $ and $\mat{A} \mat{\Sigma} $ to the equations in \eqref{eq:coeff-loopstar}
\begin{gather}
 \mat{\Lambda} \mat{\tilde{l}} = \mat{\Lambda} (\mat{\Lambda}^\mathrm{T} \mat{A}^{-1} \mat{\Lambda})^{-1} \mat{\Lambda}^\mathrm{T} \mat{A}^{-1} \mat{\alpha} \label{eq:dec1}\,,\\
\mat{A} \mat{\Sigma} \mat{\tilde{s}} = \mat{A} \mat{\Sigma} (\mat{\Sigma}^\mathrm{T} \mat{A} \mat{\Sigma})^+ \mat{\Sigma}^\mathrm{T} \mat{\alpha}\,.\label{eq:dec2}
\end{gather}
What we have obtained is a family of two complementary oblique projectors 
\begin{gather}
 \mat{P^{\Lambda}_{A^{-1}}} = \mat{\Lambda} \left( \mat{\Lambda}^\mathrm{{\mathrm{T}}} \mat{A}^{-1} \mat{\Lambda} \right)^{-1} \mat{\Lambda}^\mathrm{{\mathrm{T}}} \mat{A}^{-1}\,,\\
 \mat{P^{\Sigma}_{A^{-1}}} = \mat{A} \mat{\Sigma} \left( \mat{\Sigma}^\mathrm{{\mathrm{T}}} \mat{A} \mat{\Sigma} \right)^{+} \mat{\Sigma}^\mathrm{{\mathrm{T}}}\,,
\end{gather}
which are the oblique quasi-Helmholtz projectors we propose in this work. From \eqref{eq:obliquedec}, \eqref{eq:dec1}, and \eqref{eq:dec2} it follows that 
$
\mat{\alpha} = (\mat{P^{\Lambda}_{A^{-1}}} + \mat{P^{\Sigma}_{A^{-1}}}) \mat{\alpha}
$
and this proves the complementarity property 
$
\mat{P^{\Lambda}_{A^{-1}}} + \mat{P^{\Sigma}_{A^{-1}}} = \identity
$.
These new projectors are, in general, not symmetric and their transposes will be denoted by $\mat{P^{\Lambda \mathrm{\normalfont{T}}}_{A^{-1}}}$ and $\mat{P^{\Sigma \mathrm{\normalfont{T}}}_{A^{-1}}}$. The following properties can be easily proven
\begin{gather} 
	\mat{P^{\Lambda \mathrm{\normalfont{T}}}_{A^{-1}}} \mat{A^{-1}} \mat{P^{\Lambda}_{A^{-1}}} = \mat{A^{-1}} \mat{P^{\Lambda}_{A^{-1}}} ;~
	\mat{P^{\Lambda \mathrm{\normalfont{T}}}_{A^{-1}}} \mat{A^{-1}} \mat{P^{\Sigma}_{A^{-1}}} = \matO \label{eq:id2_scaled_proj_ns_A}\,,\\
	\mat{P^{\Sigma \mathrm{\normalfont{T}}}_{A^{-1}}} \mat{A^{-1}} \mat{P^{\Lambda}_{A^{-1}} } = \matO ;~
	\mat{P^{\Sigma \mathrm{\normalfont{T}}}_{A^{-1}}} \mat{A^{-1}} \mat{P^{\Sigma}_{A^{-1}}} =  \mat{A^{-1}} \mat{P^{\Sigma}_{A^{-1}}}\,.
	\label{eq:id4_scaled_proj_ns_A}
\end{gather}
 When $\mat{A} = \identity$, we obtain the standard Loop-Star decomposition and the associated projectors are the standard quasi-Helmholtz projectors \cite{qH_Andriulli}. The above formulas define in general oblique projectors for all formulations where a valid Loop-Star decomposition can be defined, be it 2D, 3D surface, or 3D volume. For specializing to the volume equation of interest in this work, however, we set $\mat{A} = \mat{G_\epsilon}^{-1}$ and we obtain the scaled projectors $\mat{P^{\Sigma}_{G_\epsilon}}$ and $\mat{P^{\Lambda}_{G_\epsilon}}$ defined as
\begin{gather}
	\mat{P^{\Sigma}_{G_\epsilon}} = \mat{G_{\epsilon}}^{-1} \mat{\Sigma} \left( \mat{\Sigma}^\mathrm{T} \mat{G_{\epsilon}}^{-1} \mat{\Sigma} \right)^{+} \mat{\Sigma}^\mathrm{T} \label{eq:scaled-projectors-star-G-ns}\,,\\
	\mat{P^{\Lambda}_{G_\epsilon}} = \identity - \mat{P^{\Sigma}_{G_\epsilon}} = \mat{\Lambda} \left( \mat{\Lambda}^\mathrm{T} \mat{G_{\epsilon}} \mat{\Lambda} \right)^{-1} \mat{\Lambda}^\mathrm{T} \mat{G_{\epsilon}}\,, \label{eq:scaled-projectors-loop-G-ns}
\end{gather}
for which the associated cancellation properties become
\begin{gather}
	\mat{P^{\Lambda \mathrm{\normalfont{T}}}_{G_\epsilon}} \mat{G_{\epsilon}} \mat{P^{\Lambda}_{G_\epsilon}} = \mat{G_{\epsilon}} \mat{P^{\Lambda}_{G_\epsilon}} ;~ 
	\mat{P^{\Lambda \mathrm{\normalfont{T}}}_{G_\epsilon}} \mat{G_{\epsilon}} \mat{P^{\Sigma}_{G_\epsilon}} = \matO \,, \label{eq:id1_scaled_proj_ns}\\
	\mat{P^{\Sigma \mathrm{\normalfont{T}}}_{G_\epsilon}} \mat{G_{\epsilon}} \mat{P^{\Lambda}_{G_\epsilon} } = \matO ;~ 
	\mat{P^{\Sigma \mathrm{\normalfont{T}}}_{G_\epsilon}} \mat{G_{\epsilon}} \mat{P^{\Sigma}_{G_\epsilon}} =  \mat{G_{\epsilon}} \mat{P^{\Sigma}_{G_\epsilon}}\,. \label{eq:id4_scaled_proj_ns}
\end{gather}
Moreover, when applying $\mat{P^{\Lambda}_{G_\epsilon}}$ to the scalar potential matrices, we obtain the relations
$
\mat{P^{\Lambda}_{G_\epsilon}} \mat{G_\epsilon}^{-1} \mat{Z_{\Phi, 11}}= \matO$, $\mat{P^{\Lambda}_{G_\epsilon}} \mat{G_\epsilon}^{-1}  \mat{Z_{\Phi, 1\epsilon}}= \matO$, and $\mat{Z_{\Phi, 11}} \mat{P^{\Lambda}_{G_\epsilon}}  = \matO$.

\tdg{AM: recall the cancellation properties on the scalar potentials ?}

\section{The New Regularized Volume Integral Equation} \label{sec:regularized_eq}
To find the correct parameters of a regularizer for the D-VIE based on the new oblique quasi-Helmholtz projectors just introduced, we will employ an approach often and successfully used with projectors at low-frequency: first a regularization is obtained by determining the parameters of a, impractical-to-implement but theoretically useful, normalized Loop-Star decomposition; then the same coefficients are used for linearly combining solenoidal and non-solenoidal projectors. In the present context, however, we not only have to regularize the low-frequency breakdown, but also the high-contrast breakdown. Thus, after finding the frequency-regularizing coefficients with a normalized Loop-Star analysis we will propose a formulation that also solves the high-contrast breakdown by using an ansatz whose effectiveness will then be proved in the following section. 

Consider the following normalized and re-scaled Loop-Star decomposition of the D-VIE
\begin{equation}
\label{eq:helmholtz_decomp_precond_ns}
\mat{\tilde{Z}_{\Lambda \Sigma}}\mat{{\alpha}_{\Lambda \Sigma}}=
\begin{bmatrix}
	C_1 \matN{\Lambda}^\mathrm{T} \\
	C_2 \matN{\Sigma}^\mathrm{T}
\end{bmatrix} \mat{Z} [C_3 \matN{\Lambda},~C_4 \matN{\Sigma}] \mat{{\alpha}_{\Lambda \Sigma}} 
= \begin{bmatrix}
	C_1 \matN{\Lambda}^\mathrm{T} \\
	C_2 \matN{\Sigma}^\mathrm{T} 
\end{bmatrix} \mat{v}=\mat{\tilde{v}}\,,
\end{equation}
in which the coefficients $C_1$, $C_2$, $C_3$ and $C_4$ have to be determined to make the Loop and Star parts of the preconditioned matrix ($\mat{\tilde{Z}_{\Lambda \Sigma}}$) and its right hand side $\mat{\tilde{v}}$ free from ill-conditioning and loss of accuracy. Since the numerical loss comes from the right hand side and not from the solution directly, we can set $C_3$ and $C_4$ to $1$ and attempt to leverage the two remaining coefficients to regularize the rest of the equation. Therefore, we now need to determine $C_1$ and $C_2$ for $\mat{\tilde{Z}_{\Lambda \Sigma}}$ to be well-conditioned and to avoid numerical loss. It will be now shown that, by choosing 
\begin{equation}\label{eq:props}
 C_1\propto 1 / (\mi \omega \epsilon_0);~
 C_2\propto 1\,,
\end{equation}
the matrix becomes well-conditioned and the right hand side is no longer subject to numerical loss of accuracy in the dominant terms of the solution for both plane wave and dipole excitations. 
\tdg{AM: this is not true, the dipole RHS still is}
In fact, the resulting Loop-Star matrix $\mat{\tilde{Z}_{\Lambda \Sigma}}$ in this case has the following frequency scalings
\begin{align} \label{eq:scaling_mat_precond_ns_real}
\Re\left(\mat{\tilde{Z}_{\Lambda \Sigma}}\right) &\underrel{\omega \to 0}{=} \begin{bmatrix}
	\mathcal{O}(1) &\mathcal{O}(\omega^2) \\
	\mathcal{O}(\omega^2) & \mathcal{O}(1)
\end{bmatrix} \,, \\ \label{eq:scaling_mat_precond_ns_imag}
\Im\left(\mat{\tilde{Z}_{\Lambda \Sigma}} \right) &\underrel{\omega \to 0}{=} \begin{bmatrix}
	\mathcal{O}(\omega) &\mathcal{O}(\omega) \\
	\mathcal{O}(\omega) & \mathcal{O}(\omega)
\end{bmatrix}\,,
\end{align}
from which it is evident that, while the diagonal blocks of \eqref{eq:scaling_mat_precond_ns_real} scale as $\mathcal{O}(1)$, all the other blocks get to zero at low frequencies, thus it results that $\mat{\tilde{Z}_{\Lambda \Sigma}}$ is well-conditioned. 
The preconditioned right hand side on the other hand scales as
\begin{gather} 
\Re \left(\transpose{\mat{\tilde{B}^{\tilde{G}_\epsilon}_{\Lambda \Sigma}}} \mat{v_\textnormal{PW}}\right) \underrel{\omega \to 0}{=} \begin{bmatrix}
	\mathcal{O}(1) \\
	\mathcal{O}(1) 
\end{bmatrix},
\Im \left(\transpose{\mat{\tilde{B}^{\tilde{G}_\epsilon}_{\Lambda \Sigma}}} \mat{v_\textnormal{PW}}\right) \underrel{\omega \to 0}{=} \begin{bmatrix}
	\mathcal{O}(\omega) \\
	\mathcal{O}(\omega) 
\end{bmatrix}, \label{eq:scaling_rhs_precond_pw_ns} \\
\Re \left(\transpose{\mat{\tilde{B}^{\tilde{G}_\epsilon}_{\Lambda \Sigma}}} \mat{v_\textnormal{dip}}\right) \underrel{\omega \to 0}{=} \begin{bmatrix}
	\mathcal{O}(\omega^4) \\
	\mathcal{O}(1) 
\end{bmatrix},
\Im \left(\transpose{\mat{\tilde{B}^{\tilde{G}_\epsilon}_{\Lambda \Sigma}}} \mat{v_\textnormal{dip}}\right) \underrel{\omega \to 0}{=} \begin{bmatrix}
	\mathcal{O}(\omega) \\
	\mathcal{O}(\omega^3) 
\end{bmatrix},  \label{eq:scaling_rhs_precond_dipole_ns}
\end{gather}
for plane wave and dipole excitations, respectively. From \cref{tab:table_vtilde}, it results that all the terms of the right hand side vectors for the plane wave and the dipole contributing to the dominants parts of the solutions are preserved at low frequencies. Therefore, the Loop-Star coefficients $C_1$, $C_2$, $C_3$, and $C_4$ would cure both matrix ill-conditioning and right hand side cancellations. 

From this analysis we learn the following guidelines to produce an ansatz for the D-VIE regularizer: (i) the regularization should be a left preconditioner only, (ii) a frequency scaling proportional to $\omega^{-1}$ should be applied to the solenoidal part $\mat{P^{\Lambda}_{G_\epsilon}}$, (iii) a clear source of high-contrast breakdown originates from the ($\epsilon$-dependent) Gram matrix. By noticing that the Gram matrix $\mat{G_{\epsilon}}$ is asymptotically proportional to $\omega$, guidelines (i)-(iii) are satisfied if the solenoidal part is multiplied by $\mat{G_{\epsilon}}^{-1}$. Otherwise said, the regularizer we propose is defined as
%
%
\begin{equation} \label{eq:precond_hc-ns}
  \mat{L_{G_\epsilon}} = \gamma_\Lambda \mat{P^{\Lambda }_{G_\epsilon}} \mat{G_\epsilon}^{-1} + \gamma_\Sigma \mat{P^{\Sigma }_{G_\epsilon}} \mat{G}^{-1}\,,
\end{equation} 
in which the proportionality factors $\gamma_\Lambda = {1}/{\| \mat{P^{\Lambda}_{G_\epsilon}} \|}$ and $\gamma_\Sigma = {1}/{\| \mat{P^{\Sigma}_{G_\epsilon}} \mat{G}^{-1} \mat{Z_\Phi} \mat{P^{\Sigma}_{G_\epsilon}} \|} $ are chosen to ensure that both solenoidal and non-solenoidal parts contribute with unitary weight to the final operator. Note that since the solenoidal part is re-scaled by $\mat{G_\epsilon}^{-1}$, the non-solenoidal part is re-scaled by the $\epsilon$-independent Gram matrix $\mat{G}^{-1}$ for the sake of consistency. The final formulation reads 
\begin{equation} \label{eq:precond_equation_G_ns}
 \mat{L_{G_\epsilon}} \mat{Z} \mat{\alpha} = \mat{L_{G_\epsilon}} \mat{v}\,.
\end{equation} 
Although \eqref{eq:precond_equation_G_ns} cures the low-frequency breakdown of the D-VIE by construction, it is only an ansatz regarding the cure of the HC breakdown. However, in the next section, we will prove also the effectiveness of the formulation for the HC breakdown.

\section{Theoretical Framework and Conditioning Bounds} \label{sec:bounds}
As already delineated in the previous section, given that $\gamma_\Lambda \mat{P^{\Lambda }_{G_\epsilon}} \mat{G_\epsilon}^{-1}=O(1/\omega)$ and that $ \gamma_\Sigma \mat{P^{\Sigma }_{G_\epsilon}} \mat{G}^{-1}=O(1)$, the low-frequency stability (in the limit $\omega\to 0$) follows from \eqref{eq:props}-\eqref{eq:scaling_mat_precond_ns_imag}. Let's now focus on the high-contrast regime studying the behavior of the new preconditioned operator $\mat{M} = \mat{L_{G_\epsilon}}\mat{Z}$ in \eqref{eq:precond_equation_G_ns} for high conductivity ratio. Using equations \eqref{eq:id1_scaled_proj_ns} and \eqref{eq:id4_scaled_proj_ns}, the resulting preconditioned matrix $\mat{M}$ can be decomposed as
\begin{equation} 
\begin{split}
  \mat{M} &= \mat{L_{G_\epsilon}}\mat{Z}
  = \gamma_\Lambda \mat{P^{\Lambda}_{G_\epsilon}} + \gamma_\Sigma \mat{P^{\Sigma}_{G_\epsilon}} \mat{G}^{-1} \mat{G_\epsilon} \\
 &+ \gamma_\Lambda \mat{P^{\Lambda}_{G_\epsilon}} \mat{G_\epsilon}^{-1} \mat{Z_A} 
 + \gamma_\Sigma \mat{P^{\Sigma}_{G_\epsilon}} \mat{G}^{-1}
 \mat{Z_A}\\
 &+ \gamma_\Sigma \mat{P^{\Sigma}_{G_\epsilon}} \mat{G}^{-1}
 \mat{Z_{\Phi,1 \epsilon}} 
 + \gamma_\Sigma \mat{P^{\Sigma }_{G_\epsilon}} \mat{G}^{-1}
 \mat{Z_{\Phi,1 1}} \mat{P^{\Sigma }_{G_\epsilon}}\,.
  \label{eq:discretized_eq_decomp_precond_ns}
 \end{split}
\end{equation}  
To investigate the conditioning of $\mat{M}$, we leverage the following auxiliary normalized Loop-Star decomposition matrices $\mat{Q_L}= [\mat{Q_{\Lambda L}}, \mat{Q_{\Sigma L}} ]$ and $\mat{Q_R}= [\mat{Q_{\Lambda R}}, \mat{Q_{\Sigma R}} ]$ such that
\begin{gather}
    [\mat{Q_{\Lambda L}}, \mat{Q_{\Sigma L}} ] = [ \mat{G_\epsilon} \mat{\Lambda} (\transpose{\mat{\Lambda}} \mat{G_\epsilon}^2 \mat{\Lambda})^{-\frac{1}{2}} ,~ \mat{\tilde{\Sigma}} (\transpose{\mat{\tilde{\Sigma}}} \mat{\tilde{\Sigma}})^{-\frac{1}{2}}],\\
    [\mat{Q_{\Lambda R}}, \mat{Q_{\Sigma R}} ] =[ \mat{\Lambda} (\transpose{\mat{\Lambda}} \mat{\Lambda})^{-\frac{1}{2}} 
 ,~ \mat{G_\epsilon}^{-1} \mat{\tilde{\Sigma}} (\transpose{\mat{\tilde{\Sigma}}} \mat{G_\epsilon}^{-2}\mat{\tilde{\Sigma}})^{-\frac{1}{2}}],
\end{gather}
\tdg{AM: You are taking the squre root of a singular matrix. Fine since 0 singular value is not here with Sigma truncated}
which allow decomposing $\mat{M}$ into a block matrix $\mat{M_B}$ whose diagonal blocks correspond to the solenoidal and non-solenoidal parts of $\mat{M}$ and the off-diagonal blocks to its associated cross terms. Note that the following identities hold with these normalized decomposition matrices
\begin{gather} 
	\transpose{\mat{Q_{\Lambda L}}} \mat{P^{\Lambda }_{G_\epsilon}} = \transpose{\mat{Q_{\Lambda L}}};~
	\transpose{\mat{Q_{\Sigma L}}} \mat{P^{\Sigma }_{G_\epsilon}} = \transpose{\mat{Q_{\Sigma L}}}\,,\\
	\mat{P^{\Lambda }_{G_\epsilon}} \mat{Q_{\Lambda R}} = \mat{Q_{\Lambda R}};~
	\mat{P^{\Sigma }_{G_\epsilon}} \mat{Q_{\Sigma R}} = \mat{Q_{\Sigma R}}\,.
\end{gather}
\tdg{AM: it looks like you are using $\mat{\Lambda}^\mathrm{T} \mat{G_{\epsilon}} \mat{\Lambda} \left( \mat{\Lambda}^\mathrm{T} \mat{G_{\epsilon}} \mat{\Lambda} \right)^{+} = \mat I$ (among others), are we sure this is true?: Fine since 0 singular value is not here with Sigma truncated}
The decomposed matrix then reads
\begin{gather}
 \mat{M_B} 
 = \transpose{\mat{Q_L}} \mat{M} \mat{Q_R} = \begin{pmatrix}
  \mat{M_{\Lambda \Lambda}} & \mat{M_{\Lambda \Sigma}}\\
  \mat{M_{\Sigma \Lambda}} & \mat{M_{\Sigma \Sigma}}
 \end{pmatrix},\,\text{ in which}\\
 \mat{M_{\Lambda \Lambda}} = \gamma_\Lambda \transpose{\mat{Q_{\Lambda L}}} ( \identity + \mat{G_\epsilon}^{-1} \mat{Z_A}) \mat{Q_{\Lambda R}} \label{eq:diagup}\,,\\
 \mat{M_{\Lambda \Sigma}} = \gamma_\Lambda \transpose{\mat{Q_{\Lambda L}}} \mat{G_\epsilon}^{-1} \mat{Z_A} \mat{Q_{\Sigma R}}\,,\\
 \mat{M_{\Sigma \Lambda}} = \gamma_\Sigma \transpose{\mat{Q_{\Sigma L}}} \mat{G}^{-1} (\mat{G_\epsilon} + \mat{Z_A} + \mat{Z_{\Phi,1 \epsilon}} ) \mat{Q_{\Lambda R}}\,,\\
 \mat{M_{\Sigma \Sigma}} = \gamma_\Sigma \transpose{\mat{Q_{\Sigma L}}} \mat{G}^{-1} (\mat{G_\epsilon} + \mat{Z_A} + \mat{Z_{\Phi}}) \mat{Q_{\Sigma R}}\,.\label{eq:diagdown}
\end{gather}
\tdb{AM: Check the last 2 results?}
Using the fact that the product of two square matrices $\mat{U}$ and $\mat{V}$ can be bounded as $\cond{\mat{U} \mat{V}} \leq \cond{\mat{U}}\cond{\mat{V}}$, we obtain the following upper bound for $\cond{\mat{M}}$
\begin{equation} 
\begin{alignedat}{2}
 \cond{\mat{M}} \leq \cond{\mat{Q_L}} \cond{\mat{M_B}} \cond{\mat{Q_R}}\,.\label{eq:bound_cound_M}
 \end{alignedat}
\end{equation}
In Appendix \ref{app:appB} we show that, in an object with $\sigma / (\omega \epsilon_0 \epsilon_r') \gg 1$, the condition number of $\mat{Q_L}$ ($\mat{A} = \mat{G_\epsilon}^{-1}$) can be bounded as
\begin{equation}
\begin{alignedat}{2}
 \cond{\mat{Q_L}} \leq \left(\frac{1 + \sqrt{1 - {\|\mat{P^{\Sigma}_{\mat{G_\epsilon}}}\|^{-2}}}}{1 - \sqrt{1 - {\|\mat{P^{\Sigma}_{\mat{G_\epsilon}}}\|^{-2}}}}\right)^{\frac{1}{2}}\,.
 \end{alignedat}
\end{equation}

Then, using the fact $\|\mat{P^{\Sigma}_{\mat{G_\epsilon}}}\| = \mathcal{O}(1)$ when $r_{\sigma} \rightarrow \infty$ for the geometry defined above (see Appendix \ref{app:appD}), we finally obtain that the condition number of $\mat{Q_L}$ is bounded.
Similarly, the condition number of $\mat{Q_R}$ can be shown to be bounded.

Subsequently, we investigate the conditioning of $\mat{M_B}$. To this end, $\mat{M_B}$ is first split into two matrices
\begin{equation}\label{eq:bound1-block}
 \mat{M_B} 
 = \mat{M_D} + \mat{M_O} = \begin{pmatrix}
  \mat{M_{\Lambda \Lambda}} & \mat{0}\\
  \mat{0} & \mat{M_{\Sigma \Sigma}}
 \end{pmatrix}
 +\begin{pmatrix}
  \mat{0} & \mat{M_{\Lambda \Sigma}}\\
  \mat{M_{\Sigma \Lambda}} & \mat{0}
 \end{pmatrix},
\end{equation}
representing its diagonal blocks and its off-diagonal blocks, respectively. To avoid that the off-diagonal blocks of $\mat{M_O} $ render $\mat{M_B}$ singular, the following inequality should hold true
\tdg{AM: what does it mean to perturb?}
\begin{equation}\label{eq:final-bound2}
 \|\mat{M_O}\| \leq \alpha s_\text{min}(\mat{M_D})\,,
\end{equation}
in which $ s_\text{min}(\mat{M_D})$ is the minimum singular value of $\mat{M_D}$ and the scalar $\alpha$ ($0<\alpha< 1$) determines how strictly the inequality should be respected. Using the fact that $
\|\mat{M_O}\| = \max(\|\mat{M_{\Lambda \Sigma}}\|,\|\mat{M_{\Sigma \Lambda}}\|) \label{eq:diag-bound}$ and $s_\text{min}(\mat{M_D}) = \min ( s_\text{min}(\mat{M_{\Lambda \Lambda}}), s_\text{min}(\mat{M_{\Sigma \Sigma}}))$, the inequality \eqref{eq:final-bound2} becomes
\begin{equation}\label{eq:final-bound}
 \max(\|\mat{M_{\Lambda \Sigma}}\|,\|\mat{M_{\Sigma \Lambda}}\|) \leq \alpha \min ( s_\text{min}(\mat{M_{\Lambda \Lambda}}), s_\text{min}(\mat{M_{\Sigma \Sigma}}))\,.
\end{equation}
To know under which condition \eqref{eq:final-bound} is satisfied, bounds for $\|\mat{M_{\Lambda \Sigma}}\|$, $\|\mat{M_{\Sigma \Lambda}}\|$, $ s_\text{min}(\mat{M_{\Lambda \Lambda}})$, and $ s_\text{min}(\mat{M_{\Sigma \Sigma}})$ are investigated next. The following result, proven in Appendix \ref{app:appF}, is employed to determine a lower bound for the minimum singular value of a sum of matrices
 \begin{equation} \label{eq:lb_sigmamin}
  s_\text{min}(\mat{A} + \mat{B}) \geq \|\mat{A}\| \frac{\text{cond}(\mat{A}) - 1}{\text{cond}(\mat{A})^2} \frac{ s_\text{min}(\mat{A}) - \| \mat{B} \| }{s_\text{min}(\mat{A}) + \| \mat{B} \| }\,,
\end{equation}
it applies when $\mat{A}$ is an invertible matrix and $\|\mat{B} \| < s_\text{min}(\mat{A})$.
Before employing this identity on the diagonal blocks of $\mat{M_D}$, we introduce the following matrices
\begin{gather}
  \mat{A_\Lambda} = \gamma_\Lambda \transpose{\mat{Q_{\Lambda L}}} \mat{Q_{\Lambda R}}\,,\\
  \mat{B_\Lambda} = \gamma_\Lambda \transpose{\mat{Q_{\Lambda L}}} \mat{G_\epsilon}^{-1} \mat{Z_A} \mat{Q_{\Lambda R}}\,,\\
  \mat{A_\Sigma} = \gamma_\Sigma \transpose{\mat{Q_{\Sigma L}}} \mat{G}^{-1} \mat{Z_\Phi} \mat{Q_{\Sigma R}}\,,\\
  \mat{B_\Sigma} = \gamma_\Sigma \transpose{\mat{Q_{\Sigma L}}} \mat{G}^{-1} (\mat{G_\epsilon} + \mat{Z_A})\mat{Q_{\Sigma R}}\,,
\end{gather}
which represent the terms in the diagonal blocks of $\mat{M_D}$ (\eqref{eq:diagup} and \eqref{eq:diagdown}). Then, using the fact that $\mat{A_\Lambda}$ and $\mat{A_\Sigma}$ are invertible ($\mat{G}$ and $\mat{Z_\Phi}$ being non-singular) and restricting ourselves to the case in which
\begin{equation}
  s_\text{min}(\mat{A_\Lambda}) > \|\mat{B_\Lambda}\| \label{eq:cond_diagUp};~
  s_\text{min}(\mat{A_\Sigma}) > \|\mat{B_\Sigma}\|\,,
\end{equation}
\tdg{AM: enforcing how? By playing on the $\gamma$ ? If so it is not clear, especially since they popped up out of nowhere}
the minimum singular values of $\mat{M_{\Lambda \Lambda}}$ and $\mat{M_{\Sigma \Sigma}}$ can be bounded as
\begin{gather}
  s_\text{min}(\mat{M_{\Lambda \Lambda}}) 
 \geq f(\mat{A_\Lambda}) \frac{s_\text{min}(\mat{A_\Lambda}) - \| \mat{B_\Lambda} \|}{s_\text{min}(\mat{A_\Lambda}) + \| \mat{B_\Lambda} \| }\label{eq:lb_blockL1}\,,\\
  s_\text{min}(\mat{M_{\Sigma \Sigma}}) \geq f(\mat{A_\Sigma}) \frac{s_\text{min}(\mat{A_\Sigma}) - \| \mat{B_{\Sigma}} \|}{ s_\text{min}(\mat{A_\Sigma}) + \| \mat{B_{\Sigma}} \|}\,,\label{eq:lb_blockS1}
\end{gather}
in which $f(\mat{A}) = \|\mat{A}\| (\text{cond}(\mat{A}) - 1)/(\text{cond}(\mat{A})^2)$. Next, it is shown in Appendix \ref{app:appE} that the minimum and maximum singular values of the scaled Gram matrix $\mat{G_\epsilon}$ ($s_\text{min}(\mat{G_\epsilon})$ and $\|\mat{G_\epsilon}\|$, respectively) can be bounded as $s_\text{min}(\mat{G_\epsilon}) = \| \mat{G_\epsilon}^{-1} \|^{-1} \geq { s_\text{min}(\mat{G}) \omega}/{\sigma_\text{max}}$ and $\|\mat{G_\epsilon}\| \leq {\|\mat{G}\| \omega}/{\sigma_\text{min}}$. Using these bounds and the fact that $\mat{Z_A}$ is only frequency-dependent when $\kappa \approx 1$ (high-contrast regime), we can extract the conductivity and frequency dependence from $\mat{B_\Lambda}$ and $\mat{B_\Sigma}$ \tdg{AM: is this enough B lambda and B simgma contain Z A too} in \eqref{eq:lb_blockL1} and \eqref{eq:lb_blockS1}. The norm of the matrix $\mat{B_\Lambda}$ can be bounded as
\begin{equation} \label{eq:bound_BLambda}
\begin{split}
 \|\mat{B_\Lambda} \| &\leq
  \gamma_\Lambda \|\transpose{\mat{Q_{\Lambda L}}}\| \|\mat{G_\epsilon}^{-1}\| \|\mat{Z_A} \mat{Q_{\Lambda R}}\| \\
   &\leq \gamma_\Lambda k_0^2 \frac{\sigma_\text{max} \|\mat{G}^{-1}\|}{\omega} \| \mat{Z^1_A} \mat{Q_{\Lambda R}}\|\,,
 \end{split}
\end{equation}
\tdg{AM: you are using a result on $\|\mat{G_\epsilon}^{-1}\|$ that is stated nowhere}
where $\mat{Z_A} = k_0^2 \mat{Z^1_A}$. Then, $\|\mat{B_\Sigma}\|$ can be bounded from above as
\begin{equation} \label{eq:bound_BSigma}
\begin{split}
 \|\mat{B_\Sigma} \| & \leq \gamma_\Sigma \left(\| \transpose{\mat{Q_{\Sigma L}}} \mat{G}^{-1} \mat{G_\epsilon} \mat{Q_{\Sigma R}}\| + \| \transpose{\mat{Q_{\Sigma L}}} \mat{G}^{-1} \mat{Z_A}\mat{Q_{\Sigma R}}\|\right) \\
 &\leq \gamma_\Sigma \left( \|\transpose{\mat{Q_{\Sigma L}}} \mat{G}^{-1} \|\frac{\|\mat{G}\| \omega}{\sigma_\text{min}} + k_0^2\| \transpose{\mat{Q_{\Sigma L}}} \mat{G}^{-1} \mat{Z^1_A}\mat{Q_{\Sigma R}}\|\right).
 \end{split}
\end{equation}
Leveraging \eqref{eq:bound_BLambda}, \eqref{eq:bound_BSigma}, \eqref{eq:lb_blockL1}, \eqref{eq:lb_blockS1}, and the fact that the inequalities $\|\mat{U}\| \leq c$ ($c \in \R_+$), $s \geq ({a - \|\mat{U}\|})/({a + \|\mat{U}\|})$ ($a \in \R_+^*$), and $\|\mat{U}\| < a$ imply $  s \geq {(a - c)}/{(a + c)}$, the following lower bounds for the minimum singular values of $\mat{M_{\Lambda \Lambda}}$ and $\mat{M_{\Sigma \Sigma}}$, which are a function of $r_\sigma$ and $\omega$, are obtained
\begin{gather}
  s_\text{min}(\mat{M_{\Lambda \Lambda}}) 
 \geq f(\mat{A_\Lambda}) \frac{s_\text{min}(\mat{A_\Lambda}) - r_\sigma \omega \tau_\Lambda}{s_\text{min}(\mat{A_\Lambda}) + r_\sigma \omega \tau_\Lambda }\label{eq:lb_blockL}\,,\\
  s_\text{min}(\mat{M_{\Sigma \Sigma}}) \geq f(\mat{A_\Sigma}) \frac{s_\text{min}(\mat{A_\Sigma}) - \omega^2 \tau_{\Sigma A} - \omega \tau_{\Sigma G} }{ s_\text{min}(\mat{A_\Sigma}) + \omega^2\tau_{\Sigma A} + \omega \tau_{\Sigma G} }\,, \text{ in which}\label{eq:lb_blockS} \\
 \tau_\Lambda = \gamma_\Lambda \sigma_\text{min} / c_0^2 \|\mat{G}^{-1}\| \| \mat{Z^1_A} \mat{Q_{\Lambda R}}\|\,, \\
 \tau_{\Sigma A} =\gamma_\Sigma / c_0^2 \| \transpose{\mat{Q_{\Sigma L}}} \mat{G}^{-1} \mat{Z^1_{A}} \mat{Q_{\Sigma R}}\| \,,\\
 \tau_{\Sigma G} =\gamma_\Sigma \sigma_\text{min}^{-1} \| \transpose{\mat{Q_{\Sigma L}}} \mat{G}^{-1} \| \|\mat{G}\|\,,
\end{gather}
are parameters that do not depend of $\omega$ and $r_\sigma$ ($\kappa \approx 1$ in this regime).
Provided that \eqref{eq:cond_diagUp} holds, we thus obtain lower bounds for the minimum singular values of $\mat{M_{\Lambda \Lambda}}$ and $\mat{M_{\Sigma \Sigma}}$ (\eqref{eq:lb_blockL} and \eqref{eq:lb_blockS}) \tdg{AM: you need an argument to use the bounds in the orignial inequality (equlity inside equlatity)} which translates into the non-singularity of $\mat{M_D}$ in \eqref{eq:bound1-block}. In that case, identity \eqref{eq:lb_sigmamin} can be applied to \eqref{eq:bound1-block} to find a lower bound for the minimum singular value of $\mat{M_B}$. Before applying \eqref{eq:lb_sigmamin}, we also need to enforce that the norms of $\mat{M_{\Lambda \Sigma}}$ and $\mat{M_{\Sigma \Lambda}}$ and the minimum singular values of $\mat{M_{\Lambda \Lambda}}$ and $\mat{M_{\Sigma \Sigma}}$ respect inequality \eqref{eq:final-bound}. Using a similar approach as before, the off-diagonal blocks $\mat{M_{\Lambda \Sigma}}$ and $\mat{M_{\Sigma \Lambda}}$ can be bounded as
\begin{gather}
 \|\mat{M_{\Lambda \Sigma}}\| \leq r_\sigma \omega \tau_{\Lambda \Sigma A} \label{eq:ub-blockL}\,,\\
 \|\mat{M_{\Sigma \Lambda}}\| \leq \omega \tau_{\Sigma G} + \omega^2 \tau_{\Sigma \Lambda A} + \omega \tau_{\Sigma \Lambda \Phi}\,, \text{ in which}\label{eq:ub-blockS} \\
 \tau_{\Lambda \Sigma A} = \gamma_\Lambda \sigma_\text{min} / c_0^2 \|\mat{G}^{-1}\| \| \mat{Z^1_A} \mat{Q_{\Sigma R}}\| \,,\\
 \tau_{\Sigma \Lambda A} = \gamma_\Sigma / c_0^2 \| \transpose{\mat{Q_{\Sigma L}}} \mat{G}^{-1} \mat{Z^1_{A}} \mat{Q_{\Lambda R}}\| \,,\\
 \tau_{\Sigma \Lambda \Phi} =\gamma_\Sigma \sigma_\text{min}^{-1} \| \transpose{\mat{Q_{\Sigma L}}} \mat{G}^{-1} \| \|\mat{Z^1_{\Phi,1\epsilon}}\|\,,
\end{gather}
such that $\|\mat{Z_{\Phi,1\epsilon}}\| \leq \omega/\sigma_\text{min}\|\mat{Z^1_{\Phi,1\epsilon}}\|$ in which $\mat{Z^1_{\Phi,1\epsilon}}$ is a permittivity-independent matrix that follows the relation $\mat{Z_{\Phi,1\epsilon}} = \mat{Z^1_{\Phi,1\epsilon}} \mat{\Upsilon_{\deltaKappa}}$ with $\mat{\Upsilon_{\deltaKappa}}$ being a diagonal matrix filled as $[\mat{\Upsilon_{\deltaKappa}}]_{nn} = \epsilon_0^{-1} \deltaKappa_n$. This last inequality can be obtained by observing that $\|\mat{\Upsilon_{\deltaKappa}}\| \leq \omega / \sigma_{\text{min}} $. 

Finally, we can enforce the conditions \eqref{eq:final-bound} and \eqref{eq:cond_diagUp} using the bounds derived in \eqref{eq:bound_BLambda}, \eqref{eq:bound_BSigma}, \eqref{eq:lb_blockL}, \eqref{eq:lb_blockS}, \eqref{eq:ub-blockL}, and \eqref{eq:ub-blockS}. These conditions being enforced, the resulting condition number of $\mat{M_B}$ can be bounded from above as
\begin{equation} \label{eq:cond_MB}
\begin{split}
 \cond{\mat{M_B}} 
 &\leq \cond{\mat{M_D}\left( \identity + \mat{M_D}^{-1} \mat{M_O} \right)} \\
 &\leq \cond{\mat{M_D}} \frac{1 + \|\mat{M_D}^{-1} \mat{M_O}\|}{1 - \|\mat{M_D}^{-1} \mat{M_O}\|} \\
 &\leq \frac{\max(\|\mat{M_{\Lambda \Lambda}}\|,\|\mat{M_{\Sigma \Sigma}}\|)}{\min(s_\text{min}(\mat{M_{\Lambda \Lambda}}),s_\text{min}(\mat{M_{\Sigma \Sigma}}))} \frac{1 + \alpha}{1 - \alpha}\,,
 \end{split}
\end{equation}
in which the condition \eqref{eq:final-bound} was employed. An upper bound for the condition number of $\mat{M}$ can then be obtained from \eqref{eq:cond_MB} using \eqref{eq:bound_cound_M}.

The variables in these bounds are the frequency and the maximum conductivity ratio in the object $r_\sigma$ which are dictated by the electromagnetic problem to solve, and the parameter $\alpha$ that should be chosen depending on the range of validity requirement. 
\tdg{AM: $r_\sigma$ is not really free, is it? It is dictated by the scenario} 
\td{AM: this thing on $\alpha$ is still not clear: Fixed}
The values of $\omega$, $r_\sigma$, and $\alpha$ for which the bounds above-mentioned are respected define the range in which the formulation is well behaving. Note that the parameter $\alpha$ allows tuning simultaneously the bound for the condition number of $\mat{M}$ and the range of values of $\omega$ and $r_\sigma$ for which the conditions are met, e.g., a low value of $\alpha$ results in a smaller upper bound for the condition number in \eqref{eq:cond_MB} but it also makes the condition \eqref{eq:final-bound} harder to respect.\tdb{AM: I fear this is not super clear...}

The bounds previously derived establish that, in the static limit, the conditions \eqref{eq:final-bound} and \eqref{eq:cond_diagUp} are met regardless of the maximum conductivity ratio $r_\sigma$ and of the parameter $\alpha$ ($0<\alpha< 1$) and the upper bound for the condition number of $\mat{M_B}$ given in \eqref{eq:cond_MB} also becomes independent of $r_\sigma$ and $\alpha$. Moreover, since the condition number of $\mat{Q_L}$ and $\mat{Q_R}$ is bounded when $r_\sigma \rightarrow \infty$, it results from \eqref{eq:bound_cound_M} that in the static limit the condition number of $\mat{M}$ is bounded as the maximum conductivity ratio increases. Away from that limit, $\mat{M}$ will still be properly conditioned as long as the bounds above-mentioned are enforced. To show the proper behavior of this formulation beyond the static limit, some values of $r_\sigma$ and the frequency for which the bounds are respected are given in the numerical results section. Finally, the reader should note that at high-frequency, when low-frequency regularization is not required, the standard D-VIE can be used without changing the implementation by simply setting the coefficients and matrices multiplying the projectors in \eqref{eq:precond_hc-ns} to \num{1} and $\mat{\identity}$, respectively.

\section{Implementation Related Details} \label{sec:implementation}

This section presents the details related to the implementation of the proposed preconditioner. Starting from an existing D-VIE, the quasi-Helmholtz projectors must be computed efficiently not to deteriorate the overall complexity of the solver. The main difficulty to build the projectors resides in the inversion of the stiffness matrix $\transpose{\mat{\Sigma}} \mat{G_\epsilon}^{-1} \mat{\Sigma}$, which has a condition number that grows when the discretization of the geometry is refined. To remedy this issue, this operation can be done leveraging the algebraic multigrid (AMG) method as preconditioner \cite{haase2002parallel} together with an iterative solver. Besides, since the Gram matrix $\mat{G_\epsilon}$ is sparse and symmetric, its direct inverse can be obtained in an efficient manner using a multifrontal solver \cite{10.1145/356044.356047}.

In addition to this, however, we have found also a very effective strategy that does not require the inversion of $\mat{G_{\epsilon}}$, resulting in less computational and implementation efforts. In fact, we observed numerically that using the real part or the imaginary part of the diagonal of the Gram matrix in the stiffness matrix also makes the formulation stable. The set of projectors in this case is defined as follows
\begin{gather}
 \mat{P^{\Sigma}_D} = \mat{D}^{-1} \mat{\Sigma} (\mat{\Sigma}^\mathrm{T} \mat{D}^{-1} \mat{\Sigma})^{+} \mat{\Sigma}^\mathrm{T}\,,\\
 \mat{P^{\Lambda}_D} = \identity - \mat{P^{\Sigma}_D} = \mat{\Lambda} (\mat{\Lambda}^\mathrm{T} \mat{D} \mat{\Lambda})^{-1} \mat{\Lambda}^\mathrm{T} \mat{D}\,,
\end{gather}
where $\mat{D} \in \Ci^{N_F \times N_F}$ is a diagonal matrix constructed as
\begin{equation} 
  \mat{D}= 
  \begin{cases} \phantom{\mi} \Re \left(\mat{D_{G}} \right) & \text{ if } \|\Re \left(\mat{D_{G}}^{-1} \right)\| > \|\Im \left( \mat{D_{G}}^{-1} \right) \|\\ 
  \mi \Im\left(\mat{D_{G}}\right) & \text{ if } \|\Re \left(\mat{D_{G}}^{-1} \right)\| < \|\Im \left( \mat{D_{G}}^{-1} \right) \|\,,
  \end{cases}
  \label{eq:def_D}
\end{equation} 
with $\mat{D_{G}}$ being a diagonal matrix filled with the diagonal of $\mat{G_{\epsilon}}$. Similarly as in \eqref{eq:precond_equation_G_ns}, the preconditioned D-VIE with $\mat{P^{\Sigma}_D}$ and $\mat{P^{\Lambda}_D}$ reads
\begin{gather} 
 \mat{L_D} \mat{Z} \mat{\alpha} = \mat{L_D} \mat{v}\,, \text{ in which} \label{eq:precond_equation_D}\\
 \mat{L_D} = \frac{\mat{P^{\Lambda}_D} \mat{D}^{-1}}{\| \mat{P^{\Lambda}_D} \mat{D}^{-1} \mat{G_\epsilon} \mat{P^{\Lambda}_D} \|} \ + \frac{\mat{P^{\Sigma}_D} \mat{D_0}^{-1}}{\| \mat{P^{\Sigma}_D} \mat{D_0}^{-1} \mat{Z_\Phi} \mat{P^{\Sigma}_D} \|}\,,
\end{gather}
and $\mat{D_0}$ is a diagonal matrix filled with the diagonal of $\mat{G}$.

This new preconditioner gives rise to a slightly higher condition number than with the preconditioner obtained from $\mat{G_{\epsilon}}$, but has been numerically verified to be stable in a broad frequency range and for high permittivity objects, as illustrated in next section. 
Moreover the stiffness matrix $\mat{\Sigma}^\mathrm{T} \mat{D}^{-1} \mat{\Sigma}$, which is also a weighted graph Laplacian matrix, can be inverted efficiently using an aggregation-based AMG method from \cite{AGMG, notay2010aggregation} together with a conjugate gradient (CG) algorithm.

\section{Numerical Results} \label{sec:results}


To further corroborate the theoretical developments, the new formulation has been tested in several scenarios.
Both regularization with projectors scaled with the full Gram matrix (equation \eqref{eq:precond_equation_G_ns}, referred in the following as ``Regularized D-VIE $\mat{G_{\epsilon}}$'') and with projectors scaled with the diagonal matrix $\mat{D}$ (equation \eqref{eq:precond_equation_D}, referred in the following as ``Regularized D-VIE $\mat{D}$'') will be considered. These two regularized formulations will be of course compared to the standard D-VIE but also, for the sake of completeness, to a Loop-Star D-VIE which can be obtained by selecting $\mat{A} = \identity$ in \eqref{eq:normDecomposition} and rescaling the solenoidal and non-solenoidal parts of $\mat{B^{\identity}_{\Lambda \Sigma}}$ by the coefficients introduced in \eqref{eq:props}, respectively. Note that this Loop-Star decomposition is the direct extension of the decomposition employed for surface formulations \cite{qH_Andriulli}.

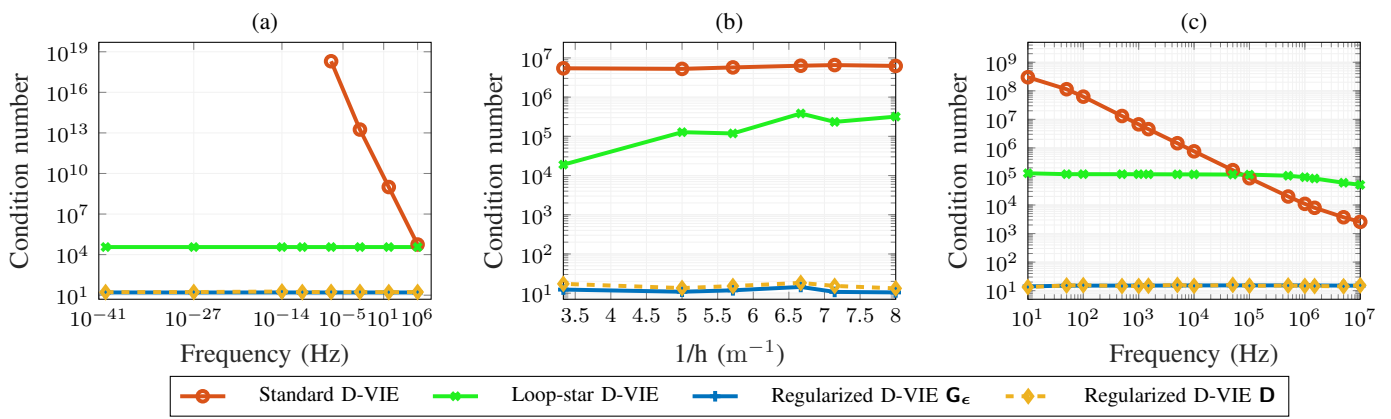
\begin{figure*} \centering
  \input{figures/gp_plot}
  \caption{Conditioning of the the standard D-VIE, the Loop-Star D-VIE, and the two proposed D-VIE computed on a 3-layer conductive sphere (a) in the low frequency regime. (b) in the dense discretization regime. (c) as a function of the frequency using frequency-dependent complex permittivities corresponding to the permittivities of the brain, the skull cortical bone, and the skin in the \num{3} layers of the spherical model.}
  \label{fig:plots_cond}
\end{figure*}

In the first test, the geometry used is composed of 3 homogeneous concentric spheres (i.e. three layer head model) with radii $\SI{87}{\milli \meter}$, $\SI{92}{\milli \meter}$, and $\SI{100}{\milli \meter}$ and respective normalized conductivities $\sigma_1=1$, $\sigma_2=1/15$, and $\sigma_3 = 1$, which represent the conductivities of brain, skull, and scalp in the quasi-static regime \cite{conductivity}. First, we verify the conditioning of the system matrices of the two regularized D-VIE, the Loop-Star D-VIE, and the standard D-VIE as a function of the frequency (Fig.~\ref{fig:plots_cond}a). The condition number of the standard D-VIE grows as the frequency decreases while the condition number of the two regularized D-VIE and the Loop-Star D-VIE remains constant until very low frequencies, which confirms the curative effects of our preconditioners. It should also be noted that, although both are constant, the condition number of the formulations we propose here is much lower than the condition number of the Loop-Star scheme, as expected.

%

\begin{figure*}
   \centering%
   \subfloat[]{\input{figures/bound_theo_sigma_f}\label{fig:bound_theo} }
   \subfloat[]{\input{figures/cond_sigma_freq_1d}\label{fig:bound_num}}
   \caption{(a) Region map showing the range (colored region) of frequency and maximum conductivity ratio $r_\sigma$ for which this new formulation is proven to be stable. The colorbar represents the theoretical upper bound for the condition number. The ranges of operation of several bioelectromagnetic applications are shown on the map. (b) Condition number (CN) of the preconditioned matrix $\mat{M}$ numerically obtained and upper bound for the condition number of $\mat{M}$ theoretically obtained as a function of the frequency and the maximum conductivity ratio.}
   \tdg{AM: I do not understand the usage of the colobars... They do not look adapted to the data}
\end{figure*}
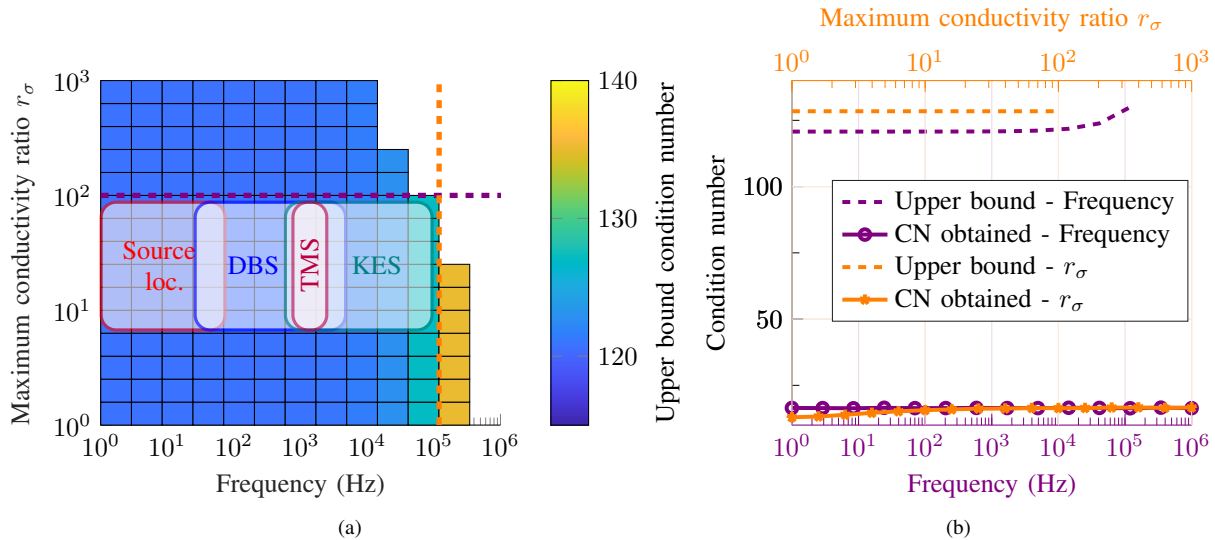

Subsequently, a second numerical example tests the range of validity of the bounds for the ``regularized D-VIE $\mat{G_{\epsilon}}$'' as a function of the maximum conductivity ratio $r_\sigma$ and the frequency (Fig.~\ref{fig:bound_theo}). The geometry used in this example is the three layer spherical model above-mentioned with $\sigma_2=1/50$ and $\sigma_1=\sigma_3 =r_\sigma \sigma_2$. Note that this scenario represents the high brain-to-skull contrast problem, which is a well-known limitation for both static and full-wave bioelectromagnetic solvers \cite{hamalainen1989realistic, genccer2005use}. The range shown in Fig.~\ref{fig:bound_theo} (colored region) corresponds to the frequency and $r_\sigma$ for which the bounds provided in \Cref{sec:bounds} are respected. Note that the parameter $\alpha$ used in condition \eqref{eq:final-bound} is set to $0.3$. The operating frequency and maximum conductivity ratio in EEG source localization and three other applications of interest, deep brain stimulation (DBS) \cite{howell2015effects}, transcranial magnetic stimulation (TMS) \cite{gomez2020conditions}, and kilohertz electrical stimulation (KES) \cite{patel2018challenges} are indicated on top of the map. 
Fig.~\ref{fig:bound_theo} also shows the upper bound for the condition number of $\mat{M}$ as a function of the frequency and $r_\sigma$ whose expression is given in \eqref{eq:cond_MB}. In the range of validity of the formulation, the condition number is bounded by \num{140}, which is reasonably low. Moreover, Fig.~\ref{fig:bound_num} shows the condition number of the ``regularized D-VIE $\mat{G_{\epsilon}}$'' obtained numerically for the values of frequency and $r_\sigma$ that are in the ranges shown with the violet and orange dashed lines in Fig.~\ref{fig:bound_theo}, respectively. The theoretical upper bound for the condition number of $\mat{M}$ as a function of these values is also shown in Fig.~\ref{fig:bound_num}. As expected, the condition number remains below the upper bound derived in \Cref{sec:bounds}. In Fig.~\ref{fig:bound_num} one can notice that the condition number remains low even when the theoretical bound does not apply.

Another numerical example illustrating the proper behavior of the regularized D-VIE formulations is provided next. In this case, the complex permittivities in the 3 layers correspond to the permittivities in the brain, the skull cortical bone, and the skin \cite{hasgall2018database}, which are a function of the frequency. Fig.~\ref{fig:plots_cond}c shows that while the Loop-Star D-VIE and the standard D-VIE are poorly conditioned due to the low-frequency breakdown and/or the high internal contrast problem, the ``regularized D-VIE $\mat{G_\epsilon}$'' and the ``regularized D-VIE $\mat{D}$'' remain stable. Note that the realistic permittivities employed here do not always have their imaginary part that is dominant as it was assumed in the theoretical treatment (\Cref{sec:bounds}), nevertheless, the proposed formulations result stable in this realistic scenario too.

The dense discretization conditioning of the two new D-VIE is verified by increasing the discretization of a homogeneous sphere of radius $\SI{1}{\meter}$, relative permittivity $15$, and normalized conductivity $1$ (Fig.~\ref{fig:plots_cond}b). Both the traditional and the regularized D-VIE do not experience a dense mesh instability, which shows that our preconditioners do not introduce a new breakdown, unlike the D-VIE preconditioned with a standard Loop-Star decomposition, which has its condition number that grows unbounded when the discretization of the geometry is refined.

%

\begin{figure} \centering
 \input{figures/potential_multilayer_sphere.tex}
 \caption{Potential obtained at the surface (in the tetrahedra of the mesh) of a 3-layer conductive sphere. The excitation is an electric point dipole.}
 \label{fig:pot-multi-sphere}
\end{figure}
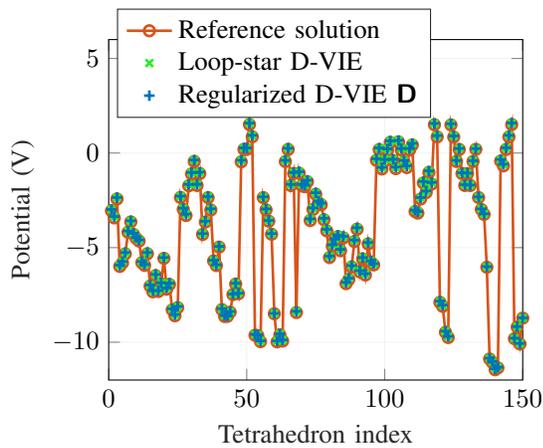

To complement our stability experiments we verify the correctness of the new formulation against references in the quasi-static regime and at higher frequencies. First, we establish the correctness of the formulation in an electroencephalography (EEG) setting by comparing the potential radiated by an electric point dipole on the surface of a conductive object since they are considered good models for focal brain activity \cite{dipole_neural}.
The electric point dipole source has a moment of $\begin{bmatrix} \num{0} & \num{0} & \num{1} \end{bmatrix}$ and an eccentricity of $\SI{43}{\percent}$ in a 3-layer sphere with the aforementioned conductivities (Fig.~\ref{fig:pot-multi-sphere}). The frequency used in the simulation is ${10^{-40}}{\SI{}{Hz}}$, although neurons operate at a frequency between $\SI{0.1}{Hz}$ and $\SI{100}{Hz}$. The reason behind this choice is to show that the formulation does not suffer from a loss of significant digits at very low frequencies unlike standard full-wave solvers. The potential obtained in the tetrahedra of the mesh with the new formulation and the Loop-Star D-VIE shows a good agreement with the well-established reference solution and confirms the applicability of the new D-VIE in a typical biomedical setting.

At higher frequencies, we use the previous 3-layer sphere with the relative permittivities $\epsilon_{r,1}=70$, $\epsilon_{r,2}=15$, and $ \epsilon_{r,3}=65$ and conductivities $\sigma_1=0.55$, $\sigma_2=0.75$, and $\sigma_3 = 1.42$. These values match the relative permittivities and conductivities of the brain, skull, and skin at $f = \SI{100}{\mega \hertz}$ \cite{Gabriel_1996}. The excitation is a plane wave and the reference solution is obtained analytically from the Mie series. This numerical test results in a relative error with respect to the reference below $\SI{0.1}{\percent}$, which shows that the regularization does not deteriorate the solution accuracy at higher frequencies.

Now that the correctness and stability of the proposed formulation have been verified in canonical settings, we verify its applicability to challenging, realistic bio-electromagnetic compatible scenarios.
To assess the low- and high-frequency versatility of the new formulation we study its applicability to the bio-electromagnetic modeling of the human head. The head geometry used for these simulations has been obtained from the segmentation of an MRI image of $256 \times 256 \times 256$ voxels in FieldTrip \cite{oostenveld2011fieldtrip}, subsequently discretized into $\num{44733}$ tetrahedra.

\begin{figure} \centering
	 \includegraphics[width=0.75\columnwidth]{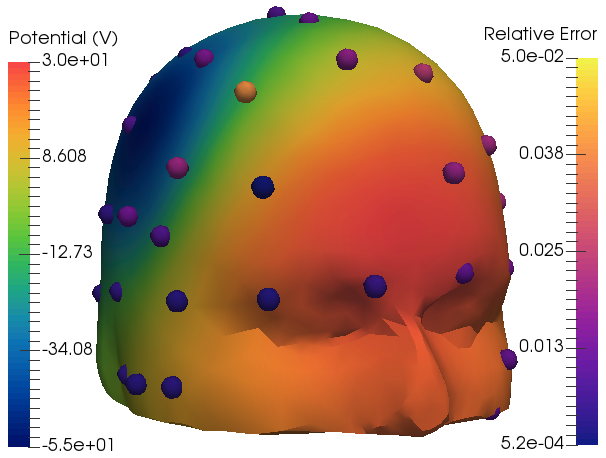}
	 \caption{3D visualization of the potential radiated from a current dipole oscillating at $\SI{1}{\hertz}$ in brain obtained on the scalp of the head (left color bar). The colored dots represent the 65 EEG electrodes with the corresponding relative error (right color bar).}
	 \label{fig:pot_scalp_3d} 
\end{figure}

At low frequencies, the formulation is applied to the problem of brain source localization which aims at retrieving the neural activity from the potential recordings on the scalp measured by EEG. This inverse problem has numerous applications ranging from epilepsy diagnostic \cite{Claassen1036} to brain computer interface \cite{CINCOTTI200831}.
Solving the inverse problem -- the mapping from EEG scalp measurement to the current distribution inside the head -- requires solving the forward problem multiple times, which is the mapping from individual current sources to scalp potentials.
In the distributed approach, the individual current sources are placed on a grid covering the parts of interest in the brain. In Fig.~\ref{fig:pot_scalp_3d}, we show the potential ``radiated'' on the scalp by a single intracranial current source oscillating at $f=\SI{1}{\hertz}$. The conductivities of the different homogeneous layers -- the scalp, the skull, and the brain -- are $\sigma_\mathrm{scalp}=1$, $\sigma_\mathrm{skull}=1/15$, and $ \sigma_\mathrm{brain} = 1$ \cite{CINCOTTI200831}. The scalp potential, measured at 65 electrodes, is compared to a reference solution obtained from a FEM solver. The relative error obtained at each electrode remains below $\SI{4}{\percent}$ (Fig.~\ref{fig:pot_scalp_3d}), which confirms the use of the regularized D-VIE to solve the forward problem in EEG source reconstruction.

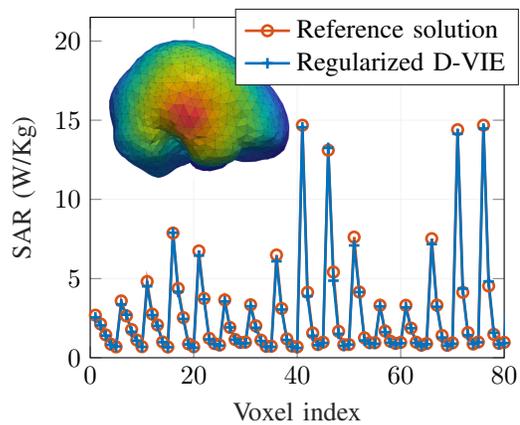
\begin{figure} \centering
	 \input{figures/sar_full_head_smoother.tex}
	 \caption{Comparison of the SAR obtained in a cubic voxel subdivision of the brain at $\SI{100}{\mega \hertz}$. The magnitude of the electric field obtained at the surface of the tetrahedral discretization of the brain is shown in the top left corner.}
	 \label{fig:sar_head_comparison} 
\end{figure}

At higher frequencies, the new formulation can be applied in the field of radiation dosimetry in the brain. It consists in the quantification of the specific absorption rate (SAR) radiated by a given source in human tissues \cite{sar_recommendation_upto300}. In this scenario, we use the same head geometry as for the EEG source localization with $\epsilon_{r,\mathrm{scalp}}=70$, $\epsilon_{r,\mathrm{skull}}=15$, and $ \epsilon_{r,\mathrm{brain}}=65$ for the relative permittivities and $\sigma_\mathrm{scalp}=0.55$, $\sigma_\mathrm{skull}=0.75$, and $\sigma_\mathrm{brain} = 1.42$ for the conductivities. The head is illuminated by an electric dipole placed at $\SI{10}{\centi \meter}$ from its right side, with dipole moment of $\begin{bmatrix}0&0&1\end{bmatrix}$, and oscillating at $f = \SI{100}{\mega \hertz}$. The brain is divided into voxels of side length $\SI{3}{\centi \meter}$. The SAR is computed in each voxel $n$ as $\sigma_n |\vt{E_n}|^2/2\rho_n$, where $|\vt{E_n}|$ is the norm of the electric field averaged over the voxel $n$ and $\rho_n$ its mass density in $\si{\kilo \gram \per \meter \cubed}$. The SAR obtained with the proposed D-VIE is consistent with the reference solution, which is obtained with a FEM solver (Fig.~\ref{fig:sar_head_comparison}).

\section{Conclusion}
We introduced a novel volume integral equation for modeling dielectric and conductive materials with high-contrast in a broad frequency range. The new D-VIE scheme leverages scaled volume quasi-Helmholtz projectors to cure both its high-contrast and low-frequency ill-conditioning without deteriorating its dense discretization behavior. The scaling of the oblique quasi-Helmholtz projectors allows us to re-scale the equation when applied to inhomogeneous objects with high complex permittivities. Numerical examples illustrate the stability and accuracy of this new method. The preconditioned D-VIE shows good accuracy in the biomedical applications presented in this paper, both in the quasi-static regime and at higher frequencies. 

\section*{Acknowledgment}
This work has been funded in part by the European Research Council (ERC) under the European Union’s Horizon 2020 research and innovation program (ERC project 321, grant No.724846) and by the French National Research Agency (ANR) through the Labex CominLabs (project CYCLE).

\begin{appendices} 


\section{Validity of the Scaled Loop-Star Decomposition} \label{app:appA}
In this appendix, it is shown that
\tdg{AM: You ar enot showing anything on the projectors (at least not directly)}
any SWG coefficient vector $\mat{\alpha}$ can be decomposed with the scaled Loop-Star decomposition
\begin{equation}\label{eq:scaled_ls_decomposition_nu}
 \mat{\alpha} =\mat{A} \mat{\Sigma} \mat{\tilde{s}}+ \mat{\Lambda} \mat{\tilde{l}}\,,
\end{equation}
where $\mat{A} \in \mathbb{C}^{N_F \times N_F}$ is an invertible complex symmetric matrix with one of the (potentially many different) square root matrices denoted by $\mat{A}^{\frac{1}{2}}$ and $\mat{\tilde{l}}$ and $\mat{\tilde{s}}$ are the coefficient vectors of the solenoidal and non-solenoidal basis functions in this new decomposition.
Without loss of generality, we show in the following that $\mat{\alpha}$ can be (uniquely) decomposed as follows
\begin{equation}\label{eq:scaled_ls_decomposition}
 \mat{\alpha} =\mat{A} \mat{\tilde{\Sigma}} \mat{\bar{s}}+ \mat{\Lambda} \mat{\tilde{l}}\,,
\end{equation}
in which we recall that $\mat{\tilde{\Sigma}}$ is obtained by removing one column (i.e. one star function) from $\mat{{\Sigma}}$ and we introduce $\mat{\bar{s}}$, the coefficient vector of the Star part in this decomposition.
%

To prove \eqref{eq:scaled_ls_decomposition}, we first show that $\mat{A}^{-\frac{1}{2}}\mat{\alpha}$ can be orthogonally decomposed as
\begin{equation} \label{eq:decomp3}
\mat{A}^{-\frac{1}{2}}\mat{\alpha}=\mat{A}^{-\frac{1}{2}} \mat{\Lambda} \mat{\tilde{l}} + \mat{A}^{\frac{1}{2}} \mat{\tilde{\Sigma}} \mat{\bar{s}}\,,
\end{equation}
the existence (and unicity) of which is equivalent to the existence (and unicity) of \eqref{eq:scaled_ls_decomposition} since $\mat{A}^{\frac{1}{2}}$ is invertible. Since $\left(\mat{A}^{-\frac{1}{2}} \mat{\Lambda}\right)^\mathrm{T} \mat{A}^{\frac{1}{2}} \mat{\tilde{\Sigma}} = \mat{0}$, the decomposition is an orthogonal one and these two scaled transformation matrices have linearly independent column vectors. Then we need to show that the rank of the sum of $\mat{A}^{-\frac{1}{2}} \mat{\Lambda}$ and $\mat{A}^{\frac{1}{2}} \mat{\tilde{\Sigma}}$ is equal to the number of SWG basis functions $N_F$. Since $\text{rank}(\mat{A}^{-\frac{1}{2}} \mat{\Lambda}) = \text{rank}(\mat{\Lambda})$ ($\mat{A}^{\frac{1}{2}}$ is invertible), $\text{rank}(\mat{A}^{\frac{1}{2}} \mat{\tilde{\Sigma}}) = \text{rank}(\mat{\tilde{\Sigma}})$, $\text{rank}(\left[\mat{\tilde{\Sigma}} ~ \mat{\Lambda}\right]) = N_F$, and $\mat{A}^{-\frac{1}{2}} \mat{\Lambda}$ and $\mat{A}^{\frac{1}{2}}\mat{\tilde{\Sigma}}$ have their column vectors linearly independent, we have
 \begin{equation}
  \begin{split}
  \text{rank}\left(\left[\mat{A}^{\frac{1}{2}} \mat{\tilde{\Sigma}} ~~ \mat{A}^{-\frac{1}{2}}\mat{\Lambda} \right]\right) 
  &= \text{rank}(\mat{A}^{\frac{1}{2}} \mat{\tilde{\Sigma}}) + \text{rank}(\mat{A}^{-\frac{1}{2}}\mat{\Lambda}) \\
  &= \text{rank}(\mat{\tilde{\Sigma}}) + \text{rank}(\mat{\Lambda}) = N_F\,,
  \end{split}
 \end{equation}
from which the existence (and unicity) of \eqref{eq:decomp3} follows, which proves the existence (and unicity) of \eqref{eq:scaled_ls_decomposition} and hence the existence of \eqref{eq:scaled_ls_decomposition_nu}.


\section{Invertibility and Conditioning of the Normalized Loop-Star Decomposition Matrix}\label{app:appB}
In this appendix, we show that the normalized version of the loop/star-to-SWG decomposition matrix $\mat{B^{A^{-1}}_{\Lambda \Sigma}}$ introduced in \eqref{eq:normDecomposition} with $\mat{A}$ being a non-singular real symmetric matrix is non-singular and that an upper bound of its condition number can be obtained. We rewrite this matrix as $\mat{B^{A^{-1}}_{\Lambda \Sigma}} = [\matH{\Lambda}, ~ \matH{\Sigma} ]$ such that $\matH{\Lambda} = \mat{\Lambda} (\transpose{\mat{\Lambda}} \mat{\Lambda})^{-\frac{1}{2}}$ and $\matH{\Sigma} = \mat{A} \mat{\tilde{\Sigma}} (\transpose{\mat{\tilde{\Sigma}}} \mat{A}^{2} \mat{\tilde{\Sigma}})^{-\frac{1}{2}}$.

First, we expand $\mat{M_{LS}} = \transpose{\mat{B^{A^{-1}}_{\Lambda \Sigma}} } \mat{B^{A^{-1}}_{\Lambda \Sigma}} $ \tdg{AM: since $\mat A$ could be complex (c.f appendix A), you need the conjugate transpose} to study the singular values of $\mat{B^{A^{-1}}_{\Lambda \Sigma}} $
\begin{equation} \label{eq:block_mat_loop-star-normalized}
	 \mat{M_{LS}}	= \begin{bmatrix}
			\matH{\Lambda}^\mathrm{T} \matH{\Lambda} & \matH{\Lambda}^\mathrm{T} \matH{\Sigma}\\
			\matH{\Sigma}^\mathrm{T} \matH{\Lambda} & \matH{\Sigma}^\mathrm{T} \matH{\Sigma} 
		\end{bmatrix} 
	 = \begin{bmatrix}
	 \identity 
	 &\mat{O}\\
	 \transpose{\mat{O}}
	 	& \identity
	 \end{bmatrix}\,,
\end{equation}
in which $\mat{O} = (\transpose{\mat{\Lambda}} \mat{\Lambda})^{-\frac{1}{2}} \transpose{\mat{\Lambda}} \mat{A} \mat{\tilde{\Sigma}} (\transpose{\mat{\tilde{\Sigma}}} \mat{A}^{2}\mat{\tilde{\Sigma}})^{-\frac{1}{2}}$.
From \eqref{eq:block_mat_loop-star-normalized}, it results that $\mat{B^{A^{-1}}_{\Lambda \Sigma}}$ has no singular value at \num{0} if the norm of the off-diagonal block $\mat{O}$ is strictly lower than \num{1}\tdg{AM: how do you show this? I think you need at least that the matrix is hermitian/symmetric (it is) but not sure this is enough}. To show that this condition is fulfilled, some properties regarding the angle between complementary subspaces \cite{szyld2006many, ipsen1995angle} are employed in the following. Here the angle of interest is the angle between the subspaces $\mathcal{R}$ and $\mathcal{N}$ in $\mathbb{C}^{N_F}$, associated to the orthonormal bases formed by the column vectors of $\matH{\Sigma}$ and $\matH{\Lambda}$, respectively. 
%
By introducing the following orthogonal (symmetric) projectors onto $\mathcal{R}$ and $\mathcal{N}$ 
\begin{gather}
 \mat{P}^\mathcal{R} = \matH{\Sigma} \transpose{\matH{\Sigma}}= \mat{A} \mat{\Sigma} (\transpose{\mat{\tilde{\Sigma}}} \mat{A}^{2} \mat{\tilde{\Sigma}})^{-1} \transpose{\mat{\tilde{\Sigma}}} \mat{A}\,,\\
 \mat{P}^\mathcal{N} = \matH{\Lambda} \transpose{\matH{\Lambda}} = \mat{\Lambda} (\transpose{\mat{\Lambda}} \mat{\Lambda})^{-1} \transpose{\mat{\Lambda}}\,,
\end{gather}
the cosine of the minimal angle $\theta$ ($0\leq \theta \leq \pi/2$) between $\mathcal{R}$ and $\mathcal{N}$ can be defined as \cite{szyld2006many,ipsen1995angle}
\begin{equation}\label{eq:costhetadef_proj}
 \cos{\theta} = \|\mat{P}^\mathcal{N}\mat{P}^\mathcal{R}\| =\|\mat{P}^\mathcal{R}\mat{P}^\mathcal{N}\|\,.
\end{equation}
Then, using the fact that $\transpose{\matH{\Lambda}} \matH{\Lambda} = \identity$ and $\transpose{\matH{\Sigma}} \matH{\Sigma} = \identity$ yields the inequalities $\|\mat{P}^\mathcal{N}\mat{P}^\mathcal{R}\| = \|\matH{\Lambda} \transpose{\matH{\Lambda}} \matH{\Sigma} \transpose{\matH{\Sigma}}\| \leq \|\transpose{\matH{\Lambda}} \matH{\Sigma} \|$ and $\|\transpose{\matH{\Lambda}} \matH{\Sigma} \| = \|\transpose{\matH{\Lambda}} \matH{\Lambda} \transpose{\matH{\Lambda}} \matH{\Sigma} \transpose{\matH{\Sigma}} \matH{\Sigma}\| =\|\transpose{\matH{\Lambda}} \mat{P}^\mathcal{N}\mat{P}^\mathcal{R} \matH{\Sigma} \| \leq \|\mat{P}^\mathcal{N}\mat{P}^\mathcal{R} \|$, which show that $ \|\mat{P}^\mathcal{N}\mat{P}^\mathcal{R}\| = \|\transpose{\matH{\Lambda}} \matH{\Sigma} \|$. From \eqref{eq:costhetadef_proj}, it follows
\begin{equation} \label{eq:costheta}
	\cos{\theta} =\|\matH{\Lambda}^\mathrm{T} \matH{\Sigma} \|\,. 
\end{equation}
Then, recalling the fact that $\mathcal{R}$ and $\mathcal{N}$ are complementary subspaces in $\mathbb{C}^{N_F}$ (see Appendix \ref{app:appA}), we have $\cos{\theta}<1$ \cite{szyld2006many,ipsen1995angle}. Finally, leveraging \eqref{eq:costheta}, it results that $\mat{B^{A^{-1}}_{\Lambda \Sigma}}$ is non-singular.

Next, we derive an upper bound for the condition numbers of $\mat{M_{LS}}$ and $\mat{B^{A^{-1}}_{\Lambda \Sigma}}$. The sine of $\theta$ can be defined as $\sin(\theta) = \left(\| \mat{P^{\Sigma}_{A^{-1}}} \|\right)^{-1}$ \cite{szyld2006many,ipsen1995angle}. Using this identity, we get $\| \mat{O} \| = \cos{\arcsin{({1}/{\|\mat{P^{\Sigma}_{\mat{A}^{-1}}}\|})}} = \sqrt{1 - {\|\mat{P^{\Sigma}_{\mat{A}^{-1}}}\|^{-2}}}$. Noticing that $\mat{M_{LS}}$ is a sum of an identity matrix and a matrix with off-diagonal blocks in which we recall that $\|\mat{O}\|<1$, we obtain 
\begin{equation} 
\begin{alignedat}{2}
 \cond{\mat{M_{LS}}} \leq \frac{1 + \|\mat{O}\|}{1 - \|\mat{O}\|} = \frac{1 + \sqrt{1 - {\|\mat{P^{\Sigma}_{\mat{A}^{-1}}}\|^{-2}}}}{1 - \sqrt{1 - {\|\mat{P^{\Sigma}_{\mat{A}^{-1}}}\|^{-2}}}}\,,
 \end{alignedat}
\end{equation}
which results in the following bound for $\mat{B^{A^{-1}}_{\Lambda \Sigma}}$
\begin{equation} \label{eq:bound_CN_LS}
 \cond{\mat{B^{A^{-1}}_{\Lambda \Sigma}}} = \sqrt{\cond{\mat{M_{LS}}}} \leq \left(\frac{1 + \sqrt{1 - {\|\mat{P^{\Sigma}_{\mat{A}^{-1}}}\|^{-2}}}}{1 - \sqrt{1 - {\|\mat{P^{\Sigma}_{\mat{A}^{-1}}}\|^{-2}}}} \right)^{\frac{1}{2}}.
\end{equation}

Note that $\mat{A}$ could also be a non-singular imaginary symmetric matrix. The procedure to derive \eqref{eq:bound_CN_LS} would remain the same except that the imaginary unit should be extracted from $\mat{A}$ in the derivations. In \Cref{sec:bounds}, the Loop-Star decomposition matrix $\mat{B^{G_\epsilon}_{\Lambda \Sigma}} =\mat{Q_L}$ is employed ($\mat{A}= \mat{G_\epsilon}^{-1}$). Using the fact that the imaginary part ($\sigma / (\omega \epsilon_0 \epsilon_r') \gg 1$) of the complex permittivity $\epsilon$ is dominant in the regime investigated, we can assume that $\mat{G_\epsilon} \approx \mi \Im(\mat{G_{\epsilon}})$ and hence the upper bound for the condition number given in \eqref{eq:bound_CN_LS} can be applied to $\mat{Q_L}$.
\tdg{You assume $\mat A$ real and now you say it's purely imaginary, it's a bit counter intuitive}

\section{Loops Functions on the Boundary of the Object}\label{app:appC}
We show in this appendix that $\mat{Z_{\Phi,11}} \mat{\Lambda} = \mat{0}$. To this aim, we first decompose $\mat{Z_{\Phi,11}}$ as $\mat{Z_{\Phi,11}} = \mat{Z^{\mathrm{full}}_{\Phi,11}} + \mat{Z^{\mathrm{hbf}}_{\Phi,11}}$ where
\begin{equation} \label{eq:Z11_full}
	\begin{split}
		&[\mat{Z^{\mathrm{full}}_{\Phi,11}}]_{m n} = \\
		& \epsilon_0^{-1}\left[ \int_\Omega \nabla \cdot \vt{f_m}(\vt{r}) \int_\Omega G_0(\vr,\vr') \kappa(\vr') \nabla \cdot \vt{f_{n}}(\vt{r'}) \diff v' \diff v \right. \\
		&\left. - \int_{\partial \Omega} \uv{n}_m \cdot \vt{f_m}(\vt{r}) \int_{\Omega} G_0(\vr,\vr') \kappa(\vr')\nabla \cdot \vt{f_{n}}(\vt{r'}) \diff v' \diff s \right]\,,
	\end{split}
\end{equation}
and $[\mat{Z^{\mathrm{hbf}}_{\Phi,11}}]_{m n} = 0$ if $\vt{f_n}$ is a full SWG basis function; instead 
\begin{equation} \label{eq:Z11_hbf}
	\begin{split}
		&[\mat{Z^{\mathrm{hbf}}_{\Phi,11}}]_{m {n}} = \\
		& \epsilon_0^{-1}\left[ \int_\Omega \nabla \cdot \vt{f_m}(\vt{r}) \int_\Omega G_0(\vr,\vr') \kappa(\vr') \nabla \cdot \vt{f_{n}}(\vt{r'}) \diff v' \diff v \right. \\
		&\left. - \int_\Omega \nabla \cdot \vt{f_m}(\vt{r}) \int_{\partial \Omega} G_0(\vr,\vr') \kappa^+_{n} \uv{n}_{n} \cdot \vt{f_{n}}(\vt{r'}) \diff s' \diff v \right.\\ 
		&\left. - \int_{\partial \Omega} \uv{n}_m \cdot \vt{f_m}(\vt{r}) \int_{\Omega} G_0(\vr,\vr') \kappa(\vr')\nabla \cdot \vt{f_{n}}(\vt{r'}) \diff v' \diff s \right.\\
		&\left.+ \int_{\partial \Omega} \uv{n}_m \cdot \vt{f_m}(\vt{r}) \int_{\partial \Omega} G_0(\vr,\vr') \kappa^+_{n} \uv{n}_{n} \cdot \vt{f_{n}}(\vt{r'}) \diff s' \diff s \right],
	\end{split}
\end{equation}
and $[\mat{Z^{\mathrm{full}}_{\Phi,11}}]_{m n} = 0$ if $\vt{f_n}$ is a half SWG basis function.
\tdg{AM: the notation for the half vs full BF int the equations is not very rigorous, why not sthg like $f^f_m$ and $f^h_m$ ?}

The property $\mat{Z^{\mathrm{full}}_{\Phi,11}} \mat{\Lambda} = \mat{0}$ can be verified trivially. However, we need a further analysis to show that $ \mat{Z^{\mathrm{hbf}}_{\Phi,11}} \mat{\Lambda} = \mat{0}$. Instead of directly proving that $\mat{Z^{\mathrm{hbf}}_{\Phi,11}} \mat{\Lambda} = \mat{0}$, we will show that $ \mat{Z^{\mathrm{hbf}}_{\Phi,11}} \mat{P^{\Lambda}_{A^{-1}}} = \mat{0}$, which implies that $\mat{Z^{\mathrm{hbf}}_{\Phi,11}} \mat{\Lambda} = \mat{0}$ ($\mat{\Lambda}$ being a  full-column-rank matrix and $(\transpose{\mat{\Lambda}} \mat{G_\epsilon} \mat{\Lambda})^{-1} \transpose{\mat{\Lambda}} \mat{G_\epsilon} \neq \mat{0} $).
\tdg{AM: I do not see why this a rguments helps.}
Recalling that the basis functions in $\mat{Z^{\mathrm{hbf}}_{\Phi,11}}$ can be reordered such that $
 \mat{Z^{\mathrm{hbf}}_{\Phi,11}} = 
\begin{bmatrix}
 \mat{0}
 & \mat{H} 
\end{bmatrix}$ in which $\mat{H} \in \mathbb{C}^{N_{F} \times N_{eF}}$ contains all the matrix entries defined in \eqref{eq:Z11_hbf} (i.e. columns of $\mat{Z^{\mathrm{hbf}}_{\Phi,11}}$ that are not zero) and that $\mat{\Sigma_s^\mathrm{T}}$ can be rearranged as $\left[\mat{0} ~ -\identity \right]$ following its definition in \eqref{eq:def_stars_hbf}, we can expand $ \mat{Z^{\mathrm{hbf}}_{\Phi,11}} \mat{P^{\Lambda}_{A^{-1}}}$ as
\begin{equation} \label{eq:expansion_Zhbf}
\begin{split}
 \mat{Z^{\mathrm{hbf}}_{\Phi,11}} \mat{P^{\Lambda}_{A^{-1}}}&= 
\begin{bmatrix}
 \mat{0}
 & \rvline & \mat{H} 
\end{bmatrix}
\begin{bmatrix}
 \mat{\identity}
 & \rvline & \mat{0} \\
\hline
 \mat{0} & \rvline &
 \mat{\identity}
\end{bmatrix}
\mat{P^{\Lambda}_{A^{-1}}} \\
&= \begin{bmatrix}
 \mat{0}
 & \rvline & \mat{H} 
\end{bmatrix}
\left[
\begin{array}{c}
\begin{array}{c|c}
\mat{\identity} &
\mat{0} 
\end{array} \\
\hline
- \mat{\Sigma_s}^\mathrm{T}
\end{array}
\right]
\mat{P^{\Lambda}_{A^{-1}}}\\
&= - \mat{H} \mat{\Sigma_s}^\mathrm{T} \mat{P^{\Lambda}_{A^{-1}}}\,.
\end{split} 
\end{equation}
Therefore, $\mat{\Sigma_s}^{\mathrm{T}} \mat{P^{\Lambda}_{A^{-1}}} = \mat{0}$ leads to $\mat{Z^{\mathrm{hbf}}_{\Phi,11}} \mat{P^{\Lambda}_{A^{-1}}} = \mat{0}$. Given that the expansion of $\mat{\Sigma}^{\mathrm{T}} \mat{P^{\Lambda}_{A^{-1}}}$ results in
\begin{equation}
\begin{split}
 \mat{{\Sigma}}^{\mathrm{T}} \mat{P^{\Lambda}_{A^{-1}}}
 &= \mat{{\Sigma}}^{\mathrm{T}} \left( \identity -\mat{P^{\Sigma}_{A^{-1}}} \right) \\
 &= \mat{{\Sigma}}^{\mathrm{T}} - \mat{{\Sigma}}^{\mathrm{T}} \mat{A} \mat{{\Sigma}} \left( \mat{\Sigma}^{\mathrm{T}} \mat{A} \mat{\Sigma} \right)^+ \mat{{\Sigma}}^{\mathrm{T}} \\
 &= \mat{{\Sigma}}^{\mathrm{T}} - \mat{{\Sigma}}^{\mathrm{T}} = \mat{0}\,,
\end{split}
\end{equation}
\tdg{AM: This is again not the pinv property}
and that the property $\mat{\Sigma}^{\mathrm{T}} \mat{P^{\Lambda}_{A^{-1}}} = \left[ \mat{\Sigma_v}~ \mat{\Sigma_s}\right]^\mathrm{T} \mat{P^{\Lambda}_{A^{-1}}} = \mat{0}$ implies that $\mat{\Sigma_s}^{\mathrm{T}} \mat{P^{\Lambda}_{A^{-1}}} = \mat{0}$ since $\mat{\Sigma_s}^{\mathrm{T}}$ is of the form $\left[\mat{0}~ - \identity \right]$, we obtain that $\mat{\Sigma_s}^{\mathrm{T}} \mat{P^{\Lambda}_{A^{-1}}} = \mat{0}$. Finally, leveraging \eqref{eq:expansion_Zhbf}, it results that $ \mat{Z^{\mathrm{hbf}}_{\Phi,11}} \mat{P^{\Lambda}_{A^{-1}}} = \mat{0}$, which in turn gives $\mat{Z_{\Phi,11}} \mat{\Lambda} = \mat{0}$.


\section{Norms of $\mat{P^{\Lambda}_{G_\epsilon}}$ and $\mat{P^{\Sigma}_{G_\epsilon}}$ when the Conductivity Contrast Goes to Infinity}\label{app:appD}

In this appendix, we prove that the norms of the scaled projectors $\mat{P^{\Lambda}_{G_\epsilon}}$ and $\mat{P^{\Sigma}_{G_\epsilon}}$ introduced in \eqref{eq:scaled-projectors-loop-G-ns} and \eqref{eq:scaled-projectors-star-G-ns} scale as $\mathcal{O}(1)$ when $r_\sigma \rightarrow \infty$. To this aim, we leverage the block structure of $\mat{G_\epsilon}$, which is provided in \eqref{eq:block-structure}, to derive the high-contrast behavior of $\mat{P^{\Lambda}_{G_\epsilon}} = \mat{\Lambda} ( \mat{\Lambda}^\mathrm{T} \mat{G_{\epsilon}} \mat{\Lambda} )^{-1} \mat{\Lambda}^\mathrm{T} \mat{G_{\epsilon}}$.  

First, we decompose $\mat{\Lambda}$ into six blocks
\begin{gather} 
\mat{\Lambda}
 = \begin{bmatrix}
 \mat{\Lambda_M} & \mat{\Lambda_{IM}} & \mat{0} \\
  \mat{0} & \mat{\Lambda_{IR}} & \mat{\Lambda_{R}} 
 \end{bmatrix}\,, \text{ in which} \label{eq:lambda_block} \\
  \mat{\Lambda_M}  = \mat{\Lambda}[a_1,\dots, a_{N_{FM}}; b_1,\dots, b_{N_{LM}}] \,,\\
  \mat{\Lambda_R}  = \mat{\Lambda}[a_{N_{FM}+1},\dots, a_{N_F};b_{N_{L} - N_{LR} +1},\dots, b_{N_{L}}]\,, \\
    \mat{\Lambda_{IM}} =  \mat{\Lambda}[a_1,\dots, a_{N_{FM}};b_{N_{LM} + 1},\dots, b_{N_{L} - N_{LR}}]\,,\\
  \mat{\Lambda_{IR}}   = \mat{\Lambda}[a_{N_{FM}+1},\dots, a_{N_F};b_{N_{LM} + 1},\dots, b_{N_{L} - N_{LR}}] \,,
\end{gather}
with $\{b_1,\dots, b_{N_{LM}}\}$, $\{b_{N_{L} - N_{LR} +1},\dots, b_{N_{L}}\}$, and $\{b_{N_{LM} + 1},\dots, b_{N_{L} - N_{LR}}\}$ being the indices of the columns of $ \mat{\Lambda}$ corresponding to the $N_{LM}$, $N_{LR}$, and $N_L - N_{LR} - N_{LM}$ Loop functions made of SWG functions having their support defined in $\Omega_M$, in $\Omega \backslash \Omega_M$, and in both $\Omega \backslash \Omega_M$ and $\Omega_M$, respectively.

From the structure of $\mat{\Lambda}$ in \eqref{eq:lambda_block} and the structure of $\mat{G_\epsilon}$ in \eqref{eq:block-structure}, we can decompose the product $\transpose{\mat{\Lambda}} \mat{G_\epsilon}$ as a 3-by-2 block matrix
\begin{equation} \label{eq:LambdaTGeps}
    \transpose{\mat{\Lambda}} \mat{G_\epsilon} = \begin{bmatrix}
 \transpose{\mat{\Lambda_M}} \mat{G_M} &  \transpose{\mat{\Lambda_M}} \transpose{\mat{B_M}} \\
  \transpose{\mat{\Lambda_{IM}}}  \mat{G_M} +  \transpose{\mat{\Lambda_{IR}}}  {\mat{B_M}} & \transpose{\mat{\Lambda_{IM}}} \transpose{\mat{B_M}} +  \transpose{\mat{\Lambda_{IR}}} \mat{G_R} \\
  \transpose{\mat{\Lambda_R}} {\mat{B_M}} & \transpose{\mat{\Lambda_R}} \mat{G_R}
 \end{bmatrix}\,.
\end{equation}

Then, from the knowledge of the high conductivity ratio behavior of the imaginary part of $\mat{G_\epsilon}$ in \eqref{eq:scaling_Gram_Im}, we obtain that
\td{AM: Why of the imag part only? Re($\mat{G_\epsilon}$(2,2)) is O(1), same as the Im part}
\begin{equation} \label{eq:LambdaTGeps_scaling}
     \transpose{\mat{\Lambda}} \mat{G_\epsilon}\underrel{r_\sigma \to \infty}{=} \begin{bmatrix}
 \mathcal{O}(1/r_\sigma)  &  \mathcal{O}(1/r_\sigma) \\
 \mathcal{O}(1/r_\sigma)  & \mathcal{O}(1) \\
  \mathcal{O}(1/r_\sigma) & \mathcal{O}(1)
 \end{bmatrix}\,,
\end{equation}
and similarly, we obtain the following behavior for $\transpose{\mat{\Lambda}} \mat{G_\epsilon} \mat{\Lambda}$ when $r_\sigma \rightarrow \infty$
\begin{equation} \label{eq:Laplacian_scaling}
    \transpose{\mat{\Lambda}} \mat{G_\epsilon} \mat{\Lambda} \underrel{r_\sigma \to \infty}{=} \left[ \begin{array}{c|c}
 \mathcal{O}(1/r_\sigma) &  \begin{matrix} \mathcal{O}(1/r_\sigma) & \mathcal{O}(1/r_\sigma)\end{matrix}  \\
 \midrule
 \begin{matrix}
 \mathcal{O}(1/r_\sigma) \\
 \mathcal{O}(1/r_\sigma)
 \end{matrix}
 &
    \begin{matrix}
 \mathcal{O}(1) & \mathcal{O}(1)\\
 \mathcal{O}(1) & \mathcal{O}(1) 
 \end{matrix}
 \end{array} \right]\,.
\end{equation}
\tdg{CH: pb pseudo inverse here}
Subsequently, since $\transpose{\mat{\Lambda}} \mat{G_\epsilon} \mat{\Lambda}$ is invertible, we can apply the Schur complement formulas to the blocks delineated in \eqref{eq:Laplacian_scaling} to retrieve the high-contrast behavior for $(\transpose{\mat{\Lambda}} \mat{G_\epsilon} \mat{\Lambda})^{-1}$
\begin{equation} \label{eq:Laplacian_inv_scaling}
    \left(\transpose{\mat{\Lambda}} \mat{G_\epsilon} \mat{\Lambda}\right)^{-1}  \underrel{r_\sigma \to \infty}{=} \left[ \begin{array}{c|c}
 \mathcal{O}(r_\sigma) &  \begin{matrix} \mathcal{O}(1) & \mathcal{O}(1)\end{matrix}  \\
 \midrule
 \begin{matrix}
 \mathcal{O}(1) \\
 \mathcal{O}(1)
 \end{matrix}
 &
    \begin{matrix}
 \mathcal{O}(1) & \mathcal{O}(1)\\
 \mathcal{O}(1) & \mathcal{O}(1) 
 \end{matrix}
 \end{array} \right]\,,
\end{equation}
\tdg{AM: I do not get the same results, can we double check?}
Finally, by combining $\eqref{eq:Laplacian_inv_scaling}$ and \eqref{eq:LambdaTGeps_scaling}, we obtain
\begin{equation}
(\transpose{\mat{\Lambda}} \mat{G_\epsilon} \mat{\Lambda})^{-1} \transpose{\mat{\Lambda}} \mat{G_\epsilon} \underrel{r_\sigma \to \infty}{=} \begin{bmatrix}
 \mathcal{O}(1) &  \mathcal{O}(1)  \\
  \mathcal{O}(1/r_\sigma) &  \mathcal{O}(1) \\
 \mathcal{O}(1/r_\sigma) &  \mathcal{O}(1)   
\end{bmatrix}\,,
\end{equation}
in which none of the blocks is diverging when the maximum conductivity ratio goes to infinity. Therefore, we obtain that $\| \mat{P^{\Lambda}_{G_\epsilon}} \| = \mathcal{O}(1)$ when $r_\sigma \rightarrow \infty$. Note that the result proven here also applies to $\mat{P^{\Sigma}_{G_\epsilon}}$ since $\|\mat{P^{\Lambda}_{G_\epsilon}}\| = \|\mat{P^{\Sigma}_{G_\epsilon}} \|$.

\tdg{AM: I need to read this}

\section{Upper and Lower Bounds for the Singular Values of $\mat{G_\epsilon}$}\label{app:appE}

In the following, the permittivity in $\Omega$ is assumed to be purely imaginary. In that case, we can assume that $\mat{G_\epsilon} \approx \mi \Im(\mat{G_\epsilon}) $ with $\Im(\mat{G_\epsilon})$ being symmetric positive-definite. The minimum and maximum purely imaginary permittivity in the object then read $\epsilon_{\text{min}} = \mi \sigma_{\text{min}} / \omega$ and $\epsilon_{\text{max}} = \mi \sigma_{\text{max}} / \omega$, respectively.
\tdg{AM: is it not in contradiction with the other hypotheses?}
Regarding the minimum singular value of $\mat{G_\epsilon}$, starting from the expression of the diagonal operator of the D-VIE, $(\operator{I}_\epsilon \vt{x})(\vr) = {\vt{x}(\vr)}/{\epsilon(\vr)}$, we define the following quantity
\begin{equation} \label{ea:def_l}
 l_{\text{min}} = \min_{\vt{x}\in L^2(\Omega)^3}\langle \vt{x}(\vr),\frac{\vt{x}(\vr)}{\Im(\epsilon(\vr))} \rangle_\Omega /\langle \vt{x}(\vr),{\vt{x}(\vr)} \rangle_\Omega\,,
\end{equation}
in which we took the imaginary part of $\epsilon$, which is purely imaginary, to handle only real eigenvalues in the following.  
From \eqref{ea:def_l}, it follows that $l_{\text{min}}$ can be bounded from below as $l_{\text{min}} \geq {1}/{\Im(\epsilon_{\text{max}})}$.
%
%
Next, $\vt{x}$ can be approximated as $\vt{x_h}$, a linear combination of $N_F$ SWG basis functions, such that $\vt{x_h}(\vr) = \sum_{i=1}^{N_F} [\mat{x}]_i \vt{f_i}(\vr)$ in which $\mat{x}\in \R^{N_F}$ is a coefficient vector. From the definition of $\vt{x_h}$, it follows that
\begin{gather} \label{eq:cont-disc}
 \langle \vt{x_h}(\vr), \frac{\vt{x_h}(\vr)}{\Im(\epsilon(\vr))}\rangle_\Omega = \transpose{\mat{x}} \Im \left(\mat{G_\epsilon}\right) \mat{x}\,.
\end{gather}
%
%
Then, by normalizing \eqref{eq:cont-disc} with $\langle \vt{x_h}(\vr), {\vt{x_h}(\vr)}\rangle_\Omega = \transpose{\mat{x}} \mat{G} \mat{x}$ and leveraging the Courant–Fischer–Weyl min-max principle, it results that
%
\begin{equation}\label{eq:bound-min-max}
 \frac{\transpose{\mat{x}} \Im(\mat{G_\epsilon}) \mat{x}}{\transpose{\mat{x}} \mat{G} \mat{x}} = \frac{\langle \vt{x_h}(\vr),\frac{\vt{x_h}(\vr)}{\Im(\epsilon(\vr))} \rangle_\Omega} {\langle \vt{x_h}(\vr),{\vt{x_h}(\vr)} \rangle_\Omega} \geq l_{\text{min}} \geq \frac{1}{\Im(\epsilon_\text{max})}\,.
\end{equation}
Note that ${\transpose{\mat{x}} \Im \left(\mat{G_\epsilon}\right) \mat{x}}/{\transpose{\mat{x}} \mat{G} \mat{x}}$ is the generalized Rayleigh quotient associated to the generalized eigenvalue problem $\Im\left(\mat{G_\epsilon}\right) \mat{x} = \lambda_G \mat{G} \mat{x}$ in which $\lambda_G$ are the generalized eigenvalues of $\Im \left(\mat{G_\epsilon}\right)$ and $\mat{G} $. The next step is to find a bound for $\lambda_\text{min}(\Im \left(\mat{G_\epsilon}\right)) = \min (\lambda)$ (eigenvalues associated to the eigenvalue problem $\Im \left(\mat{G_\epsilon}\right) \mat{x} = \lambda \mat{x}$) from \eqref{eq:bound-min-max}. Using the fact that $\lambda_\text{min}(\mat{G}) = \min_{\mat{x}\in \R^{N_F}} ~ ({\transpose{\mat{x}} \mat{G} \mat{x}}) / ({\transpose{\mat{x}} \mat{x}})$ (i.e. Rayleigh principle) allows rewriting \eqref{eq:bound-min-max} as
\begin{equation}
  \frac{\transpose{\mat{x}} \Im\left(\mat{G_\epsilon}\right) \mat{x}}{\transpose{\mat{x}}\mat{x}} \geq \lambda_\text{min}(\mat{G}) \frac{\transpose{\mat{x}} \Im \left(\mat{G_\epsilon}\right) \mat{x}}{\transpose{\mat{x}} \mat{G} \mat{x}} \geq \lambda_\text{min}(\mat{G}) \frac{1}{\Im \left(\epsilon_\text{max}\right)}\,,
\end{equation}
and thus $\lambda_\text{min}\left(\Im\left(\mat{G_\epsilon}\right)\right) = \min_{\mat{x}\in \R^{N_F}} ~ ({\transpose{\mat{x}} \Im\left(\mat{G_\epsilon}\right) \mat{x}})/({\transpose{\mat{x}}\mat{x}}) \geq \lambda_\text{min}(\mat{G})/{\Im\left(\epsilon_\text{max}\right)}$.
Since $\mat{G_\epsilon}$ is symmetric, the inequality $ s_\text{min}(\mat{G_\epsilon}) \geq { s_\text{min}(\mat{G})} \omega /{\sigma_\text{max}}$ also holds true.

Similarly, an upper bound for the maximum singular value of $\mat{G_\epsilon}$ can be derived. Instead of $l_{\text{min}}$, $l_{\text{max}}$ is introduced
\begin{equation} \label{ea:def_l_max}
 l_{\text{max}} = \max_{\vt{x}\in L^2(\Omega)^3}\langle \vt{x}(\vr),\frac{\vt{x}(\vr)}{\Im(\epsilon(\vr))} \rangle_\Omega /\langle \vt{x}(\vr),{\vt{x}(\vr)} \rangle_\Omega \,.
\end{equation}
Then, following the same procedure that was used to derive $s_\text{min}(\mat{G_\epsilon})$, we obtain $s_\text{max}(\mat{G_\epsilon}) \leq { s_\text{max}(\mat{G})} \omega /{\sigma_\text{min}}$.
\tdg{AM: I am not seeing it...}
\tdg{AM: @clement can we see this together?}
%

\section{Bound for the Minimum Singular Value of a Sum of Two Matrices}\label{app:appF}

Given two matrices $\mat{A}$ and $\mat{B}$, with $\mat{A}$ being invertible, if $\|\mat{B}\| < s_\text{min}(\mat{A})$, the condition number of $\mat{A} + \mat{B}$ can be bounded from above as
%
\begin{equation} \label{eq:bound-CN1}
\begin{split}
 \text{cond} (\mat{A}+\mat{B}) &\leq \text{cond}(\mat{A}) \text{cond}(\identity + \mat{A}^{-1} \mat{B}) \\
 &\leq \text{cond}(\mat{A}) \frac{1 + \|\mat{A}^{-1} \mat{B} \|}{1 - \|\mat{A}^{-1} \mat{B} \|}\,.
\end{split}
\end{equation}
\tdg{AM: For the last passage, don't you need that all e.v.s are on the same axis? def cond number + triangular ineq}

Then using the fact that $ \|\mat{A}^{-1} \mat{B} \| \leq \| \mat{B} \| / s_\text{min}(\mat{A})$, inequality \eqref{eq:bound-CN1} becomes
$
  \text{cond} (\mat{A}+\mat{B}) \leq \text{cond}(\mat{A}) \frac{1 + \| \mat{B} \| / s_\text{min}(\mat{A})}{1 - \| \mat{B} \| / s_\text{min}(\mat{A})} = \gamma\,.
$
Since $\text{cond} (\mat{A}+\mat{B}) = {\|\mat{A}+\mat{B}\|}/{ s_\text{min}(\mat{A}+\mat{B})} $, the next step is to find a lower bound for $ \|\mat{A}+\mat{B}\| / \gamma$, which can be expressed as $
 s_\text{min}(\mat{A}+\mat{B}) \geq \|\mat{A}+\mat{B}\| / \gamma \geq |(\|\mat{A}\|-\|\mat{B}\|)| / \gamma$. Subsequently, leveraging the fact that $\|\mat{B}\|\leq s_\text{min}(\mat{A})$, we can bound $|(\|\mat{A}\|-\|\mat{B}\|)|$ from below with $\| \mat{A}\| (1 - 1 / \text{cond}(\mat{A}))$.
Finally, the bound for $ s_\text{min}(\mat{A} + \mat{B})$ reads
$s_\text{min}(\mat{A} + \mat{B}) \geq \|\mat{A}\| \frac{\text{cond}(\mat{A}) - 1}{\text{cond}(\mat{A}) \gamma}$.

\end{appendices}

\bibliographystyle{IEEEtran}
\bibliography{biblio.bib}

\end{document}

%% file: figures/geogebra-swg2.tex
\definecolor{black}{rgb}{0,0,0}
\begin{tikzpicture}[line cap=round,line join=round,>=triangle 45,x=1cm,y=1cm]
\clip(-5.5,-1.7) rectangle (0.5,2);
\fill[line width=1.5pt,color=black,fill=black,fill opacity=0.25] (-3.13,-0.15) -- (-2.43,1.66) -- (-2.59,-1.47) -- cycle;
\fill[line width=1.5pt,color=black,fill=black,fill opacity=0.1] (-5.2,0.6) -- (-2.43,1.66) -- (-3.13,-0.15) -- cycle;
\fill[line width=1.5pt,color=black,fill=black,fill opacity=0.1] (-5.2,0.6) -- (-3.13,-0.15) -- (-2.59,-1.47) -- cycle;
\fill[line width=1.5pt,color=black,fill=black,fill opacity=0.1] (-3.13,-0.15) -- (0.11,0.48) -- (-2.43,1.66) -- cycle;
\fill[line width=1.5pt,color=black,fill=black,fill opacity=0.1] (-2.59,-1.47) -- (0.11,0.48) -- (-3.13,-0.15) -- cycle;
\draw [line width=1.5pt,color=black] (-3.13,-0.15)-- (-2.43,1.66);
\draw [line width=1.5pt,color=black] (-2.43,1.66)-- (-2.59,-1.47);
\draw [line width=1.5pt,color=black] (-2.59,-1.47)-- (-3.13,-0.15);
\draw [line width=1.5pt,color=black] (-5.2,0.6)-- (-2.43,1.66);
\draw [line width=1.5pt,color=black] (-2.43,1.66)-- (-3.13,-0.15);
\draw [line width=1.5pt,color=black] (-3.13,-0.15)-- (-5.2,0.6);
\draw [line width=1.5pt,color=black] (-5.2,0.6)-- (-3.13,-0.15);
\draw [line width=1.5pt,color=black] (-3.13,-0.15)-- (-2.59,-1.47);
\draw [line width=1.5pt,color=black] (-2.59,-1.47)-- (-5.2,0.6);
\draw [line width=1.5pt,color=black] (-3.13,-0.15)-- (0.11,0.48);
\draw [line width=1.5pt,color=black] (0.11,0.48)-- (-2.43,1.66);
\draw [line width=1.5pt,color=black] (-2.43,1.66)-- (-3.13,-0.15);
\draw [line width=1.5pt,color=black] (-2.59,-1.47)-- (0.11,0.48);
\draw [line width=1.5pt,color=black] (0.11,0.48)-- (-3.13,-0.15);
\draw [line width=1.5pt,color=black] (-3.13,-0.15)-- (-2.59,-1.47);
\begin{scriptsize}
\fill [color=black] (-5.2,0.6) circle (3pt);
\draw[color=black] (-5.28,0.92) node {$\vr_n^+$};
\fill [color=black] (-2.43,1.66) circle (3pt);
\fill [color=black] (-2.59,-1.47) circle (3pt);
\fill [color=black] (-3.13,-0.15) circle (3pt);
\fill [color=black] (0.11,0.48) circle (3pt);
\draw[color=black] (0.31,0.72) node {$\vr_n^-$};
\draw[color=black] (-2.7,0.28) node {$\Gamma_n$};
\draw[color=black] (-3.67,0.58) node {$T_n^+$};
\draw[color=black] (-1.63,0.58) node {$T_n^-$};
\end{scriptsize}
\end{tikzpicture}

%% file: figures/bound_theo_sigma_f.tex
%
%
\begin{tikzpicture}

\begin{axis}[%
width=0.6\columnwidth,
at={(0.687in,0.486in)},
scale only axis,
point meta min=115,
point meta max=140,
unbounded coords=jump,
xmode=log,
xmin=1,
xmax=1000000,
xminorticks=true,
xlabel style={font=\color{white!15!black}},
xlabel={Frequency (Hz)},
ymode=log,
ymin=1,
ymax=1000,
yminorticks=true,
ylabel style={font=\color{white!15!black}},
ylabel={Maximum conductivity ratio $r_\sigma$},
axis background/.style={fill=white},
axis x line*=bottom,
axis y line*=left,
legend style={legend cell align=left, align=left, draw=white!15!black},
colormap={mymap}{[1pt] rgb(0pt)=(0.2422,0.1504,0.6603); rgb(1pt)=(0.2444,0.1534,0.6728); rgb(2pt)=(0.2464,0.1569,0.6847); rgb(3pt)=(0.2484,0.1607,0.6961); rgb(4pt)=(0.2503,0.1648,0.7071); rgb(5pt)=(0.2522,0.1689,0.7179); rgb(6pt)=(0.254,0.1732,0.7286); rgb(7pt)=(0.2558,0.1773,0.7393); rgb(8pt)=(0.2576,0.1814,0.7501); rgb(9pt)=(0.2594,0.1854,0.761); rgb(11pt)=(0.2628,0.1932,0.7828); rgb(12pt)=(0.2645,0.1972,0.7937); rgb(13pt)=(0.2661,0.2011,0.8043); rgb(14pt)=(0.2676,0.2052,0.8148); rgb(15pt)=(0.2691,0.2094,0.8249); rgb(16pt)=(0.2704,0.2138,0.8346); rgb(17pt)=(0.2717,0.2184,0.8439); rgb(18pt)=(0.2729,0.2231,0.8528); rgb(19pt)=(0.274,0.228,0.8612); rgb(20pt)=(0.2749,0.233,0.8692); rgb(21pt)=(0.2758,0.2382,0.8767); rgb(22pt)=(0.2766,0.2435,0.884); rgb(23pt)=(0.2774,0.2489,0.8908); rgb(24pt)=(0.2781,0.2543,0.8973); rgb(25pt)=(0.2788,0.2598,0.9035); rgb(26pt)=(0.2794,0.2653,0.9094); rgb(27pt)=(0.2798,0.2708,0.915); rgb(28pt)=(0.2802,0.2764,0.9204); rgb(29pt)=(0.2806,0.2819,0.9255); rgb(30pt)=(0.2809,0.2875,0.9305); rgb(31pt)=(0.2811,0.293,0.9352); rgb(32pt)=(0.2813,0.2985,0.9397); rgb(33pt)=(0.2814,0.304,0.9441); rgb(34pt)=(0.2814,0.3095,0.9483); rgb(35pt)=(0.2813,0.315,0.9524); rgb(36pt)=(0.2811,0.3204,0.9563); rgb(37pt)=(0.2809,0.3259,0.96); rgb(38pt)=(0.2807,0.3313,0.9636); rgb(39pt)=(0.2803,0.3367,0.967); rgb(40pt)=(0.2798,0.3421,0.9702); rgb(41pt)=(0.2791,0.3475,0.9733); rgb(42pt)=(0.2784,0.3529,0.9763); rgb(43pt)=(0.2776,0.3583,0.9791); rgb(44pt)=(0.2766,0.3638,0.9817); rgb(45pt)=(0.2754,0.3693,0.984); rgb(46pt)=(0.2741,0.3748,0.9862); rgb(47pt)=(0.2726,0.3804,0.9881); rgb(48pt)=(0.271,0.386,0.9898); rgb(49pt)=(0.2691,0.3916,0.9912); rgb(50pt)=(0.267,0.3973,0.9924); rgb(51pt)=(0.2647,0.403,0.9935); rgb(52pt)=(0.2621,0.4088,0.9946); rgb(53pt)=(0.2591,0.4145,0.9955); rgb(54pt)=(0.2556,0.4203,0.9965); rgb(55pt)=(0.2517,0.4261,0.9974); rgb(56pt)=(0.2473,0.4319,0.9983); rgb(57pt)=(0.2424,0.4378,0.9991); rgb(58pt)=(0.2369,0.4437,0.9996); rgb(59pt)=(0.2311,0.4497,0.9995); rgb(60pt)=(0.225,0.4559,0.9985); rgb(61pt)=(0.2189,0.462,0.9968); rgb(62pt)=(0.2128,0.4682,0.9948); rgb(63pt)=(0.2066,0.4743,0.9926); rgb(64pt)=(0.2006,0.4803,0.9906); rgb(65pt)=(0.195,0.4861,0.9887); rgb(66pt)=(0.1903,0.4919,0.9867); rgb(67pt)=(0.1869,0.4975,0.9844); rgb(68pt)=(0.1847,0.503,0.9819); rgb(69pt)=(0.1831,0.5084,0.9793); rgb(70pt)=(0.1818,0.5138,0.9766); rgb(71pt)=(0.1806,0.5191,0.9738); rgb(72pt)=(0.1795,0.5244,0.9709); rgb(73pt)=(0.1785,0.5296,0.9677); rgb(74pt)=(0.1778,0.5349,0.9641); rgb(75pt)=(0.1773,0.5401,0.9602); rgb(76pt)=(0.1768,0.5452,0.956); rgb(77pt)=(0.1764,0.5504,0.9516); rgb(78pt)=(0.1755,0.5554,0.9473); rgb(79pt)=(0.174,0.5605,0.9432); rgb(80pt)=(0.1716,0.5655,0.9393); rgb(81pt)=(0.1686,0.5705,0.9357); rgb(82pt)=(0.1649,0.5755,0.9323); rgb(83pt)=(0.161,0.5805,0.9289); rgb(84pt)=(0.1573,0.5854,0.9254); rgb(85pt)=(0.154,0.5902,0.9218); rgb(86pt)=(0.1513,0.595,0.9182); rgb(87pt)=(0.1492,0.5997,0.9147); rgb(88pt)=(0.1475,0.6043,0.9113); rgb(89pt)=(0.1461,0.6089,0.908); rgb(90pt)=(0.1446,0.6135,0.905); rgb(91pt)=(0.1429,0.618,0.9022); rgb(92pt)=(0.1408,0.6226,0.8998); rgb(93pt)=(0.1383,0.6272,0.8975); rgb(94pt)=(0.1354,0.6317,0.8953); rgb(95pt)=(0.1321,0.6363,0.8932); rgb(96pt)=(0.1288,0.6408,0.891); rgb(97pt)=(0.1253,0.6453,0.8887); rgb(98pt)=(0.1219,0.6497,0.8862); rgb(99pt)=(0.1185,0.6541,0.8834); rgb(100pt)=(0.1152,0.6584,0.8804); rgb(101pt)=(0.1119,0.6627,0.877); rgb(102pt)=(0.1085,0.6669,0.8734); rgb(103pt)=(0.1048,0.671,0.8695); rgb(104pt)=(0.1009,0.675,0.8653); rgb(105pt)=(0.0964,0.6789,0.8609); rgb(106pt)=(0.0914,0.6828,0.8562); rgb(107pt)=(0.0855,0.6865,0.8513); rgb(108pt)=(0.0789,0.6902,0.8462); rgb(109pt)=(0.0713,0.6938,0.8409); rgb(110pt)=(0.0628,0.6972,0.8355); rgb(111pt)=(0.0535,0.7006,0.8299); rgb(112pt)=(0.0433,0.7039,0.8242); rgb(113pt)=(0.0328,0.7071,0.8183); rgb(114pt)=(0.0234,0.7103,0.8124); rgb(115pt)=(0.0155,0.7133,0.8064); rgb(116pt)=(0.0091,0.7163,0.8003); rgb(117pt)=(0.0046,0.7192,0.7941); rgb(118pt)=(0.0019,0.722,0.7878); rgb(119pt)=(0.0009,0.7248,0.7815); rgb(120pt)=(0.0018,0.7275,0.7752); rgb(121pt)=(0.0046,0.7301,0.7688); rgb(122pt)=(0.0094,0.7327,0.7623); rgb(123pt)=(0.0162,0.7352,0.7558); rgb(124pt)=(0.0253,0.7376,0.7492); rgb(125pt)=(0.0369,0.74,0.7426); rgb(126pt)=(0.0504,0.7423,0.7359); rgb(127pt)=(0.0638,0.7446,0.7292); rgb(128pt)=(0.077,0.7468,0.7224); rgb(129pt)=(0.0899,0.7489,0.7156); rgb(130pt)=(0.1023,0.751,0.7088); rgb(131pt)=(0.1141,0.7531,0.7019); rgb(132pt)=(0.1252,0.7552,0.695); rgb(133pt)=(0.1354,0.7572,0.6881); rgb(134pt)=(0.1448,0.7593,0.6812); rgb(135pt)=(0.1532,0.7614,0.6741); rgb(136pt)=(0.1609,0.7635,0.6671); rgb(137pt)=(0.1678,0.7656,0.6599); rgb(138pt)=(0.1741,0.7678,0.6527); rgb(139pt)=(0.1799,0.7699,0.6454); rgb(140pt)=(0.1853,0.7721,0.6379); rgb(141pt)=(0.1905,0.7743,0.6303); rgb(142pt)=(0.1954,0.7765,0.6225); rgb(143pt)=(0.2003,0.7787,0.6146); rgb(144pt)=(0.2061,0.7808,0.6065); rgb(145pt)=(0.2118,0.7828,0.5983); rgb(146pt)=(0.2178,0.7849,0.5899); rgb(147pt)=(0.2244,0.7869,0.5813); rgb(148pt)=(0.2318,0.7887,0.5725); rgb(149pt)=(0.2401,0.7905,0.5636); rgb(150pt)=(0.2491,0.7922,0.5546); rgb(151pt)=(0.2589,0.7937,0.5454); rgb(152pt)=(0.2695,0.7951,0.536); rgb(153pt)=(0.2809,0.7964,0.5266); rgb(154pt)=(0.2929,0.7975,0.517); rgb(155pt)=(0.3052,0.7985,0.5074); rgb(156pt)=(0.3176,0.7994,0.4975); rgb(157pt)=(0.3301,0.8002,0.4876); rgb(158pt)=(0.3424,0.8009,0.4774); rgb(159pt)=(0.3548,0.8016,0.4669); rgb(160pt)=(0.3671,0.8021,0.4563); rgb(161pt)=(0.3795,0.8026,0.4454); rgb(162pt)=(0.3921,0.8029,0.4344); rgb(163pt)=(0.405,0.8031,0.4233); rgb(164pt)=(0.4184,0.803,0.4122); rgb(165pt)=(0.4322,0.8028,0.4013); rgb(166pt)=(0.4463,0.8024,0.3904); rgb(167pt)=(0.4608,0.8018,0.3797); rgb(168pt)=(0.4753,0.8011,0.3691); rgb(169pt)=(0.4899,0.8002,0.3586); rgb(170pt)=(0.5044,0.7993,0.348); rgb(171pt)=(0.5187,0.7982,0.3374); rgb(172pt)=(0.5329,0.797,0.3267); rgb(173pt)=(0.547,0.7957,0.3159); rgb(175pt)=(0.5748,0.7929,0.2941); rgb(176pt)=(0.5886,0.7913,0.2833); rgb(177pt)=(0.6024,0.7896,0.2726); rgb(178pt)=(0.6161,0.7878,0.2622); rgb(179pt)=(0.6297,0.7859,0.2521); rgb(180pt)=(0.6433,0.7839,0.2423); rgb(181pt)=(0.6567,0.7818,0.2329); rgb(182pt)=(0.6701,0.7796,0.2239); rgb(183pt)=(0.6833,0.7773,0.2155); rgb(184pt)=(0.6963,0.775,0.2075); rgb(185pt)=(0.7091,0.7727,0.1998); rgb(186pt)=(0.7218,0.7703,0.1924); rgb(187pt)=(0.7344,0.7679,0.1852); rgb(188pt)=(0.7468,0.7654,0.1782); rgb(189pt)=(0.759,0.7629,0.1717); rgb(190pt)=(0.771,0.7604,0.1658); rgb(191pt)=(0.7829,0.7579,0.1608); rgb(192pt)=(0.7945,0.7554,0.157); rgb(193pt)=(0.806,0.7529,0.1546); rgb(194pt)=(0.8172,0.7505,0.1535); rgb(195pt)=(0.8281,0.7481,0.1536); rgb(196pt)=(0.8389,0.7457,0.1546); rgb(197pt)=(0.8495,0.7435,0.1564); rgb(198pt)=(0.86,0.7413,0.1587); rgb(199pt)=(0.8703,0.7392,0.1615); rgb(200pt)=(0.8804,0.7372,0.165); rgb(201pt)=(0.8903,0.7353,0.1695); rgb(202pt)=(0.9,0.7336,0.1749); rgb(203pt)=(0.9093,0.7321,0.1815); rgb(204pt)=(0.9184,0.7308,0.189); rgb(205pt)=(0.9272,0.7298,0.1973); rgb(206pt)=(0.9357,0.729,0.2061); rgb(207pt)=(0.944,0.7285,0.2151); rgb(208pt)=(0.9523,0.7284,0.2237); rgb(209pt)=(0.9606,0.7285,0.2312); rgb(210pt)=(0.9689,0.7292,0.2373); rgb(211pt)=(0.977,0.7304,0.2418); rgb(212pt)=(0.9842,0.733,0.2446); rgb(213pt)=(0.99,0.7365,0.2429); rgb(214pt)=(0.9946,0.7407,0.2394); rgb(215pt)=(0.9966,0.7458,0.2351); rgb(216pt)=(0.9971,0.7513,0.2309); rgb(217pt)=(0.9972,0.7569,0.2267); rgb(218pt)=(0.9971,0.7626,0.2224); rgb(219pt)=(0.9969,0.7683,0.2181); rgb(220pt)=(0.9966,0.774,0.2138); rgb(221pt)=(0.9962,0.7798,0.2095); rgb(222pt)=(0.9957,0.7856,0.2053); rgb(223pt)=(0.9949,0.7915,0.2012); rgb(224pt)=(0.9938,0.7974,0.1974); rgb(225pt)=(0.9923,0.8034,0.1939); rgb(226pt)=(0.9906,0.8095,0.1906); rgb(227pt)=(0.9885,0.8156,0.1875); rgb(228pt)=(0.9861,0.8218,0.1846); rgb(229pt)=(0.9835,0.828,0.1817); rgb(230pt)=(0.9807,0.8342,0.1787); rgb(231pt)=(0.9778,0.8404,0.1757); rgb(232pt)=(0.9748,0.8467,0.1726); rgb(233pt)=(0.972,0.8529,0.1695); rgb(234pt)=(0.9694,0.8591,0.1665); rgb(235pt)=(0.9671,0.8654,0.1636); rgb(236pt)=(0.9651,0.8716,0.1608); rgb(237pt)=(0.9634,0.8778,0.1582); rgb(238pt)=(0.9619,0.884,0.1557); rgb(239pt)=(0.9608,0.8902,0.1532); rgb(240pt)=(0.9601,0.8963,0.1507); rgb(241pt)=(0.9596,0.9023,0.148); rgb(242pt)=(0.9595,0.9084,0.145); rgb(243pt)=(0.9597,0.9143,0.1418); rgb(244pt)=(0.9601,0.9203,0.1382); rgb(245pt)=(0.9608,0.9262,0.1344); rgb(246pt)=(0.9618,0.932,0.1304); rgb(247pt)=(0.9629,0.9379,0.1261); rgb(248pt)=(0.9642,0.9437,0.1216); rgb(249pt)=(0.9657,0.9494,0.1168); rgb(250pt)=(0.9674,0.9552,0.1116); rgb(251pt)=(0.9692,0.9609,0.1061); rgb(252pt)=(0.9711,0.9667,0.1001); rgb(253pt)=(0.973,0.9724,0.0938); rgb(254pt)=(0.9749,0.9782,0.0872); rgb(255pt)=(0.9769,0.9839,0.0805)},
colorbar,
colorbar style={ ylabel=Upper bound condition number}
]

\addplot[%
surf,
shader=flat, draw=black, colormap={mymap}{[1pt] rgb(0pt)=(0.2422,0.1504,0.6603); rgb(1pt)=(0.2444,0.1534,0.6728); rgb(2pt)=(0.2464,0.1569,0.6847); rgb(3pt)=(0.2484,0.1607,0.6961); rgb(4pt)=(0.2503,0.1648,0.7071); rgb(5pt)=(0.2522,0.1689,0.7179); rgb(6pt)=(0.254,0.1732,0.7286); rgb(7pt)=(0.2558,0.1773,0.7393); rgb(8pt)=(0.2576,0.1814,0.7501); rgb(9pt)=(0.2594,0.1854,0.761); rgb(11pt)=(0.2628,0.1932,0.7828); rgb(12pt)=(0.2645,0.1972,0.7937); rgb(13pt)=(0.2661,0.2011,0.8043); rgb(14pt)=(0.2676,0.2052,0.8148); rgb(15pt)=(0.2691,0.2094,0.8249); rgb(16pt)=(0.2704,0.2138,0.8346); rgb(17pt)=(0.2717,0.2184,0.8439); rgb(18pt)=(0.2729,0.2231,0.8528); rgb(19pt)=(0.274,0.228,0.8612); rgb(20pt)=(0.2749,0.233,0.8692); rgb(21pt)=(0.2758,0.2382,0.8767); rgb(22pt)=(0.2766,0.2435,0.884); rgb(23pt)=(0.2774,0.2489,0.8908); rgb(24pt)=(0.2781,0.2543,0.8973); rgb(25pt)=(0.2788,0.2598,0.9035); rgb(26pt)=(0.2794,0.2653,0.9094); rgb(27pt)=(0.2798,0.2708,0.915); rgb(28pt)=(0.2802,0.2764,0.9204); rgb(29pt)=(0.2806,0.2819,0.9255); rgb(30pt)=(0.2809,0.2875,0.9305); rgb(31pt)=(0.2811,0.293,0.9352); rgb(32pt)=(0.2813,0.2985,0.9397); rgb(33pt)=(0.2814,0.304,0.9441); rgb(34pt)=(0.2814,0.3095,0.9483); rgb(35pt)=(0.2813,0.315,0.9524); rgb(36pt)=(0.2811,0.3204,0.9563); rgb(37pt)=(0.2809,0.3259,0.96); rgb(38pt)=(0.2807,0.3313,0.9636); rgb(39pt)=(0.2803,0.3367,0.967); rgb(40pt)=(0.2798,0.3421,0.9702); rgb(41pt)=(0.2791,0.3475,0.9733); rgb(42pt)=(0.2784,0.3529,0.9763); rgb(43pt)=(0.2776,0.3583,0.9791); rgb(44pt)=(0.2766,0.3638,0.9817); rgb(45pt)=(0.2754,0.3693,0.984); rgb(46pt)=(0.2741,0.3748,0.9862); rgb(47pt)=(0.2726,0.3804,0.9881); rgb(48pt)=(0.271,0.386,0.9898); rgb(49pt)=(0.2691,0.3916,0.9912); rgb(50pt)=(0.267,0.3973,0.9924); rgb(51pt)=(0.2647,0.403,0.9935); rgb(52pt)=(0.2621,0.4088,0.9946); rgb(53pt)=(0.2591,0.4145,0.9955); rgb(54pt)=(0.2556,0.4203,0.9965); rgb(55pt)=(0.2517,0.4261,0.9974); rgb(56pt)=(0.2473,0.4319,0.9983); rgb(57pt)=(0.2424,0.4378,0.9991); rgb(58pt)=(0.2369,0.4437,0.9996); rgb(59pt)=(0.2311,0.4497,0.9995); rgb(60pt)=(0.225,0.4559,0.9985); rgb(61pt)=(0.2189,0.462,0.9968); rgb(62pt)=(0.2128,0.4682,0.9948); rgb(63pt)=(0.2066,0.4743,0.9926); rgb(64pt)=(0.2006,0.4803,0.9906); rgb(65pt)=(0.195,0.4861,0.9887); rgb(66pt)=(0.1903,0.4919,0.9867); rgb(67pt)=(0.1869,0.4975,0.9844); rgb(68pt)=(0.1847,0.503,0.9819); rgb(69pt)=(0.1831,0.5084,0.9793); rgb(70pt)=(0.1818,0.5138,0.9766); rgb(71pt)=(0.1806,0.5191,0.9738); rgb(72pt)=(0.1795,0.5244,0.9709); rgb(73pt)=(0.1785,0.5296,0.9677); rgb(74pt)=(0.1778,0.5349,0.9641); rgb(75pt)=(0.1773,0.5401,0.9602); rgb(76pt)=(0.1768,0.5452,0.956); rgb(77pt)=(0.1764,0.5504,0.9516); rgb(78pt)=(0.1755,0.5554,0.9473); rgb(79pt)=(0.174,0.5605,0.9432); rgb(80pt)=(0.1716,0.5655,0.9393); rgb(81pt)=(0.1686,0.5705,0.9357); rgb(82pt)=(0.1649,0.5755,0.9323); rgb(83pt)=(0.161,0.5805,0.9289); rgb(84pt)=(0.1573,0.5854,0.9254); rgb(85pt)=(0.154,0.5902,0.9218); rgb(86pt)=(0.1513,0.595,0.9182); rgb(87pt)=(0.1492,0.5997,0.9147); rgb(88pt)=(0.1475,0.6043,0.9113); rgb(89pt)=(0.1461,0.6089,0.908); rgb(90pt)=(0.1446,0.6135,0.905); rgb(91pt)=(0.1429,0.618,0.9022); rgb(92pt)=(0.1408,0.6226,0.8998); rgb(93pt)=(0.1383,0.6272,0.8975); rgb(94pt)=(0.1354,0.6317,0.8953); rgb(95pt)=(0.1321,0.6363,0.8932); rgb(96pt)=(0.1288,0.6408,0.891); rgb(97pt)=(0.1253,0.6453,0.8887); rgb(98pt)=(0.1219,0.6497,0.8862); rgb(99pt)=(0.1185,0.6541,0.8834); rgb(100pt)=(0.1152,0.6584,0.8804); rgb(101pt)=(0.1119,0.6627,0.877); rgb(102pt)=(0.1085,0.6669,0.8734); rgb(103pt)=(0.1048,0.671,0.8695); rgb(104pt)=(0.1009,0.675,0.8653); rgb(105pt)=(0.0964,0.6789,0.8609); rgb(106pt)=(0.0914,0.6828,0.8562); rgb(107pt)=(0.0855,0.6865,0.8513); rgb(108pt)=(0.0789,0.6902,0.8462); rgb(109pt)=(0.0713,0.6938,0.8409); rgb(110pt)=(0.0628,0.6972,0.8355); rgb(111pt)=(0.0535,0.7006,0.8299); rgb(112pt)=(0.0433,0.7039,0.8242); rgb(113pt)=(0.0328,0.7071,0.8183); rgb(114pt)=(0.0234,0.7103,0.8124); rgb(115pt)=(0.0155,0.7133,0.8064); rgb(116pt)=(0.0091,0.7163,0.8003); rgb(117pt)=(0.0046,0.7192,0.7941); rgb(118pt)=(0.0019,0.722,0.7878); rgb(119pt)=(0.0009,0.7248,0.7815); rgb(120pt)=(0.0018,0.7275,0.7752); rgb(121pt)=(0.0046,0.7301,0.7688); rgb(122pt)=(0.0094,0.7327,0.7623); rgb(123pt)=(0.0162,0.7352,0.7558); rgb(124pt)=(0.0253,0.7376,0.7492); rgb(125pt)=(0.0369,0.74,0.7426); rgb(126pt)=(0.0504,0.7423,0.7359); rgb(127pt)=(0.0638,0.7446,0.7292); rgb(128pt)=(0.077,0.7468,0.7224); rgb(129pt)=(0.0899,0.7489,0.7156); rgb(130pt)=(0.1023,0.751,0.7088); rgb(131pt)=(0.1141,0.7531,0.7019); rgb(132pt)=(0.1252,0.7552,0.695); rgb(133pt)=(0.1354,0.7572,0.6881); rgb(134pt)=(0.1448,0.7593,0.6812); rgb(135pt)=(0.1532,0.7614,0.6741); rgb(136pt)=(0.1609,0.7635,0.6671); rgb(137pt)=(0.1678,0.7656,0.6599); rgb(138pt)=(0.1741,0.7678,0.6527); rgb(139pt)=(0.1799,0.7699,0.6454); rgb(140pt)=(0.1853,0.7721,0.6379); rgb(141pt)=(0.1905,0.7743,0.6303); rgb(142pt)=(0.1954,0.7765,0.6225); rgb(143pt)=(0.2003,0.7787,0.6146); rgb(144pt)=(0.2061,0.7808,0.6065); rgb(145pt)=(0.2118,0.7828,0.5983); rgb(146pt)=(0.2178,0.7849,0.5899); rgb(147pt)=(0.2244,0.7869,0.5813); rgb(148pt)=(0.2318,0.7887,0.5725); rgb(149pt)=(0.2401,0.7905,0.5636); rgb(150pt)=(0.2491,0.7922,0.5546); rgb(151pt)=(0.2589,0.7937,0.5454); rgb(152pt)=(0.2695,0.7951,0.536); rgb(153pt)=(0.2809,0.7964,0.5266); rgb(154pt)=(0.2929,0.7975,0.517); rgb(155pt)=(0.3052,0.7985,0.5074); rgb(156pt)=(0.3176,0.7994,0.4975); rgb(157pt)=(0.3301,0.8002,0.4876); rgb(158pt)=(0.3424,0.8009,0.4774); rgb(159pt)=(0.3548,0.8016,0.4669); rgb(160pt)=(0.3671,0.8021,0.4563); rgb(161pt)=(0.3795,0.8026,0.4454); rgb(162pt)=(0.3921,0.8029,0.4344); rgb(163pt)=(0.405,0.8031,0.4233); rgb(164pt)=(0.4184,0.803,0.4122); rgb(165pt)=(0.4322,0.8028,0.4013); rgb(166pt)=(0.4463,0.8024,0.3904); rgb(167pt)=(0.4608,0.8018,0.3797); rgb(168pt)=(0.4753,0.8011,0.3691); rgb(169pt)=(0.4899,0.8002,0.3586); rgb(170pt)=(0.5044,0.7993,0.348); rgb(171pt)=(0.5187,0.7982,0.3374); rgb(172pt)=(0.5329,0.797,0.3267); rgb(173pt)=(0.547,0.7957,0.3159); rgb(175pt)=(0.5748,0.7929,0.2941); rgb(176pt)=(0.5886,0.7913,0.2833); rgb(177pt)=(0.6024,0.7896,0.2726); rgb(178pt)=(0.6161,0.7878,0.2622); rgb(179pt)=(0.6297,0.7859,0.2521); rgb(180pt)=(0.6433,0.7839,0.2423); rgb(181pt)=(0.6567,0.7818,0.2329); rgb(182pt)=(0.6701,0.7796,0.2239); rgb(183pt)=(0.6833,0.7773,0.2155); rgb(184pt)=(0.6963,0.775,0.2075); rgb(185pt)=(0.7091,0.7727,0.1998); rgb(186pt)=(0.7218,0.7703,0.1924); rgb(187pt)=(0.7344,0.7679,0.1852); rgb(188pt)=(0.7468,0.7654,0.1782); rgb(189pt)=(0.759,0.7629,0.1717); rgb(190pt)=(0.771,0.7604,0.1658); rgb(191pt)=(0.7829,0.7579,0.1608); rgb(192pt)=(0.7945,0.7554,0.157); rgb(193pt)=(0.806,0.7529,0.1546); rgb(194pt)=(0.8172,0.7505,0.1535); rgb(195pt)=(0.8281,0.7481,0.1536); rgb(196pt)=(0.8389,0.7457,0.1546); rgb(197pt)=(0.8495,0.7435,0.1564); rgb(198pt)=(0.86,0.7413,0.1587); rgb(199pt)=(0.8703,0.7392,0.1615); rgb(200pt)=(0.8804,0.7372,0.165); rgb(201pt)=(0.8903,0.7353,0.1695); rgb(202pt)=(0.9,0.7336,0.1749); rgb(203pt)=(0.9093,0.7321,0.1815); rgb(204pt)=(0.9184,0.7308,0.189); rgb(205pt)=(0.9272,0.7298,0.1973); rgb(206pt)=(0.9357,0.729,0.2061); rgb(207pt)=(0.944,0.7285,0.2151); rgb(208pt)=(0.9523,0.7284,0.2237); rgb(209pt)=(0.9606,0.7285,0.2312); rgb(210pt)=(0.9689,0.7292,0.2373); rgb(211pt)=(0.977,0.7304,0.2418); rgb(212pt)=(0.9842,0.733,0.2446); rgb(213pt)=(0.99,0.7365,0.2429); rgb(214pt)=(0.9946,0.7407,0.2394); rgb(215pt)=(0.9966,0.7458,0.2351); rgb(216pt)=(0.9971,0.7513,0.2309); rgb(217pt)=(0.9972,0.7569,0.2267); rgb(218pt)=(0.9971,0.7626,0.2224); rgb(219pt)=(0.9969,0.7683,0.2181); rgb(220pt)=(0.9966,0.774,0.2138); rgb(221pt)=(0.9962,0.7798,0.2095); rgb(222pt)=(0.9957,0.7856,0.2053); rgb(223pt)=(0.9949,0.7915,0.2012); rgb(224pt)=(0.9938,0.7974,0.1974); rgb(225pt)=(0.9923,0.8034,0.1939); rgb(226pt)=(0.9906,0.8095,0.1906); rgb(227pt)=(0.9885,0.8156,0.1875); rgb(228pt)=(0.9861,0.8218,0.1846); rgb(229pt)=(0.9835,0.828,0.1817); rgb(230pt)=(0.9807,0.8342,0.1787); rgb(231pt)=(0.9778,0.8404,0.1757); rgb(232pt)=(0.9748,0.8467,0.1726); rgb(233pt)=(0.972,0.8529,0.1695); rgb(234pt)=(0.9694,0.8591,0.1665); rgb(235pt)=(0.9671,0.8654,0.1636); rgb(236pt)=(0.9651,0.8716,0.1608); rgb(237pt)=(0.9634,0.8778,0.1582); rgb(238pt)=(0.9619,0.884,0.1557); rgb(239pt)=(0.9608,0.8902,0.1532); rgb(240pt)=(0.9601,0.8963,0.1507); rgb(241pt)=(0.9596,0.9023,0.148); rgb(242pt)=(0.9595,0.9084,0.145); rgb(243pt)=(0.9597,0.9143,0.1418); rgb(244pt)=(0.9601,0.9203,0.1382); rgb(245pt)=(0.9608,0.9262,0.1344); rgb(246pt)=(0.9618,0.932,0.1304); rgb(247pt)=(0.9629,0.9379,0.1261); rgb(248pt)=(0.9642,0.9437,0.1216); rgb(249pt)=(0.9657,0.9494,0.1168); rgb(250pt)=(0.9674,0.9552,0.1116); rgb(251pt)=(0.9692,0.9609,0.1061); rgb(252pt)=(0.9711,0.9667,0.1001); rgb(253pt)=(0.973,0.9724,0.0938); rgb(254pt)=(0.9749,0.9782,0.0872); rgb(255pt)=(0.9769,0.9839,0.0805)}, mesh/rows=14]
table[row sep=crcr, point meta=\thisrow{c}] {%
x	y	c\\
1	1	120.749815600025\\
1	1.58489319246111	120.749815600025\\
1	2.51188643150958	120.749815600025\\
1	3.98107170553497	120.749815600025\\
1	6.30957344480193	120.749815600025\\
1	10	120.749815600025\\
1	15.8489319246111	120.749815600025\\
1	25.1188643150958	120.749815600025\\
1	39.8107170553497	120.749815600025\\
1	63.0957344480193	120.749815600025\\
1	100	120.749815600025\\
1	158.489319246111	120.749815600025\\
1	251.188643150958	120.749815600025\\
1	398.107170553497	120.749815600025\\
1	630.957344480193	120.749815600025\\
1	1000	120.749815600025\\
2.89426612471675	1	120.749956547846\\
2.89426612471675	1.58489319246111	120.749956547846\\
2.89426612471675	2.51188643150958	120.749956547846\\
2.89426612471675	3.98107170553497	120.749956547846\\
2.89426612471675	6.30957344480193	120.749956547846\\
2.89426612471675	10	120.749956547846\\
2.89426612471675	15.8489319246111	120.749956547846\\
2.89426612471675	25.1188643150958	120.749956547846\\
2.89426612471675	39.8107170553497	120.749956547846\\
2.89426612471675	63.0957344480193	120.749956547846\\
2.89426612471675	100	120.749956547846\\
2.89426612471675	158.489319246111	120.749956547846\\
2.89426612471675	251.188643150958	120.749956547846\\
2.89426612471675	398.107170553497	120.749956547846\\
2.89426612471675	630.957344480193	120.749956547846\\
2.89426612471675	1000	120.749956547846\\
8.37677640068292	1	120.750364489274\\
8.37677640068292	1.58489319246111	120.750364489274\\
8.37677640068292	2.51188643150958	120.750364489274\\
8.37677640068292	3.98107170553497	120.750364489274\\
8.37677640068292	6.30957344480193	120.750364489274\\
8.37677640068292	10	120.750364489274\\
8.37677640068292	15.8489319246111	120.750364489274\\
8.37677640068292	25.1188643150958	120.750364489274\\
8.37677640068292	39.8107170553497	120.750364489274\\
8.37677640068292	63.0957344480193	120.750364489274\\
8.37677640068292	100	120.750364489274\\
8.37677640068292	158.489319246111	120.750364489274\\
8.37677640068292	251.188643150958	120.750364489274\\
8.37677640068292	398.107170553497	120.750364489274\\
8.37677640068292	630.957344480193	120.750364489274\\
8.37677640068292	1000	120.750364489274\\
24.2446201708233	1	120.751545188091\\
24.2446201708233	1.58489319246111	120.751545188091\\
24.2446201708233	2.51188643150958	120.751545188091\\
24.2446201708233	3.98107170553497	120.751545188091\\
24.2446201708233	6.30957344480193	120.751545188091\\
24.2446201708233	10	120.751545188091\\
24.2446201708233	15.8489319246111	120.751545188091\\
24.2446201708233	25.1188643150958	120.751545188091\\
24.2446201708233	39.8107170553497	120.751545188091\\
24.2446201708233	63.0957344480193	120.751545188091\\
24.2446201708233	100	120.751545188091\\
24.2446201708233	158.489319246111	120.751545188091\\
24.2446201708233	251.188643150958	120.751545188091\\
24.2446201708233	398.107170553497	120.751545188091\\
24.2446201708233	630.957344480193	120.751545188091\\
24.2446201708233	1000	120.751545188091\\
70.1703828670383	1	120.754962509676\\
70.1703828670383	1.58489319246111	120.754962509676\\
70.1703828670383	2.51188643150958	120.754962509676\\
70.1703828670383	3.98107170553497	120.754962509676\\
70.1703828670383	6.30957344480193	120.754962509676\\
70.1703828670383	10	120.754962509676\\
70.1703828670383	15.8489319246111	120.754962509676\\
70.1703828670383	25.1188643150958	120.754962509676\\
70.1703828670383	39.8107170553497	120.754962509676\\
70.1703828670383	63.0957344480193	120.754962509676\\
70.1703828670383	100	120.754962509676\\
70.1703828670383	158.489319246111	120.754962509676\\
70.1703828670383	251.188643150958	120.754962509676\\
70.1703828670383	398.107170553497	120.754962509676\\
70.1703828670383	630.957344480193	120.754962509676\\
70.1703828670383	1000	120.754962509676\\
203.091762090473	1	120.76485369227\\
203.091762090473	1.58489319246111	120.76485369227\\
203.091762090473	2.51188643150958	120.76485369227\\
203.091762090473	3.98107170553497	120.76485369227\\
203.091762090473	6.30957344480193	120.76485369227\\
203.091762090473	10	120.76485369227\\
203.091762090473	15.8489319246111	120.76485369227\\
203.091762090473	25.1188643150958	120.76485369227\\
203.091762090473	39.8107170553497	120.76485369227\\
203.091762090473	63.0957344480193	120.76485369227\\
203.091762090473	100	120.76485369227\\
203.091762090473	158.489319246111	120.76485369227\\
203.091762090473	251.188643150958	120.76485369227\\
203.091762090473	398.107170553497	120.76485369227\\
203.091762090473	630.957344480193	120.76485369227\\
203.091762090473	1000	120.76485369227\\
587.801607227491	1	120.793485968837\\
587.801607227491	1.58489319246111	120.793485968837\\
587.801607227491	2.51188643150958	120.793485968837\\
587.801607227491	3.98107170553497	120.793485968837\\
587.801607227491	6.30957344480193	120.793485968837\\
587.801607227491	10	120.793485968837\\
587.801607227491	15.8489319246111	120.793485968837\\
587.801607227491	25.1188643150958	120.793485968837\\
587.801607227491	39.8107170553497	120.793485968837\\
587.801607227491	63.0957344480193	120.793485968837\\
587.801607227491	100	120.793485968837\\
587.801607227491	158.489319246111	120.793485968837\\
587.801607227491	251.188643150958	120.793485968837\\
587.801607227491	398.107170553497	120.793485968837\\
587.801607227491	630.957344480193	120.793485968837\\
587.801607227491	1000	120.793485968837\\
1701.25427985259	1	120.876393629149\\
1701.25427985259	1.58489319246111	120.876393629149\\
1701.25427985259	2.51188643150958	120.876393629149\\
1701.25427985259	3.98107170553497	120.876393629149\\
1701.25427985259	6.30957344480193	120.876393629149\\
1701.25427985259	10	120.876393629149\\
1701.25427985259	15.8489319246111	120.876393629149\\
1701.25427985259	25.1188643150958	120.876393629149\\
1701.25427985259	39.8107170553497	120.876393629149\\
1701.25427985259	63.0957344480193	120.876393629149\\
1701.25427985259	100	120.876393629149\\
1701.25427985259	158.489319246111	120.876393629149\\
1701.25427985259	251.188643150958	120.876393629149\\
1701.25427985259	398.107170553497	120.876393629149\\
1701.25427985259	630.957344480193	120.876393629149\\
1701.25427985259	1000	120.876393629149\\
4923.88263170674	1	121.116671174217\\
4923.88263170674	1.58489319246111	121.116671174217\\
4923.88263170674	2.51188643150958	121.116671174217\\
4923.88263170674	3.98107170553497	121.116671174217\\
4923.88263170674	6.30957344480193	121.116671174217\\
4923.88263170674	10	121.116671174217\\
4923.88263170674	15.8489319246111	121.116671174217\\
4923.88263170674	25.1188643150958	121.116671174217\\
4923.88263170674	39.8107170553497	121.116671174217\\
4923.88263170674	63.0957344480193	121.116671174217\\
4923.88263170674	100	121.116671174217\\
4923.88263170674	158.489319246111	121.116671174217\\
4923.88263170674	251.188643150958	121.116671174217\\
4923.88263170674	398.107170553497	121.116671174217\\
4923.88263170674	630.957344480193	121.116671174217\\
4923.88263170674	1000	121.116671174217\\
14251.02670303	1	121.814795825343\\
14251.02670303	1.58489319246111	121.814795825343\\
14251.02670303	2.51188643150958	121.814795825343\\
14251.02670303	3.98107170553497	121.814795825343\\
14251.02670303	6.30957344480193	121.814795825343\\
14251.02670303	10	121.814795825343\\
14251.02670303	15.8489319246111	121.814795825343\\
14251.02670303	25.1188643150958	121.814795825343\\
14251.02670303	39.8107170553497	121.814795825343\\
14251.02670303	63.0957344480193	121.814795825343\\
14251.02670303	100	121.814795825343\\
14251.02670303	158.489319246111	121.814795825343\\
14251.02670303	251.188643150958	121.814795825343\\
14251.02670303	398.107170553497	121.814795825343\\
14251.02670303	630.957344480193	121.814795825343\\
14251.02670303	1000	121.814795825343\\
41246.2638290135	1	123.858219520835\\
41246.2638290135	1.58489319246111	123.858219520835\\
41246.2638290135	2.51188643150958	123.858219520835\\
41246.2638290135	3.98107170553497	123.858219520835\\
41246.2638290135	6.30957344480193	123.858219520835\\
41246.2638290135	10	123.858219520835\\
41246.2638290135	15.8489319246111	123.858219520835\\
41246.2638290135	25.1188643150958	123.858219520835\\
41246.2638290135	39.8107170553497	123.858219520835\\
41246.2638290135	63.0957344480193	123.858219520835\\
41246.2638290135	100	123.858219520835\\
41246.2638290135	158.489319246111	123.858219520835\\
41246.2638290135	251.188643150958	123.858219520835\\
41246.2638290135	398.107170553497	nan\\
41246.2638290135	630.957344480193	nan\\
41246.2638290135	1000	nan\\
119377.664171444	1	129.970723181268\\
119377.664171444	1.58489319246111	129.970723181268\\
119377.664171444	2.51188643150958	129.970723181268\\
119377.664171444	3.98107170553497	129.970723181268\\
119377.664171444	6.30957344480193	129.970723181268\\
119377.664171444	10	129.970723181268\\
119377.664171444	15.8489319246111	129.970723181268\\
119377.664171444	25.1188643150958	129.970723181268\\
119377.664171444	39.8107170553497	129.970723181268\\
119377.664171444	63.0957344480193	129.970723181268\\
119377.664171444	100	129.970723181268\\
119377.664171444	158.489319246111	nan\\
119377.664171444	251.188643150958	nan\\
119377.664171444	398.107170553497	nan\\
119377.664171444	630.957344480193	nan\\
119377.664171444	1000	nan\\
345510.729459222	1	149.505236243139\\
345510.729459222	1.58489319246111	149.505236243139\\
345510.729459222	2.51188643150958	149.505236243139\\
345510.729459222	3.98107170553497	149.505236243139\\
345510.729459222	6.30957344480193	149.505236243139\\
345510.729459222	10	149.505236243139\\
345510.729459222	15.8489319246111	149.505236243139\\
345510.729459222	25.1188643150958	149.505236243139\\
345510.729459222	39.8107170553497	nan\\
345510.729459222	63.0957344480193	nan\\
345510.729459222	100	nan\\
345510.729459222	158.489319246111	nan\\
345510.729459222	251.188643150958	nan\\
345510.729459222	398.107170553497	nan\\
345510.729459222	630.957344480193	nan\\
345510.729459222	1000	nan\\
1000000	1	nan\\
1000000	1.58489319246111	nan\\
1000000	2.51188643150958	nan\\
1000000	3.98107170553497	nan\\
1000000	6.30957344480193	nan\\
1000000	10	nan\\
1000000	15.8489319246111	nan\\
1000000	25.1188643150958	nan\\
1000000	39.8107170553497	nan\\
1000000	63.0957344480193	nan\\
1000000	100	nan\\
1000000	158.489319246111	nan\\
1000000	251.188643150958	nan\\
1000000	398.107170553497	nan\\
1000000	630.957344480193	nan\\
1000000	1000	nan\\
};

\addplot [mark=none,violet, line width=2pt,dashed]
table[row sep=crcr]{
1 1e2 \\
1e7 1e2 \\};

\addplot [mark=none,orange, line width=2pt,dashed]
table[row sep=crcr]{
1.2e5 1 \\
1.2e5 1e3 \\};

\end{axis}
\filldraw[color=red!100, fill=red!5, very thick,, opacity=0.6,rounded corners=0.2cm](1.75,2.5) rectangle (3.4,4.2);
\path[draw=white,line join=miter,line cap=butt,line width=0.956pt]
    (1.9,3.3) node[text=red!100,above right] (text837-3-9-9) {Source};
\path[draw=white,line join=miter,line cap=butt,line width=0.956pt]
(2.2,2.9) node[text=red!100,above right] (text837-3-9-9) {loc.};
\filldraw[color=blue!100, fill=blue!5, very thick,, opacity=0.6,rounded corners=0.2cm](3,2.5) rectangle (5,4.2);
\path[draw=white,line join=miter,line cap=butt,line width=0.956pt]
    (3.3,3.1) node[text=blue!100,above right] (text837-3-9-9) {DBS};
\filldraw[color=teal!100, fill=teal!5, very thick,, opacity=0.6,rounded corners=0.2cm](4.2,2.5) rectangle (6.15,4.2);
\path[draw=white,line join=miter,line cap=butt,line width=0.956pt]
    (4.95,3.1) node[text=teal!100,above right] (text837-3-9-9) {KES};
\filldraw[color=purple!100, fill=purple!5, very thick,, opacity=0.6,rounded corners=0.2cm](4.3,2.5) rectangle (4.75,4.2);
\path[draw=white,line join=miter,line cap=butt,line width=0.956pt]
    (4.78,2.87) node[text=purple!100,above right,rotate=90] (text837-3-9-9) {TMS};

\end{tikzpicture}%

%% file: figures/cond_sigma_freq_1d.tex
%
%
\begin{tikzpicture}

\begin{axis}[%
width=0.6\columnwidth,
at={(0.758in,0.576in)},
scale only axis,
unbounded coords=jump,
every outer x axis line/.append style={violet},
every x tick label/.append style={font=\color{violet}},
every x tick/.append style={violet},
xmode=log,
xmin=1,
xmax=1000000,
xminorticks=true,
grid,
grid style={violet!15},
minor tick num=1,
minor x tick num=1,
xlabel style={font=\color{violet}},
xlabel={Frequency (Hz)},
every outer y axis line/.append style={black},
every y tick label/.append style={font=\color{black}},
every y tick/.append style={black},
ymin=10,
ymax=140,
ylabel={Condition number},
axis background/.style={fill=white},
axis x line*=bottom,
axis y line*=left,
legend style={legend cell align=left, align=left, draw=white!15!black}
]
\addplot [line width=1.5pt, color=violet, dashed, mark options={solid, violet}]
  table[row sep=crcr]{%
1	120.749815600025\\
2.89426612471675	120.749956547846\\
8.37677640068292	120.750364489274\\
24.2446201708233	120.751545188091\\
70.1703828670383	120.754962509676\\
203.091762090473	120.76485369227\\
587.801607227491	120.793485968837\\
1701.25427985259	120.876393629149\\
4923.88263170674	121.116671174217\\
14251.02670303	121.814795825343\\
41246.2638290135	123.858219520835\\
119377.664171444	129.970723181268\\
345510.729459222	nan\\
1000000	nan\\
};\label{plot_one}

\addplot [line width=1.5pt, color=violet, mark=o, mark options={solid, violet}]
  table[row sep=crcr]{%
1	16.4672515110062\\
2.89426612471675	16.4204383811487\\
8.37677640068292	16.3493345059601\\
24.2446201708233	16.5309695878317\\
70.1703828670383	16.4022123201404\\
203.091762090473	16.4222455674738\\
587.801607227491	16.4796228543013\\
1701.25427985259	16.4334137834324\\
4923.88263170674	16.4175326505043\\
14251.02670303	16.5387848077961\\
41246.2638290135	16.4800370243043\\
119377.664171444	16.3477736099887\\
345510.729459222	16.5898799428757\\
1000000	16.3995859655385\\
}; \label{plot_two}

\end{axis}

\begin{axis}[%
width=0.6\columnwidth,
at={(0.758in,0.576in)},
scale only axis,
unbounded coords=jump,
every outer x axis line/.append style={orange},
every x tick label/.append style={font=\color{orange}},
every x tick/.append style={orange},
xmode=log,
xmin=1,
xmax=1000,
grid,
grid style={orange!15},
minor tick num=1,
minor x tick num=1,
xminorticks=true,
xlabel style={font=\color{orange}},
xlabel={Maximum conductivity ratio $r_\sigma$},
every outer y axis line/.append style={draw=none},
every y tick/.append style={draw=none},
ymin=10,
ymax=140,
axis x line*=top,
legend style={at={(0.10,0.3)}, anchor=south west, legend cell align=left, align=left, draw=white!15!black}
]

\addlegendimage{/pgfplots/refstyle=plot_one}\addlegendentry{Upper bound - Frequency}
\addlegendimage{/pgfplots/refstyle=plot_two}\addlegendentry{CN obtained - Frequency}

\addplot [line width=1.5pt, color=orange, dashed, mark options={solid, orange}]
  table[row sep=crcr]{%
1	128.426530610987\\
1.58489319246111	128.426530610987\\
2.51188643150958	128.426530610987\\
3.98107170553497	128.426530610987\\
6.30957344480193	128.426530610987\\
10	128.426530610987\\
15.8489319246111	128.426530610987\\
25.1188643150958	128.426530610987\\
39.8107170553497	128.426530610987\\
63.0957344480193	128.426530610987\\
100	128.426530610987\\
158.489319246111	nan\\
251.188643150958	nan\\
398.107170553497	nan\\
630.957344480193	nan\\
1000	nan\\
}; \addlegendentry{Upper bound - $r_\sigma$}

\addplot [line width=1.5pt, color=orange, mark=star, mark options={solid, orange}]
  table[row sep=crcr]{%
1	12.9479219036366\\
1.58489319246111	13.2342169734667\\
2.51188643150958	13.8842396980802\\
3.98107170553497	14.6133300548835\\
6.30957344480193	15.1814163575795\\
10	15.6283360909864\\
15.8489319246111	15.9166326887118\\
25.1188643150958	16.188264779248\\
39.8107170553497	16.2465339267419\\
63.0957344480193	16.411579679113\\
100	16.3773210647567\\
158.489319246111	16.4258811726515\\
251.188643150958	16.5971856019691\\
398.107170553497	16.6451799621649\\
630.957344480193	16.4772021014815\\
1000	16.6377757718555\\
}; \addlegendentry{CN obtained - $r_\sigma$}

\end{axis}
\end{tikzpicture}%

%% file: figures/gp_plot.tex
\definecolor{mycolor1}{rgb}{0.00000,0.44700,0.74100}%
\definecolor{mycolor2}{rgb}{0.85000,0.32500,0.09800}%
\definecolor{mycolor3}{rgb}{0.92900,0.69400,0.12500}%
\definecolor{mycolor4}{rgb}{0.12900,0.9400,0.12500}%

\begin{tikzpicture}
\pgfplotsset{footnotesize,samples=10}

\begin{groupplot}[group style = {group size = 3 by 1, horizontal sep = 50pt}, width = 6.0cm, height = 5.0cm]
\nextgroupplot[title={(a)},legend style = { column sep = 10pt, legend columns = -1, legend to name = grouplegend},xmode=log,
xmin=0.1e-40,
xmax=100000000,
xminorticks=true,
xlabel style={font=\color{white!15!black}},
xlabel={Frequency (Hz)},
ymode=log,
ymin=5,
ymax=5e+19,
yminorticks=true,
ylabel style={font=\color{white!15!black}},
ylabel={Condition number},
grid=both,
grid style={line width=.1pt, draw=gray!10},
xtick={0.1e-40, 1e-27,1e-14 ,1e-5, 1e1, 1000000}]

\addplot [line width=1.5pt, color=mycolor2, mark=o, mark options={solid, mycolor2}]
  table[row sep=crcr]{%
1.77827941003892e-07	1.9851e+18\\
0.00316227766016838	1.7645e+13\\
56.2341325190349	 9.9224e+08\\
1000000	5.5798e+04\\
}; \addlegendentry{Standard D-VIE}

\addplot [line width=1.5pt, color=mycolor4, mark=x, mark options={solid, mycolor4}]
  table[row sep=crcr]{%
1e-40	3.619505340387529e+04\\
1e-27	3.619505340387529e+04\\
1e-14	3.619505340387529e+04\\
1e-11	3.619505340387529e+04\\
1.77827941003892e-07	3.619505340387529e+04\\
0.00316227766016838	3.619505340387529e+04\\
56.2341325190349	3.619505340387529e+04\\
1000000	3.619505340387529e+04\\
}; \addlegendentry{Loop-star D-VIE}

\addplot [line width=1.5pt, color=mycolor1, mark=+, mark options={solid, mycolor1}]
  table[row sep=crcr]{%
1e-40	15.9453860554636\\
1e-27	15.8434161763263\\
1e-14	15.9337209991274\\
1e-11	15.8789357843667\\
1.77827941003892e-07	15.8146017424083\\
0.00316227766016838	15.9076613430996\\
56.2341325190349	15.8356313235237\\
1000000	15.92649919883\\
}; \addlegendentry{Regularized D-VIE $\mat{G_\epsilon}$}

\addplot [line width=1.5pt, color=mycolor3, mark=diamond,dashed, mark options={solid, mycolor3}]
  table[row sep=crcr]{%
1e-40	16.7343838443863\\
1e-27	17.0979690592484\\
1e-14	18.5413733805133\\
1e-11	16.8563856484957\\
1.77827941003892e-07	17.3474147721712\\
0.00316227766016838	17.282564198998\\
56.2341325190349	16.6918818053481\\
1000000	17.0803533596877\\
}; \addlegendentry{Regularized D-VIE $\mat{D}$}

\nextgroupplot[title={(b)},xmin=3.33333333333333,
xmode=linear,
xmax=8,
xminorticks=true,
xlabel style={font=\color{white!15!black}},
xlabel={1/h (\SI{}{\meter^{-1}})},
ymode=log,
ymin=7,
ymax=2.5e7,
yminorticks=true,
ylabel style={font=\color{white!15!black}},
ylabel={Condition number},
grid=both,
grid style={line width=.1pt, draw=gray!10},]

\addplot [line width=1.5pt, color=mycolor2, mark=o, mark options={solid, mycolor2}]
  table[row sep=crcr]{%
8	6243567.40627175\\
7.14285714285714	6568513.92225659\\
6.66666666666667	6321777.66672259\\
5.71428571428571	5729862.02312065\\
5	5249695.10110248\\
3.33333333333333	5437112.10302476\\
};

\addplot [line width=1.5pt, color=mycolor4, mark=x, mark options={solid, mycolor4}]
  table[row sep=crcr]{%
8	3.182765189905446e+05\\
7.14285714285714	2.335320406396675e+05\\
6.66666666666667	3.828918030705945e+05\\
5.71428571428571    1.185353328755122e+05\\
5	1.276293072296501e+05\\
3.33333333333333	1.895248731579215e+04\\
};

\addplot [line width=1.5pt, color=mycolor1, mark=+, mark options={solid, mycolor1}]
  table[row sep=crcr]{%
8	10.4702026815072\\
7.14285714285714	 10.77266252392\\
6.66666666666667	14.419890812820253\\
5.71428571428571	11.825842257206\\
5	10.8840597764\\
3.33333333333333	12.349568597804 \\
};

\addplot [line width=1.5pt, dashed, color=mycolor3, mark=diamond, mark options={solid, mycolor3}]
  table[row sep=crcr]{%
8	13.26307960632\\
7.14285714285714	15.31094615405 \\
6.66666666666667	18.2738363920039\\
5.71428571428571	15.19722618873 \\
5	13.490867452663 \\
3.33333333333333	17.29774614831 \\
};
\nextgroupplot[title={(c)},xmode=log,
xmin=10,
xmax=10^7,
xminorticks=true,
xlabel style={font=\color{white!15!black}},
xlabel={Frequency (Hz)},
ymode=log,
ymin=5,
ymax=5e+09,
yminorticks=true,
ylabel style={font=\color{white!15!black}},
ylabel={Condition number},
grid=both,
grid style={line width=.1pt, draw=gray!10},] 
\addplot [line width=1.5pt, color=mycolor2, mark=o, mark options={solid, mycolor2}]
  table[row sep=crcr]{%
1	2301573924.45596\\
5	500418610.552312\\
10	299354114.681537\\
50	113665746.586258\\
100	62472443.5174948\\
500	13109642.1729509\\
1000	6692383.59079628\\
1500	4542763.62712215\\
5000	1460002.64340187\\
10000	759880.23607655\\
50000	166378.523022797\\
100000	86830.4152246658\\
500000	19863.5851646966\\
1000000	10979.9531781566\\
1500000	7985.58008182501\\
5000000	3728.80325642883\\
10000000	2565.74723951367\\
};

\addplot [line width=1.5pt, color=mycolor4, mark=x, mark options={solid, mycolor4}]
  table[row sep=crcr]{%
1	1.603062117158918e+05\\
5	1.481221250225792e+05\\
10	1.290756764119125e+05\\
50	1.204318177996962e+05\\
100	1.203863741920267e+05\\
500	1.201250861209044e+05\\
1000	1.197856481854059e+05\\
1500	1.195163137490773e+05\\
5000	1.186342706034502e+05\\
10000	1.181500257613528e+05\\
50000	1.168633446463013e+05\\
100000	1.158787097982674e+05\\
500000	1.063694068146925e+05\\
1000000	9.411319269442139e+04\\
1500000	8.483911270336700e+04\\
5000000	6.031207210978344e+04\\
10000000	5.080412866228961e+04\\
};

\addplot [line width=1.5pt, color=mycolor1, mark=+, mark options={solid, mycolor1}]
  table[row sep=crcr]{%
1	13.5825356045716\\
5	13.8839091690168\\
10	14.0786596532754\\
50	14.9826714694881\\
100	14.9952636953412\\
500	15.0735634232054\\
1000	15.1100379384118\\
1500	15.0490802548392\\
5000	15.2022956991191\\
10000	15.2357588651701\\
50000	15.3022291276987\\
100000	15.2999634696425\\
500000	15.3070971154027\\
1000000	15.2793143504228\\
1500000	15.1726303436627\\
5000000	15.0599511650733\\
10000000	14.8970555075692\\
};

\addplot [line width=1.5pt, color=mycolor3, mark=diamond,dashed, mark options={solid, mycolor3}]
  table[row sep=crcr]{%
1	13.2271045828294\\
5	13.161068538931\\
10	13.3983658625684\\
50	15.1980766664188\\
100	15.5355776885939\\
500	14.5036302171652\\
1000	14.5300208114777\\
1500	14.8590639673735\\
5000	15.8535977167755\\
10000	14.7215889218714\\
50000	15.9514696855939\\
100000	14.8233637677733\\
500000	15.1222449432589\\
1000000	14.6131432255965\\
1500000	14.5214670710566\\
5000000	14.2048022123316\\
10000000	15.339292465583\\
};

\end{groupplot}
    \node at ($(group c2r1) + (0,-3.0cm)$) {\ref{grouplegend}}; 
\end{tikzpicture}

%% file: figures/potential_multilayer_sphere.tex
%
%
\definecolor{mycolor1}{rgb}{0.00000,0.44700,0.74100}%
\definecolor{mycolor2}{rgb}{0.85000,0.32500,0.09800}%
\definecolor{mycolor4}{rgb}{0.12900,0.9400,0.12500}%
\begin{tikzpicture}

\begin{axis}[%
width=0.8\columnwidth,
xmin=0,
xmax=150,
xlabel style={font=\color{white!15!black}},
xlabel={Tetrahedron index},
ymin=-12,
ymax=6,
ylabel style={font=\color{white!15!black}},
ylabel={Potential (V)},
grid=both,
grid style={line width=.1pt, draw=gray!10},
axis background/.style={fill=white},
legend style={at={(0.02,0.77)}, anchor=south west, legend cell align=left, align=left, draw=white!15!black}
]
\addplot [line width=1pt, color=mycolor2, mark=o, mark options={solid, mycolor2}]
  table[row sep=crcr]{%
1	-3.05309971666613\\
2	-3.42733569777619\\
3	-2.42749770039954\\
4	-5.99482690455957\\
5	-5.80885055447323\\
6	-5.32880738457117\\
7	-4.18890098546675\\
8	-3.64218381336998\\
9	-4.30419536247055\\
10	-4.51832385215795\\
11	-4.65358862009154\\
12	-5.79514965381022\\
13	-5.93947792069905\\
14	-5.32496209492352\\
15	-7.03277737776239\\
16	-7.3358197610622\\
17	-6.48039755666735\\
18	-7.31858970433815\\
19	-6.90600823295084\\
20	-5.56804193838863\\
21	-7.19497229640447\\
22	-6.94530724525672\\
23	-8.29023424452086\\
24	-8.60231235630568\\
25	-8.17205382841165\\
26	-2.34992673109709\\
27	-2.97854018620306\\
28	-3.30249095917869\\
29	-1.73086459211906\\
30	-1.10358290025244\\
31	-0.44504892637551\\
32	-1.71269891188258\\
33	-1.08194957628539\\
34	-4.30423856402427\\
35	-3.65056731141697\\
36	-2.35871253427728\\
37	-3.00914703973584\\
38	-5.70378050405838\\
39	-5.94783876358415\\
40	-4.97124730398419\\
41	-8.27717527591386\\
42	-8.6397883886992\\
43	-8.632712578541\\
44	-8.41269788135343\\
45	-7.48178740736726\\
46	-6.92602244546003\\
47	-7.43844175830274\\
48	-0.463409878467016\\
49	0.191969745259633\\
50	0.191749619862201\\
51	1.50775623102363\\
52	0.858874664891198\\
53	-9.63282871152364\\
54	-9.65198533314679\\
55	-9.96932762887522\\
56	-2.35756634756318\\
57	-3.00164145752995\\
58	-3.60388736495159\\
59	-4.2916691544084\\
60	-8.49501848330314\\
61	-9.98428758262863\\
62	-9.61159430498377\\
63	-9.93746079274564\\
64	-0.449572921258592\\
65	0.191084458100632\\
66	-1.7167012439769\\
67	-1.08026557451575\\
68	-8.43255608800899\\
69	-1.06854112462211\\
70	-1.7316857570252\\
71	-1.77887915641601\\
72	-1.52170433980524\\
73	-3.56486612748233\\
74	-2.98545387031077\\
75	-2.17924248374605\\
76	-2.62434089471102\\
77	-2.73645743685838\\
78	-3.52714918699213\\
79	-4.11372354185923\\
80	-5.52141208034339\\
81	-4.92248129738676\\
82	-4.60007946775825\\
83	-4.4206460087575\\
84	-5.16612960850935\\
85	-4.48586188352607\\
86	-6.90405722773626\\
87	-6.68637128185821\\
88	-6.02351955977569\\
89	-4.67757896908234\\
90	-4.02847873036305\\
91	-6.28702234319534\\
92	-5.59843832109359\\
93	-6.4478219985849\\
94	-4.80841381891084\\
95	-5.76415973539625\\
96	-5.91330421541288\\
97	-0.384534632557341\\
98	0.18011825720489\\
99	-0.840558995578813\\
100	-0.420234533218102\\
101	0.140276512881438\\
102	0.560049009816757\\
103	-0.401598258136325\\
104	-0.82196531800795\\
105	0.585072461128943\\
106	0.165282771169464\\
107	-0.481985619558992\\
108	-0.767277645531528\\
109	0.155822357553445\\
110	0.423885751162065\\
111	-3.09303905726056\\
112	-3.17436766335461\\
113	-2.43124611249865\\
114	-1.58341976892387\\
115	-2.09071777150614\\
116	-1.02800701715701\\
117	-1.6387966234751\\
118	1.49809794370814\\
119	0.854116812309461\\
120	-7.89367598775029\\
121	-8.10380278518593\\
122	-9.47569944871444\\
123	-9.75398214317226\\
124	1.48888000303849\\
125	0.848611741701034\\
126	-0.443135920944953\\
127	0.190828099625678\\
128	-1.08153770271195\\
129	-1.71223671067007\\
130	-1.08333828926829\\
131	-1.71390224279493\\
132	-0.443506241024843\\
133	0.19713238416657\\
134	-2.35025047797114\\
135	-2.97931796256902\\
136	-3.24343276404489\\
137	-6.0373839185387\\
138	-10.8873718040696\\
139	-11.0673820296438\\
140	-11.4560269472643\\
141	-11.3574548801597\\
142	-0.425980424005461\\
143	-0.676806382462178\\
144	0.190414652053516\\
145	0.853750537859483\\
146	1.50596214479598\\
147	-9.81427271093338\\
148	-9.19857167191166\\
149	-10.0936860503228\\
150	-8.7263400776264\\
};
\addlegendentry{Reference solution}

\addplot [line width=1pt, mark=x, mark options={solid, mycolor4}, only marks]
  table[row sep=crcr]{%
1	-2.98222732212687\\
2	-3.35460511984271\\
3	-2.3530630276477\\
4	-5.92967372174665\\
5	-5.74408893219255\\
6	-5.26254827116824\\
7	-4.12731638432907\\
8	-3.58645371735282\\
9	-4.2464307872281\\
10	-4.45825586799462\\
11	-4.60095784330336\\
12	-5.73765662677592\\
13	-5.88650643530197\\
14	-5.26341121723391\\
15	-6.9772203987623\\
16	-7.27599196787221\\
17	-6.42102038030707\\
18	-7.26573618014617\\
19	-6.8563651933291\\
20	-5.52570101578372\\
21	-7.13073871817867\\
22	-6.88412667585658\\
23	-8.23473107882596\\
24	-8.54456548486642\\
25	-8.11198878565261\\
26	-2.28405285525601\\
27	-2.91545965704609\\
28	-3.23514583392411\\
29	-1.65910348792203\\
30	-1.03114178049905\\
31	-0.374103102627191\\
32	-1.6488946974122\\
33	-1.01674898230182\\
34	-4.25853893646936\\
35	-3.60373991232513\\
36	-2.30241024617513\\
37	-2.9519352238643\\
38	-5.66786909397878\\
39	-5.90750723196365\\
40	-4.93730122673901\\
41	-8.23264233559271\\
42	-8.58914076320319\\
43	-8.59639366887458\\
44	-8.38061894268356\\
45	-7.45690294553108\\
46	-6.88439903530908\\
47	-7.40678533190772\\
48	-0.384678165953727\\
49	0.270326078142209\\
50	0.26249739394953\\
51	1.58088922169531\\
52	0.934312357011515\\
53	-9.58588158162404\\
54	-9.61645070605754\\
55	-9.92878902053727\\
56	-2.3125645949823\\
57	-2.95586827502437\\
58	-3.56751188152977\\
59	-4.25535125135049\\
60	-8.47674861000344\\
61	-9.93397972997738\\
62	-9.55917198993887\\
63	-9.88264005371016\\
64	-0.387934713094973\\
65	0.252018557611067\\
66	-1.66292631340074\\
67	-1.02641097891609\\
68	-8.3711626783164\\
69	-0.9899614754349\\
70	-1.65008995055006\\
71	-1.70095876892192\\
72	-1.44077696468471\\
73	-3.48467841813251\\
74	-2.90035424750191\\
75	-2.10076507089952\\
76	-2.54223424074135\\
77	-2.65580952409909\\
78	-3.45391273973097\\
79	-4.03712477755295\\
80	-5.4534704613707\\
81	-4.84901642256877\\
82	-4.53064871317352\\
83	-4.34716734414896\\
84	-5.0934937190653\\
85	-4.40974683454724\\
86	-6.83739119929347\\
87	-6.62083878002499\\
88	-5.9557600964801\\
89	-4.60431568221757\\
90	-3.95172352122893\\
91	-6.2207156166146\\
92	-5.52771256586986\\
93	-6.38126824488728\\
94	-4.73627290886414\\
95	-5.69869867154993\\
96	-5.85223045961887\\
97	-0.304018164303715\\
98	0.265266447188161\\
99	-0.763274131211212\\
100	-0.340890766890881\\
101	0.221979313037625\\
102	0.640830217473817\\
103	-0.33044437731816\\
104	-0.747599070197659\\
105	0.665264925350267\\
106	0.242218950808225\\
107	-0.39538787766236\\
108	-0.682168545411069\\
109	0.242357872026498\\
110	0.508248229262621\\
111	-3.01976750245361\\
112	-3.09979151152786\\
113	-2.35678914365481\\
114	-1.50783538962589\\
115	-2.01233586986621\\
116	-0.948375279135184\\
117	-1.55808737079611\\
118	1.56302890562943\\
119	0.920535094073341\\
120	-7.83186167133738\\
121	-8.0429841065065\\
122	-9.41906077980889\\
123	-9.69875673932514\\
124	1.54211726662278\\
125	0.904098542716981\\
126	-0.394564443067566\\
127	0.239122131100064\\
128	-1.04134341834348\\
129	-1.67231839685742\\
130	-1.06151981887675\\
131	-1.69333341238276\\
132	-0.410084791515533\\
133	0.231505005020429\\
134	-2.32121452094836\\
135	-2.94748215915124\\
136	-3.21459439530308\\
137	-6.00855516965757\\
138	-10.8511904091908\\
139	-11.0187322729452\\
140	-11.4052483563868\\
141	-11.3167192581186\\
142	-0.344842632806124\\
143	-0.591683120588322\\
144	0.273038971205901\\
145	0.935453002725204\\
146	1.58526430328228\\
147	-9.78960542267306\\
148	-9.17510072060669\\
149	-10.064638533481\\
150	-8.71457878135541\\
};
\addlegendentry{Loop-star D-VIE}

\addplot [line width=1pt, mark=+, mark options={solid, mycolor1}, only marks]
  table[row sep=crcr]{%
1	-2.98222732212687\\
2	-3.35460511984271\\
3	-2.3530630276477\\
4	-5.92967372174665\\
5	-5.74408893219255\\
6	-5.26254827116824\\
7	-4.12731638432907\\
8	-3.58645371735282\\
9	-4.2464307872281\\
10	-4.45825586799462\\
11	-4.60095784330336\\
12	-5.73765662677592\\
13	-5.88650643530197\\
14	-5.26341121723391\\
15	-6.9772203987623\\
16	-7.27599196787221\\
17	-6.42102038030707\\
18	-7.26573618014617\\
19	-6.8563651933291\\
20	-5.52570101578372\\
21	-7.13073871817867\\
22	-6.88412667585658\\
23	-8.23473107882596\\
24	-8.54456548486642\\
25	-8.11198878565261\\
26	-2.28405285525601\\
27	-2.91545965704609\\
28	-3.23514583392411\\
29	-1.65910348792203\\
30	-1.03114178049905\\
31	-0.374103102627191\\
32	-1.6488946974122\\
33	-1.01674898230182\\
34	-4.25853893646936\\
35	-3.60373991232513\\
36	-2.30241024617513\\
37	-2.9519352238643\\
38	-5.66786909397878\\
39	-5.90750723196365\\
40	-4.93730122673901\\
41	-8.23264233559271\\
42	-8.58914076320319\\
43	-8.59639366887458\\
44	-8.38061894268356\\
45	-7.45690294553108\\
46	-6.88439903530908\\
47	-7.40678533190772\\
48	-0.384678165953727\\
49	0.270326078142209\\
50	0.26249739394953\\
51	1.58088922169531\\
52	0.934312357011515\\
53	-9.58588158162404\\
54	-9.61645070605754\\
55	-9.92878902053727\\
56	-2.3125645949823\\
57	-2.95586827502437\\
58	-3.56751188152977\\
59	-4.25535125135049\\
60	-8.47674861000344\\
61	-9.93397972997738\\
62	-9.55917198993887\\
63	-9.88264005371016\\
64	-0.387934713094973\\
65	0.252018557611067\\
66	-1.66292631340074\\
67	-1.02641097891609\\
68	-8.3711626783164\\
69	-0.9899614754349\\
70	-1.65008995055006\\
71	-1.70095876892192\\
72	-1.44077696468471\\
73	-3.48467841813251\\
74	-2.90035424750191\\
75	-2.10076507089952\\
76	-2.54223424074135\\
77	-2.65580952409909\\
78	-3.45391273973097\\
79	-4.03712477755295\\
80	-5.4534704613707\\
81	-4.84901642256877\\
82	-4.53064871317352\\
83	-4.34716734414896\\
84	-5.0934937190653\\
85	-4.40974683454724\\
86	-6.83739119929347\\
87	-6.62083878002499\\
88	-5.9557600964801\\
89	-4.60431568221757\\
90	-3.95172352122893\\
91	-6.2207156166146\\
92	-5.52771256586986\\
93	-6.38126824488728\\
94	-4.73627290886414\\
95	-5.69869867154993\\
96	-5.85223045961887\\
97	-0.304018164303715\\
98	0.265266447188161\\
99	-0.763274131211212\\
100	-0.340890766890881\\
101	0.221979313037625\\
102	0.640830217473817\\
103	-0.33044437731816\\
104	-0.747599070197659\\
105	0.665264925350267\\
106	0.242218950808225\\
107	-0.39538787766236\\
108	-0.682168545411069\\
109	0.242357872026498\\
110	0.508248229262621\\
111	-3.01976750245361\\
112	-3.09979151152786\\
113	-2.35678914365481\\
114	-1.50783538962589\\
115	-2.01233586986621\\
116	-0.948375279135184\\
117	-1.55808737079611\\
118	1.56302890562943\\
119	0.920535094073341\\
120	-7.83186167133738\\
121	-8.0429841065065\\
122	-9.41906077980889\\
123	-9.69875673932514\\
124	1.54211726662278\\
125	0.904098542716981\\
126	-0.394564443067566\\
127	0.239122131100064\\
128	-1.04134341834348\\
129	-1.67231839685742\\
130	-1.06151981887675\\
131	-1.69333341238276\\
132	-0.410084791515533\\
133	0.231505005020429\\
134	-2.32121452094836\\
135	-2.94748215915124\\
136	-3.21459439530308\\
137	-6.00855516965757\\
138	-10.8511904091908\\
139	-11.0187322729452\\
140	-11.4052483563868\\
141	-11.3167192581186\\
142	-0.344842632806124\\
143	-0.591683120588322\\
144	0.273038971205901\\
145	0.935453002725204\\
146	1.58526430328228\\
147	-9.78960542267306\\
148	-9.17510072060669\\
149	-10.064638533481\\
150	-8.71457878135541\\
};
\addlegendentry{Regularized D-VIE $\mat{D}$}

\end{axis}
\end{tikzpicture}%

%% file: figures/sar_full_head_smoother.tex
%
%
\definecolor{mycolor1}{rgb}{0.00000,0.44700,0.74100}%
\definecolor{mycolor2}{rgb}{0.85000,0.32500,0.09800}%
\begin{tikzpicture}

\begin{axis}[%
width=0.8\columnwidth,
xmin=0,
xmax=80,
xlabel style={font=\color{white!15!black}},
xlabel={Voxel index},
ymin=0,
ymax=21.5,
ylabel style={font=\color{white!15!black}},
ylabel={SAR (W/Kg)},
axis background/.style={fill=white},
grid=both,
grid style={line width=.1pt, draw=gray!10},
legend style={at={(0.35,0.8)}, anchor=south west, legend cell align=left, align=left, draw=white!15!black}
]
\addplot [line width=1pt, color=mycolor2, mark=o, mark options={solid, mycolor2}]
  table[row sep=crcr]{%
1	2.68627835043709\\
2	2.13474200499101\\
3	1.44318738328084\\
4	0.852197532620137\\
5	0.6856820224703\\
6	3.58833600518843\\
7	2.64344580389756\\
8	1.78020690111354\\
9	1.11734293852857\\
10	0.699222060358618\\
11	4.8115918793046\\
12	2.75285721904091\\
13	2.07005547086045\\
14	0.993993061779791\\
15	0.67722493847324\\
16	7.87391652419656\\
17	4.37617056235202\\
18	2.53132209918865\\
19	0.874353033609438\\
20	0.695258552583551\\
21	6.74301880103707\\
22	3.74444181441273\\
23	1.2260360856967\\
24	0.895234452152351\\
25	0.799223003950645\\
26	3.65005281746185\\
27	1.91471422554481\\
28	1.15132937815019\\
29	0.9631056031516\\
30	0.984484508240834\\
31	3.34315562315211\\
32	2.01643392908453\\
33	1.1086588244624\\
34	0.724839636339109\\
35	0.72947454943887\\
36	6.48622535735166\\
37	3.11065506304514\\
38	1.18971295560144\\
39	0.738585289232344\\
40	0.686006063074339\\
41	14.6814833111015\\
42	4.12958007095377\\
43	1.5598366675173\\
44	0.831897214142712\\
45	0.984968983133996\\
46	13.1065154407455\\
47	5.41018578506905\\
48	1.66571191811918\\
49	0.821176030005134\\
50	0.835223866650859\\
51	7.61684196403869\\
52	4.15049858717213\\
53	1.25352867257666\\
54	0.952780365289116\\
55	0.947803204523919\\
56	3.31331987773473\\
57	1.68526196145191\\
58	1.03973786417051\\
59	0.903663538674282\\
60	0.968131883089231\\
61	3.30981713999615\\
62	1.86819823021944\\
63	0.981000291425872\\
64	0.791925490142373\\
65	0.896249955109216\\
66	7.51949595505346\\
67	3.32190437022016\\
68	1.40537793275452\\
69	0.783127736676119\\
70	0.941422792721642\\
71	14.3991550820872\\
72	4.1324563284218\\
73	1.58649473618147\\
74	0.866385178665737\\
75	0.991335483044948\\
76	14.6827649242038\\
77	4.53302179713139\\
78	1.47657575905458\\
79	0.908915247035585\\
80	0.972279931462319\\
81	6.11510202047097\\
82	3.64886404681074\\
83	1.17189158578657\\
84	0.916377865790511\\
85	1.00812494415999\\
86	2.86932798421783\\
87	1.76951992084089\\
88	0.928922188365096\\
89	0.817454147390057\\
90	0.924225407264098\\
91	2.95001496926141\\
92	1.48267197378905\\
93	0.903168976401707\\
94	0.701705703025346\\
95	0.775250041113048\\
96	4.92461437860185\\
97	2.55038733416726\\
98	1.16219151714377\\
99	0.768082847025424\\
100	0.821106327900488\\
101	7.62069695367663\\
102	3.01685508620372\\
103	1.63762935787209\\
104	0.872209579258775\\
105	0.8662689844222\\
106	8.65177248598767\\
107	3.4248945595475\\
108	1.43369478415996\\
109	0.910000285183837\\
110	0.859438096277051\\
111	5.07379201430815\\
112	2.92617065554107\\
113	1.17127911995022\\
114	0.812435318025217\\
115	0.841510828180262\\
116	2.799416416382\\
117	1.40936915776328\\
118	0.770956864727081\\
119	0.701440118244136\\
120	0.807287802050995\\
121	2.99681923025463\\
122	1.44408894929486\\
123	0.85846414762988\\
124	0.678362536531299\\
125	0.695902463274409\\
126	3.22056701246605\\
127	1.75184112320085\\
128	0.929950763546039\\
129	0.69244624284317\\
130	0.729364812267349\\
131	4.47871924837947\\
132	2.14027298259035\\
133	0.994329700482453\\
134	0.739969720596865\\
135	0.719667432348669\\
136	4.52646908243725\\
137	2.15596451350123\\
138	1.08256243749931\\
139	0.761304541181802\\
140	0.780211892607155\\
141	3.59591855419133\\
142	1.73781081222148\\
143	0.948876531351682\\
144	0.742172562203791\\
145	0.772695566913767\\
146	2.32537995346184\\
147	1.36907339917415\\
148	0.952681276824838\\
149	0.697812678232711\\
150	0.746846400437305\\
};
\addlegendentry{Reference solution}

\addplot [line width=1pt, color=mycolor1, mark=+, mark options={solid, mycolor1}]
  table[row sep=crcr]{%
1	2.54299162281114\\
2	2.04335437582043\\
3	1.40002890770198\\
4	0.810180417652701\\
5	0.713281936626006\\
6	3.36523102015739\\
7	2.71727581574906\\
8	1.6362094235704\\
9	1.07986815906241\\
10	0.682554455634876\\
11	4.51189422956606\\
12	2.69074240198957\\
13	2.01191487414136\\
14	0.983055409243363\\
15	0.661261118783755\\
16	7.8928398805106\\
17	4.16642980057854\\
18	2.45217599122919\\
19	0.845249927872905\\
20	0.639650738189238\\
21	6.4519107056499\\
22	3.68337787471749\\
23	1.15599885951408\\
24	0.852740116403865\\
25	0.764514233863388\\
26	3.55317599856865\\
27	1.91441234329399\\
28	1.13385045141596\\
29	0.916452417845227\\
30	0.918738509129953\\
31	3.25105022045517\\
32	1.89576777794154\\
33	1.05792667872275\\
34	0.681982670182727\\
35	0.685771270228822\\
36	6.09106928710828\\
37	3.04272810190499\\
38	1.12549508385757\\
39	0.69585748853065\\
40	0.618558586417913\\
41	14.5523672286761\\
42	3.8912363979243\\
43	1.38701291558737\\
44	0.791111372004014\\
45	0.931866712038973\\
46	13.2535914537344\\
47	4.86629378766948\\
48	1.50506227310548\\
49	0.752324348521528\\
50	0.832772615838031\\
51	7.09285831091188\\
52	4.14437679935076\\
53	1.0887438771284\\
54	0.881527893778032\\
55	0.881990726840336\\
56	3.23597997369148\\
57	1.61749173400315\\
58	0.949674360368512\\
59	0.850820584304386\\
60	0.915842113291593\\
61	3.17816355615005\\
62	1.85977202971128\\
63	0.933997062520226\\
64	0.748480024806142\\
65	0.857567940161466\\
66	7.17053349143614\\
67	3.21911127113688\\
68	1.2720451842692\\
69	0.758571539004854\\
70	0.883589082372019\\
71	14.1391525383528\\
72	4.36282316902906\\
73	1.41732798847812\\
74	0.876399252302321\\
75	0.972787266734916\\
76	14.4777306016537\\
77	4.81774955167525\\
78	1.57754321833995\\
79	0.84482276577371\\
80	0.92496660622832\\
81	6.24087348361095\\
82	3.53316096482361\\
83	1.20194245306574\\
84	0.88031500387213\\
85	0.96100339489159\\
86	2.75148437649361\\
87	1.69788873768488\\
88	0.906732027673641\\
89	0.764547416284008\\
90	0.880687663040877\\
91	2.77190294998163\\
92	1.47917444077909\\
93	0.924000141775819\\
94	0.71747715294324\\
95	0.741583352229533\\
96	4.79541150022192\\
97	2.54816151239088\\
98	1.18874161495288\\
99	0.735787102643029\\
100	0.838296595911117\\
101	7.79065572938158\\
102	2.74320204926782\\
103	1.4840551158005\\
104	0.852863638690947\\
105	0.845153359289666\\
106	8.35700388950942\\
107	3.62021445056788\\
108	1.33320388546883\\
109	0.829447870258306\\
110	0.840530154377569\\
111	4.85474464683355\\
112	3.0824689899481\\
113	1.16960283789418\\
114	0.74531700910178\\
115	0.782219381221305\\
116	2.68446698351077\\
117	1.33017317308266\\
118	0.74831004712056\\
119	0.69336345964093\\
120	0.774047187175623\\
121	2.99176685215837\\
122	1.47322785709868\\
123	0.880179709006587\\
124	0.689349965166981\\
125	0.674438486164251\\
126	3.28795427731917\\
127	1.83770285310329\\
128	1.00355522402174\\
129	0.699740039221795\\
130	0.703564081910164\\
131	4.47561849788381\\
132	2.19672277322172\\
133	0.98246175548884\\
134	0.739490341352603\\
135	0.688070095549163\\
136	4.62070009363419\\
137	2.12733912912637\\
138	1.12689066044355\\
139	0.759007305895113\\
140	0.776700157524692\\
141	3.72353616778325\\
142	1.778399936981\\
143	0.977613102666044\\
144	0.753039812022311\\
145	0.776443413968826\\
146	2.36353468620443\\
147	1.40948889447548\\
148	0.977701151569873\\
149	0.730117896998325\\
150	0.759161420834528\\
};
\addlegendentry{Regularized D-VIE}
            \node [above right] at (rel axis cs:0.01,0.5) {\includegraphics[width=2.5cm]{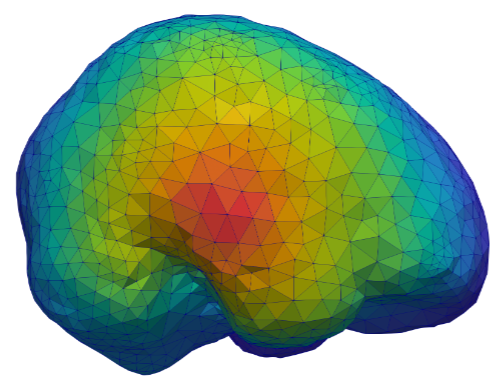}};
\end{axis}
\end{tikzpicture}%